\documentclass[12pt,a4paper]{article}
\usepackage[utf8x]{inputenc}
\usepackage{amsmath}      
\usepackage{amssymb}      
\usepackage{graphicx}
\usepackage[left=1.5cm,right=1.5cm,top=2cm,bottom=2cm]{geometry}
\usepackage[numbers,sort&compress]{natbib}
\usepackage{bibunits}

\usepackage{caption} 

\usepackage[raggedright]{titlesec} 

\usepackage[bookmarks=true,colorlinks=true,citecolor=blue,linkcolor=blue,urlcolor=magenta]{hyperref}

\usepackage{color}

\title{\textbf{Axion mechanism of Sun luminosity,\\dark matter and extragalactic background light}}

\author{V.D. Rusov$^1$\footnote{Corresponding author: Vitaliy D. Rusov, E-mail: siiis@te.net.ua}, I.V. Sharf$^1$, V.A. Tarasov$^1$, M.V. Eingorn$^{1,2}$, V.P. Smolyar$^1$,\\ D.S. Vlasenko$^1$, T.N. Zelentsova$^1$, E.P. Linnik$^1$,  M.E. Beglaryan$^1$}

\date{}

\begin{document}

\captionsetup{font=footnotesize}

\maketitle

%

\begin{center}
$^1$\textit{Department of Theoretical and Experimental Nuclear Physics, \\Odessa National Polytechnic University, 1 Shevchenko ave., Odessa 65044, Ukraine}

\vspace{0.5cm}

$^2$\textit{North Carolina Central University, \\1801 Fayetteville st., Durham, North Carolina 27707, USA}
\end{center}

\abstract{We show the existence of the strong inverse correlation between the temporal variations of the toroidal component of the magnetic field in the solar tachocline (the bottom of the convective zone) and the Earth magnetic field (the Y-component). The possibility that the hypothetical solar axions, which can transform into photons in external electric or magnetic fields (the inverse Primakoff effect), can be the instrument by which the magnetic field of the Sun convective zone modulates  the magnetic field of the Earth is considered.

We propose the axion mechanism of Sun luminosity and "solar dynamo -- geodynamo"  connection, where the energy of one of the solar axion flux components emitted in M1 transition  in $^{57}$Fe  nuclei is modulated at first by the magnetic field of the solar tachocline zone (due to the inverse coherent Primakoff effect) and after that is resonantly absorbed in the core of the Earth, thereby playing the role of the energy modulator of the Earth magnetic field. Within the framework of this mechanism estimations of the strength of the axion coupling to a photon ($g_{a \gamma} = 7.07 \cdot 10^{-11} ~GeV^{-1}$), the axion-nucleon coupling ($g _{an} = 3.20 \cdot 10^{-7}$), the axion-electron coupling ($g_{ae} = 5.28 \cdot 10^{-11}$) and the axion mass ($m_a = 17~eV$) have been obtained. It is also shown that the claimed axion parameters do not contradict any known experimental and theoretical model-independent limitations.

We consider the effect of dark matter in the form of 17~eV axions on the extragalactic back-ground light. Our treatment is based on theoretical results by Overduin and Wesson (Phys. Rep. 402 (2004) 267), who described the axion halos as a luminous element of a pressureless perfect fluid in the standard Friedman-Robertson-Walker universe basing on the assumption that axions are clustered in Galactic halos with nonzero velocity dispersions. We find that the spectral intensity $I_{\lambda }(\lambda_0)$ of the extragalactic background radiation from decaying axions ($m_a = 17 ~eV$, $c_{a \gamma \gamma} = 0.02$) as a function of the observed wavelength $\lambda_0$ is in good agreement with the known experimental data for the near ultraviolet, optical and near infrared bands (1500-20000~\AA). In the framework of such approach it is shown that the present density parameter $\Omega_a$ of thermal axions satisfies the inequality $0.12 \leqslant \Omega _a \leqslant 0.25$ and is comparable to the density parameter of dark matter.}

\newpage

\tableofcontents

\begin{bibunit}[unsrt]

\section{Introduction}

\begin{figure}[b!]
    \begin{center}
        \includegraphics[width=10cm]{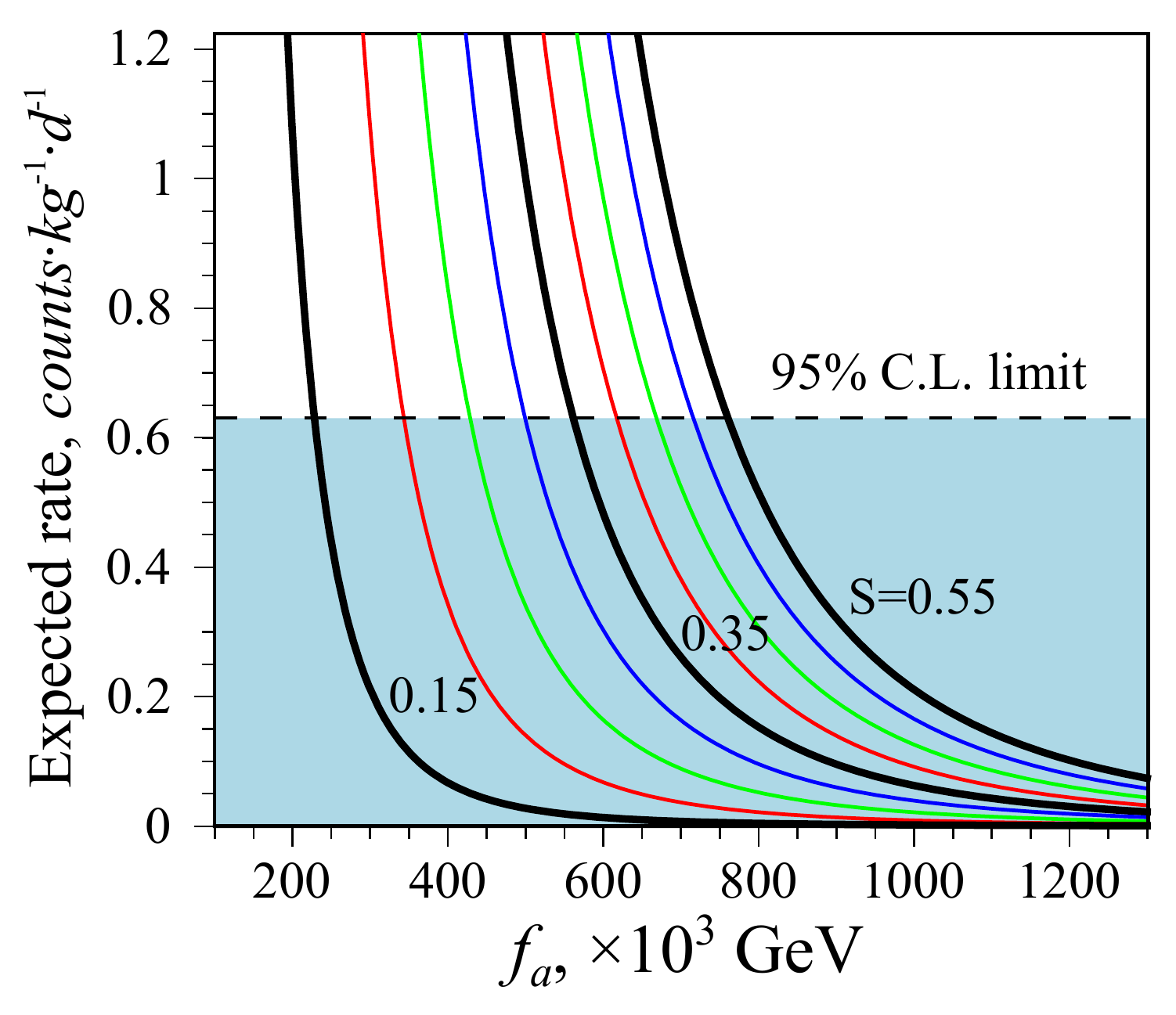}
    \end{center}
\caption{Expected rate in the axion region as a function of the $f_a$ axion constant for different values of the nuclear $S$ parameter. The horizontal line indicates the upper limit obtained in CUORE experiment ($f_a \sim 0.76 \cdot 10^6 ~GeV$ for $S$ = 0.55) \cite{ref001}.}
\label{fig01}
\end{figure}

In the recent paper by Alessandria et al. the results of CUORE experimental search for axions from the solar core from the 14.4~keV M1 ground-state nuclear transition in $^{57}$Fe were presented~\cite{ref001}. The detection technique employed a search for a peak in the energy spectrum at 14.4~keV when an axion is absorbed by an electron via the axio-electric effect. The cross-section for this process is proportional to the photoelectric absorption cross-section for photons. In this pilot experiment 43.65~$kg \cdot d$ of data were analyzed resulting in a lower bound on the Peccei-Quinn energy scale of $f_a \geqslant 0.76 \cdot 10^6 ~GeV$ for the value for the flavor-singlet axial vector matrix element of $S = 0.55$; bounds are presented in the graph for values $0.15 \leqslant S \leqslant 0.55$ (Fig.~\ref{fig01}). With the numbers quoted in the text, the limit on $f_a$ translates into the axion mass limit $m_a < 8 ~eV$, significantly more stringent than in the recent results obtained with $^{57}$Fe detectors \cite{ref002,ref003} and by the Borexino experiment \cite{ref004,ref005}.

Despite the fine and elegant experimental implementation of the idea of detecting the solar axions through the axio-electric effect in TeO$_2$ bolometers (CUORE detection technique~\cite{ref001}), a number of fundamental questions regarding the appropriateness of some assumptions used in the problem statement arises immediately.

\begin{figure}[tb]
    \begin{center}
         \includegraphics[width=18cm]{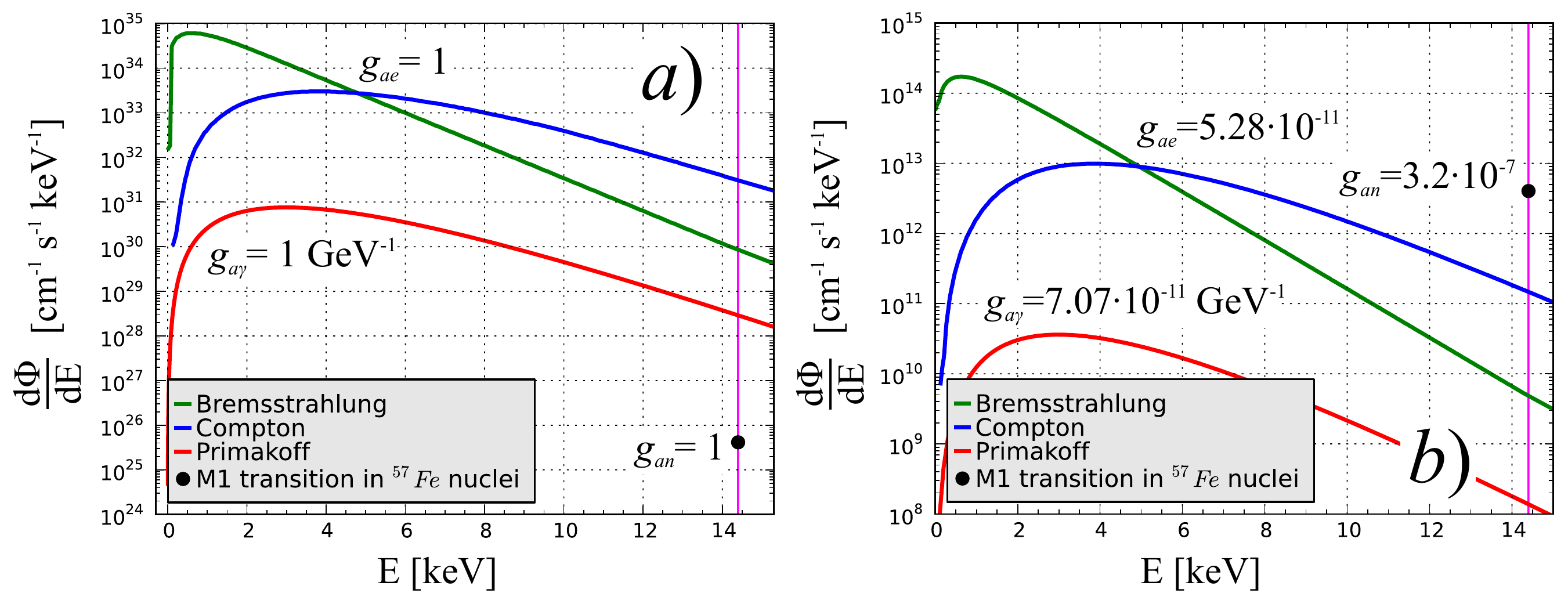}
    \end{center}
\caption{Spectra of solar axions at the ground produced by Primakov effect ($g_{a\gamma}$), M1 ground-state nuclear transition in solar $^{57}$Fe ($g_{an}$), bremsstrahlung ($g_{ae}$, green line) and  the Compton  process ($g_{ae}$, blue line) correspondingly: \textbf{(a)} $g_{a \gamma} = 1$, $g_{an} = 1$ ($S = 0.5$), $g_{ae} = 1$ ($S=0.5$); \textbf{(b)} $g_{a \gamma} = 7.07 \cdot 10^{-11}$~GeV, $g_{an} = 3.2 \cdot 10^{-7}$, $g_{ae} = 5.28 \cdot 10^{-11}$. The data by Derbin et~al.~\cite{ref007} were used in order to plot the bremsstrahlung and Compton spectra. 14.4 keV axions are marked with the pink line.}
\label{fig02}
\end{figure}

The first one is rather obvious and lies in the following. Why is the 14.4~keV M1 ground-state nuclear transition in solar $^{57}$Fe chosen as the main mechanism of solar axions production in CUORE experiment, whereas there are other solar axion production mechanisms discussed in scientific literature in detail which also make their respective contributions into the 14.4~keV axions flux, such as the so called Primakoff effect (e.g.~\cite{ref002,ref003}), bremsstrahlung and the Compton process (e.g.~\cite{ref006,ref007}) (see Fig.~\ref{fig02})? From the analysis of Fig.~\ref{fig02}a, where the spectra of the processes under discussion normalized by the corresponding constants are shown, it follows that this question is absolutely nontrivial, and the answer depends on the knowledge of the values of all these constants simultaneously. In fact, as it will be shown below, the real solar axions spectra may look like the ones depicted on Fig.~\ref{fig02}b. Therefore the question asked above may be reformulated as follows: "What must be the basic physical criterion of the accepted problem statement justification, for example, for the experiment on 14.4~keV solar axions detection?"

In our opinion, one of the most effective ways of establishing such a criterion is the search for the models which would describe some experimentally observed phenomena in the framework of standard or non-standard solar physics using these properties of axions. If such a model is found, then the pivotal estimates of e.g. the axion mass or the upper limits on the axion coupling constants to photons ($g_{a \gamma}$), nucleons ($g_{an}$) and electrons ($g_{ae}$), obtained in the framework of the given model, may play a role of the main physical justification criterion for the accepted problem statement in the 14.4~keV solar axions detection experiment.

\begin{figure}
    \begin{center}
        \includegraphics[width=15cm]{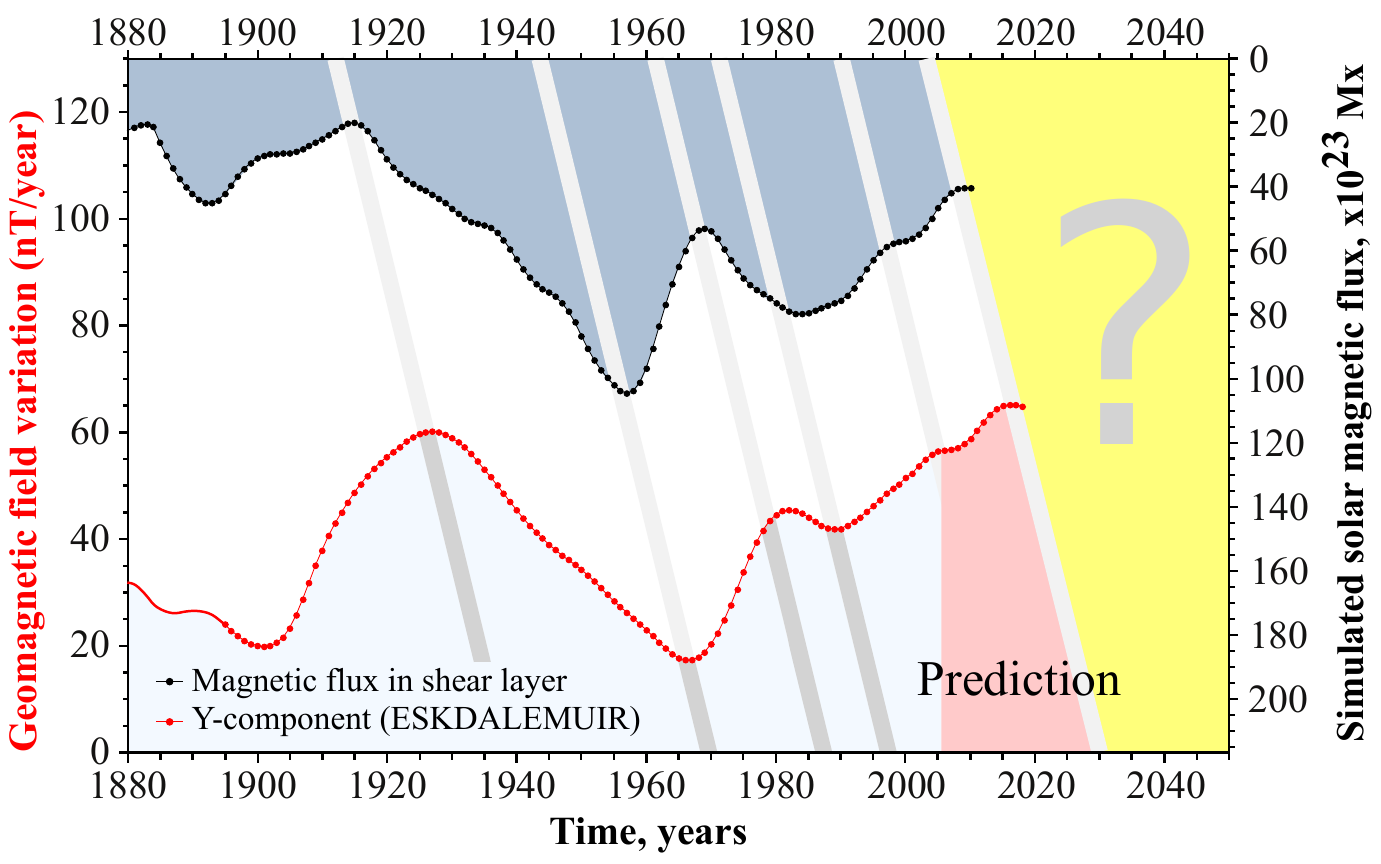}
    \end{center}
\caption{Time evolution of \textbf{(a)} the variations of the magnetic flux at the bottom (the tachocline zone) of the Sun convective zone (see Fig.~7f in Ref.~\cite{ref012}), \textbf{(b)} the geomagnetic field secular variations (the Y-component, nT / year) measured at the Eskdalemuir observatory (England)~\cite{ref013}. Curves are smoothed by the sliding intervals in 5 and 11 years. The pink area is a prediction region.}
\label{fig03}
\end{figure}

In order to justify such a criterion for the future experiments (e.g. CAST, CUORE, XMASS etc.) we decided to create a modified model of the axion mechanism of Sun luminosity\footnote{It should be noted here that the axion mechanism of Sun luminosity, which served as a basis for one of the first axion mass estimates, was described for the first time in 1978 in the paper~\cite{ref008}.} and solar dynamo -- geodynamo  connection, which had been described in our previous paper~\cite{ref009}. The basic idea of such a mechanism, which may be split into two stages for convenience, is the following. At the first stage the solar axions flux variations produced by the previously mentioned processes are modulated by the solar tachocline magnetic field variations through the inverse Primakoff effect~\cite{ref010}. At the second stage the "modulated" solar axion flux travels to the Earth, where its "iron" component containing the 14.4~keV solar axions is resonantly absorbed in the iron-nickel core of the Earth. If the energy of the axions supplied to the Earth core is enough for generation and maintaining the geomagnetic field, then this process will result in a persistent anticorrelation between the variations of the solar magnetic field and the geomagnetic field (the Y-component)\footnote{Note that the strong (inverse) correlation between the temporal variations of the magnetic flux in the tachocline zone and the Earth magnetic field (the Y-component) are observed only for experimental data obtained at that observatories where the temporal variations of declination ($\delta D / \delta t$) or the closely associated east component ($\delta Y / \delta t$) are directly proportional to the westward drift of magnetic features~\cite{ref011}. This condition is very important for understanding of the physical nature of the indicated above correlation since it is known that it is only the motions of the top layers of the Earth's core that are responsible for most magnetic variations and, in particular, for the westward drift of magnetic features seen on the Earth surface on the decade time scale. Europe and Australia are geographical places, where this condition is fulfilled (see Fig.~2 in~\cite{ref011}). For more detailed discussion of this question see below (Section~\ref{sec-02}).}
 (Fig.~\ref{fig03}). This is extremely important, because such effect of anticorrelation was discovered recently~\cite{ref009}, and has a strong experimental basis.

It should be added that the solar axion flux modulated by the inverse Primakoff effect in the magnetic field of the solar tachocline must not only explain the value of solar luminosity, but also describe the solar photon spectrum from the Active Sun, which in its turn must be equivalent to the data from accumulated observations~\cite{ref014}.

Thus, the main purpose of the present report was, on the one hand, to develop a modified axion model of the Sun luminosity and solar dynamo -- geodynamo connection mechanism; and on the other hand, to obtain the consistent estimates for the axion mass and the axion coupling constants to photons ($g_{a \gamma}$), nucleons ($g_{an}$) and electrons ($g_{ae}$) through the comparison and generalization of the model results and the known experiments including CAST, CUORE and XMASS.

\section{Magnetic field of solar tachocline zone and axion mechanism of the solar dynamo -- geodynamo connection}
\label{sec-02}

It is known that in spite of a long history, the nature of the energy source maintaining a convection in the liquid core of the Earth, or more exactly the mechanism of the magnetohydrodynamic dynamo (MHD) generating the magnetic field of the Earth, still has no clear and unambiguous physical interpretation~\cite{ref015,ref016,ref017,ref018,ref019}. The problem is aggravated by the fact that none of the candidates for an energy source of the Earth magnetic-field~\cite{ref015} (secular cooling due to the heat transfer from the core to the mantle, internal heating by radiogenic isotopes, e.g. $^{40}$K, latent heat due to the inner core solidification, compositional buoyancy due to the ejection of light elements at the inner core surface) can in principle explain one of the most remarkable phenomena in solar-terrestrial physics, which consists in strong (inverse) correlation between the temporal variations of the magnetic flux in the tachocline zone (the bottom of the Sun convective zone)~\cite{ref012} and the Earth magnetic field (the Y-component)~\cite{ref013} (Fig.~\ref{fig03}).

\begin{figure}[tb!]
    \begin{center}
        \includegraphics[width=18cm]{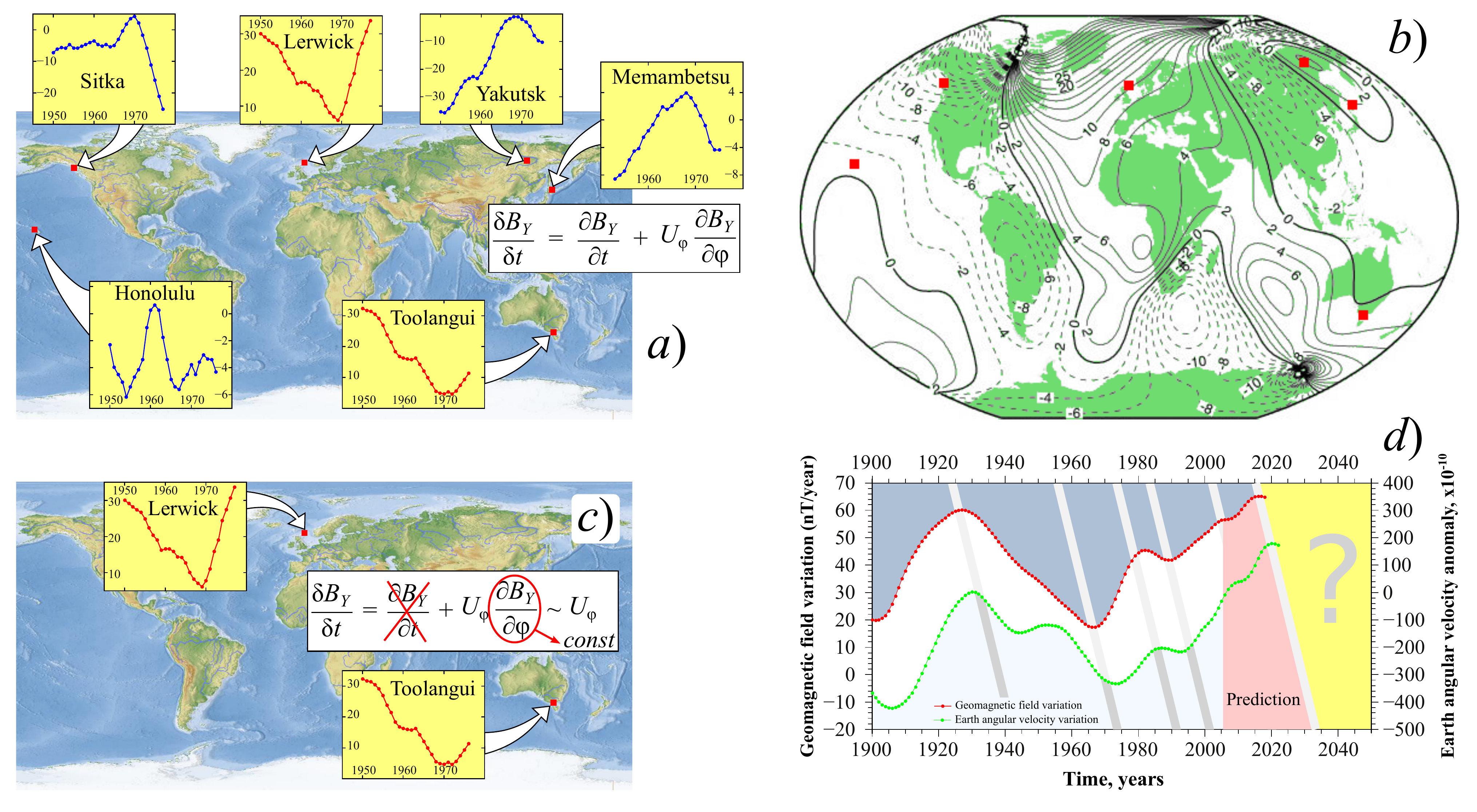}
    \end{center}
\caption{\textbf{(a)} Geomagnetic filed Y-component variations at different observatories~\cite{ref011}; \textbf{(b)} Secular variation of declination for 2005-2010~\cite{ref020} (the closely associated east Y-component of the geomagnetic field). \textbf{(c)} Direct impact of the westward drift on the geomagnetic field Y-component in Europe (e.g. Lerwick) and Australia (e.g. Toolangui). Here $\delta B_Y / \delta t$ is the magnetic variation seen at magnetic observatories; the $\partial B_Y / \partial t$ term accounts for the effects of non-uniform and north-south motions (as well as the effects of magnetic diffusion), $U_{\varphi}$ is a magnitude of the westward drift as seen at the Earth surface  and $\partial B_Y / \partial \varphi$ is the longitudinal gradient of the magnetic field as seen at the surface. \textbf{(d)} Time evolution of the geomagnetic field secular variations (Y-component, $nT / year$),~\cite{ref013} and the variation of the Earth rotation velocity~\cite{ref021} (green line). All curves are smoothed by the sliding intervals in 5 and 11 years. The pink area is a prediction region.}
\label{fig04}
\end{figure}

At the same time, supposing that the transversal (radial) surface area of tachocline zone, through which the magnetic flux passes, is constant in the first approximation, we can assume that magnetic flux variations also describe the temporal variations of the magnetic field in the tachocline zone of the Sun. In this sense, it is obvious that a future candidate for an energy source of the Earth magnetic field must not only play the role of a natural trigger of solar-terrestrial connection, but also directly generate the solar-terrestrial magnetic correlation by its own participation.

At this point a question about the physical nature of such correlation arises. Let us turn to the concept of the westward drift of the Earth magnetic field.
The nondipole part of the main field has a characteristic feature -- it drifts westward with time. The phenomenon of the westward drift was noticed as far back as the XVII century. Each component of the geomagnetic filed has its own drift speed with the average westward drift speed 0.2$^{\circ}$ per year. It means that the nondipole field makes one complete revolution around the Earth rotation axis in 1800 years. A higher rotation velocity of the mantle in comparison with the outer core is supposed to be the physical mechanism of the westward drift. Because of electromagnetic forces, the solid mantle of the Earth is coupled to the core as a whole, and the outer part of the core therefore travels westward relative to the mantle, carrying the minor features of the field with it~\cite{ref022}.

To explain the westward drift of magnetic features we have to distinguish between the drift effect and other causes of magnetic variations~\cite{ref011}. The magnetic secular variation can be written as:

\begin{equation}
\frac{\delta B_Y}{\delta t} = \frac{\partial B_Y}{\partial t} + U_{\varphi} \frac{\partial B_Y}{\partial \varphi}
\label{eq001}
\end{equation}

\noindent where $\delta B_Y / \delta t$ is the magnetic variation (the Y-component) seen at magnetic observatories; the $\partial B_Y / \partial t$ term accounts for effects of non-uniform and north-south motions (as well as the less important effects of magnetic diffusion), $U_{\varphi}$ is the magnitude of the westward drift as seen at the Earth surface and $\partial B_Y / \partial \varphi$ is the longitudinal gradient of the magnetic field as seen at the surface.

It is known that most of the early magnetic observatories are located in Europe (Fig.~\ref{fig04}a) and fortunately, the $\partial B_Y / \partial \varphi$ term here is large~\cite{ref011} and smoothly varying~\cite{ref011,ref023} (Fig.~\ref{fig04}b). Moreover, the term $\delta B_Y / \delta t$ is dominant and directly proportional to the magnitude of the westward drift at the geographical places where the term $\partial B_Y / \partial \varphi$  is large and smoothly varying and the term $\partial B_Y / \partial t \to 0$ (for example, Europe and Australia (see Fig.~\ref{fig04}c)):

\begin{equation}
\frac{\delta B_Y}{\delta t} \sim U_{\varphi}.
\label{eq002}
\end{equation}

\begin{figure}[tb]
    \begin{center}
        \includegraphics[width=10cm]{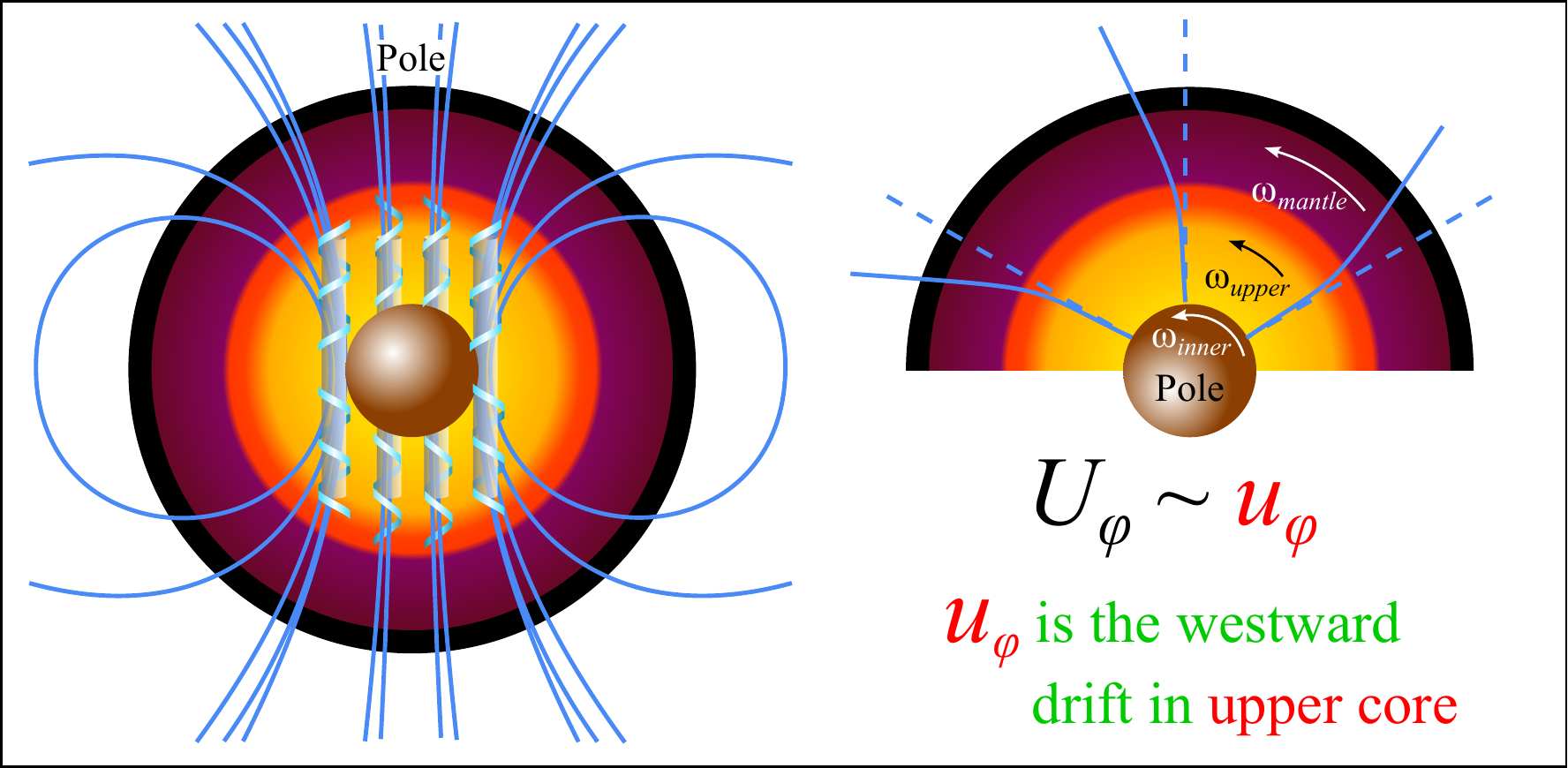}
    \end{center}
\caption{Sketch of Earth magnetic field and westward drift of Earth magnetic field on the core-mantle boundary ($u_{\varphi}$) and at the Earth surface ($U_{\varphi}$).}
\label{fig05}
\end{figure}

Consequently, if the westward drift of the magnetic field on the core-mantle boundary (Fig.~\ref{fig05}) is caused by the core-mantle coupling, which induces the corresponding westward drift of magnetic feature at the Earth surface (Fig.~\ref{fig05}), then

\begin{equation}
\frac{\delta B_Y}{\delta t} \sim U_{\varphi} \sim u_{\varphi}.
\label{eq003}
\end{equation}

\noindent where $u_{\varphi}$ is the westward drift of the magnetic field on the core-mantle boundary.

On the other hand, as far back as 1953 basing on the Bullard's model~\cite{ref022} analysis, Vestine~\cite{ref024} came to a conclusion that if the core-mantle coupling mechanism exists, it should also cause a correlation between the westward drift of the eccentric dipole (the magnetic centre in the Earth core) and the variations of the Earth rotation velocity $\Omega$. As the further analysis of the magnetic observations and the Earth rotation variations~\cite{ref011} reveals, such kind of correlation indeed takes place (Fig.~\ref{fig04}d), which is an obvious sign of core-mantle coupling mechanism existence producing the westward drift of magnetic features at the Earth surface.

It becomes clear therefore that a question about the physical nature of the strong anticorrelation between the Solar magnetic field and the geomagnetic field (the Y-component) variations (Fig.~\ref{fig03}) comes to a question: "\textit{How can the Sun know about the processes in the Earth liquid core?}".

The answer is very simple: \textit{it governs them!} And it governs the magnetic field in the Earth core by means of some unknown interaction carrier! More precisely speaking, the Sun governs the processes in the Earth liquid core through some kind of interaction which must be transmitted by some unknown particles with their flux controlled (modulated) by the Solar magnetic field.

According to our supposition, these particles may be the axions born primarily inside the Sun core and may be converted into $\gamma$-quanta in the tachocline magnetic field. This supposition is the leading idea of the present paper.

The fact that the solar-terrestrial magnetic correlation has the undoubtedly fundamental importance for evolution of all the geospheres is confirmed by existence of stable and strong correlation between temporal variations of the Earth magnetic field, the Earth angular velocity, the average ocean level and the number of large earthquakes (with the magnitude M$\geqslant$7), which are apparently driven by a common physical cause of unknown nature (see e.g.~\cite{ref009}).

In this section we consider the hypothetical particles (solar axions) as the main carriers of the solar-terrestrial connection, which by virtue of the inverse coherent Primakoff effect can transform into photons in external fluctuating electric or magnetic fields~\cite{ref010}. At the same time we ground and develop the axion mechanism of solar dynamo -- geodynamo connection, where the energy of axions, which originate from the Sun core, is modulated at first by the magnetic field of the solar tachocline zone (due to the inverse coherent Primakoff effect), and after that is resonantly ($^{57}$Fe solar axions) absorbed in the iron core of the Earth, thereby playing the role of an energy source and a modulator of the Earth magnetic field. Justification of the axion mechanism of the Sun luminosity and solar dynamo -- geodynamo connection is the goal of the current section.

\subsection{Implication from "axion helioscope" technique (axion-photon interaction)}

As it is seen from the Earth, the most important astrophysical source of axions is the core of the Sun. There, pseudoscalar particles like axions would be continuously produced in the fluctuating electric and magnetic fields of the plasma via their coupling to two photons (the Primakoff effect~\cite{ref010}). After production the axions would freely stream out of the Sun without any further interaction. The resulting differential solar axion flux on the Earth would be~\cite{ref025,ref026}

\begin{equation}
\frac{d \Phi _a}{d E} = 6.02 \cdot 10^{10} g_{10}^2 E^{2.481} \exp \left( - \frac{E}{1.205} \right) ~~ cm^{-2} s^{-1} keV^{-1},
\label{eq004}
\end{equation}

\noindent where $E$ is in keV and $g_{10} = g_{a \gamma} / (10^{-10} ~GeV^{-1}$). 

The spectral energy of the axions (\ref{eq004}) follows the thermal energy distribution between 1 and 100~keV, which peaks at $\approx$3~keV and the average energy $\langle E_a \rangle = 4.2$~keV. To be able to compare the expected axion flux in a specific energy range with available data, by integrating the spectrum (\ref{eq004}) over the energy range of 1 to 100~keV we find the solar axion flux at the Earth to be

\begin{equation}
\Phi_a \approx 3.75 \cdot 10^{11} g_{10}^2 ~~ cm^{-2}s^{-1}.
\label{eq005}
\end{equation}

In the case of the coherent Primakoff effect the number of photons leaving the magnetic field towards the detector is determined by the probability $P_{a \to \gamma}$ that an axion converts back to an "observable" photon inside the magnetic field~\cite{ref027}

\begin{equation}
P_{a \rightarrow \gamma} = \left( \frac{B g_{a \gamma}}{2} \right) ^2 \frac{1}{q^2 + \Gamma^2 / 4} \left[ 1 + e^{-\Gamma L} - 2 e^{-\Gamma L / 2} \cos (qL) \right],
\label{eq006}
\end{equation}

\noindent where $B$ is the strength of the transverse magnetic field along the axion path, $L$ is the path length traveled by the axion in the magnetic region, $l = 2 \pi / q$ is the oscillation length, $\Gamma = \lambda ^{-1}$ is the absorption coefficient for the X-rays in the medium, $\lambda$ is the absorption length for the X-rays in the medium and the longitudinal momentum difference $q$ between the axion and the X-rays energy $E_{\gamma} = E_a$ is 

\begin{equation}
q = \frac{\left \vert m_{\gamma}^2 - m_{a}^2 \right \vert}{2E_a}
\label{eq007}
\end{equation}

\noindent with the effective photon mass

\begin{equation}
m_{\gamma} \cong \sqrt{\frac{4 \pi \alpha n_e}{m_e}} = 28.9 \sqrt{\frac{Z}{A} \rho},
\label{eq008}
\end{equation}

\noindent where $m_{\gamma} = m_a$ is the axion mass, $\alpha$ is the fine-structure constant, $n_e$ is the number of electrons in the medium, $m_e$ is the electron mass, $Z$ is the atomic number of the buffer medium, $A$ is atomic mass of the medium and its density $\rho$ in $g / cm^3$.

On the other hand, it is known that the axion is a neutral pseudoscalar particle that was introduced in the particle theory to explain the absence of CP violation in strong interactions~\cite{ref028,ref029,ref030}. The most natural solution to the CP-violation problem was obtained by introducing a new chiral symmetry, known as Peccei-Quinn (PQ) symmetry~\cite{ref001}, the spontaneous breakdown of which at the energy $f_a$ fully compensates the CP-nonivariant term in the QCD Lagrangian and leads to the appearance of the axion~\cite{ref029,ref030}. The axion is not massless because the chiral U(1) PQ-symmetry is anomalous. As a result, the axion gets a mass of the order~\cite{ref031}

\begin{equation}
m_a \sim \frac{\Lambda_{QCD}^2}{f_a},
\label{eq009}
\end{equation}

\noindent where $\Lambda_{QCD}$ is the confining QCD-scale and $f_a$ is the energy scale associated with the breakdown of the U(1) PQ symmetry.

At the same time it is necessary to mention the axion mass estimates obtained in the framework of the so-called invisible axion models (KSVZ~\cite{ref032,ref032a} and DFSZ~\cite{ref033,ref033a}), which restrict the allowed range for $f_a$, or equivalently the range for the axion mass

\begin{equation}
m_a = 6 \cdot \frac{10^6 ~GeV}{f_a} ~~eV,
\label{eq010}
\end{equation}

\subsection{Axion conversion in the Sun magnetic field and the plasma mass of photon}

\begin{figure}
    \begin{center}
        \includegraphics[width=10cm]{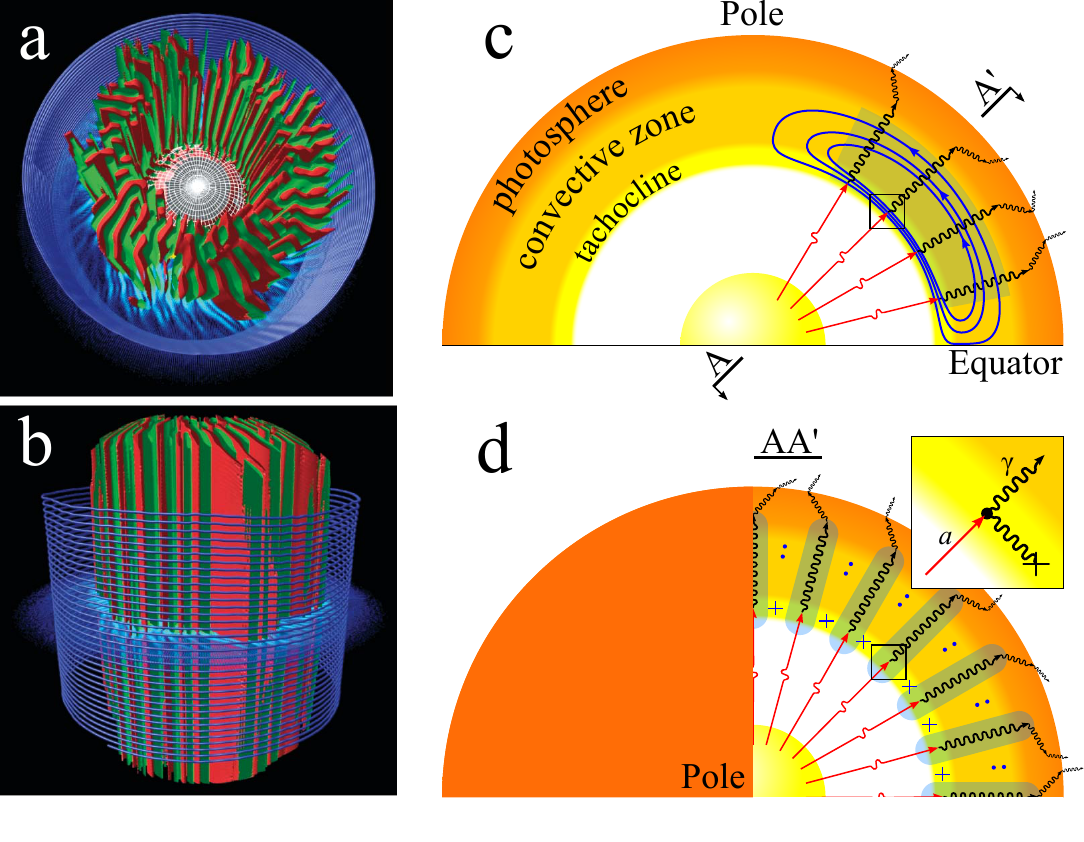}
    \end{center}
\caption{Examples of simulation of the periodic alternation of layers (the zonal flow) in the convective structures of the Earth outer core (a, b~\cite{ref038}) and the convective zone of the Sun (c, d). \textbf{a)} View from the north. Isosurfaces of the axial vorticity, $\omega_z$, are shown in red ($\omega_z = 0.4$) and green ($\omega_z = -0.4$) to illustrate the sheet plumes. Each line forms a closed ring, indicating that the flow is nearly purely westward; \textbf{b)} Same as a), but viewed from a different angle; \textbf{c)} The section $AA'$ along one of the alternate convective layers of the Sun. In the tachocline axions are converted into $\gamma$-quanta (see the inset in d)) channeling in the green area. In the photosphere $\gamma$-quanta are scattering due to the Compton effect. \textbf{d)} Same as c), but viewed from a different angle (see the section $AA'$ in c)). Alternate layers in the convective zone, where layers in which the channeling takes place are shown in green, are also presented. Blue points (in the upper convective zone) and crosses (in the tachocline) show the direction of output and input of the magnetic field in the convective zone of the Sun.}
\label{fig06}
\end{figure}

Let us consider the modulation of the axion flux emerging from the Sun core and passing through the solar tachocline region (ST) located at the base of the solar convective zone (Fig.~\ref{fig06}c,d). As is known~\cite{ref014,ref034}, the equatorial thickness of ST, where the toroidal magnetic field $B \sim 10 \div 50$~T dominates~\cite{ref034,ref035,ref036}, attains $L_{ST} \sim 0.039 R_S$ (where $R_S = 6.96 \cdot 10^8$~m~\cite{ref037} is the Sun radius). At the same time the values of pressure, temperature and density for the ST are $P_{ST} \sim 6.0 \cdot 10^{12}$~Pa, $T_{ST} \sim 2.0 \cdot 10^6$~K and $\rho \sim 0.2 ~g \cdot cm^{-3}$, respectively.

To estimate the plasma mass of a photon $m_\gamma$ in the hydrogen-helium medium of ST it is possible, without loss of generality, to use the modified Eq.~(\ref{eq008}) in the form~\cite{ref025}

\begin{equation}
m_{\gamma} (eV) = m_a \cong \sqrt{0.02 \frac{P_{ST} (mbar)}{T_{ST}(K)}} \cong 25 ~~ eV,
\label{eq011}
\end{equation}

\noindent where we use the corresponding parameters $P_{ST} \sim 6.0 \cdot 10^{12}$ Pa and $T_{ST} \sim 2.0 \cdot 10^6$ K for the hydrogen-helium medium of ST obtained by Bahcall \& Pinsonneault for the standard model of the Sun~\cite{ref079}.

Thus, the axion mass in the standard model of the Sun is $\sim$25~eV. However, it will be shown below that in the framework of the axion mechanism of Sun luminosity the axion mass will be different, since the total energy balance of the Sun is not violated, but indicates a substantial change in radiation transport through the radiative zone and the convective zone with respect to the standard model of the Sun. The energy portion of the axion-independent radiation transport is rather small here and equals to $\sim 0.015 \Lambda_{Sun}$ (see (\ref{eq015})). Since the total energy balance of the Sun is not violated in the axion model, one may suppose that the basic parameters of the solar core -- the region of the energy generation -- remain approximately the same in both the standard and the "axion" models. Meanwhile, the thermodynamic parameters (temperature, pressure, plasma density, electron density etc.) \textbf{\textit{outside}} the solar core (between the core and photosphere) are \textbf{\textit{substantially smaller}} in the axion model as compared to the corresponding parameters in the standard model of the Sun.

It should be noted here that the calculation of these parameters for the axion model is a rather nontrivial task, which, because of its complexity, will be performed in a separate publication. For this reason from now on let us use the "experimental" value of the axion mass found during the extragalactic background light investigation (see Section~\ref{sec-05}).

\begin{equation}
m_a \sim 17 ~~ eV.
\label{eq011a}
\end{equation}

Taking into account (\ref{eq011}) and (\ref{eq010}) it is easy to derive the value of the energy $f_a$:

\begin{equation}
f_a \cong 0.353 \cdot 10^6 ~~GeV
\label{eq012}
\end{equation}

Now we make an important assumption that the axion mass is equal to the plasma mass of a photon, i.e., $m_{\gamma} = m_a \sim 17$~eV. It is obvious, that by virtue of Eq.~(\ref{eq007}) $q \to 0$, whence it follows that the oscillation length $l$ becomes an infinite quantity, i.e. $l = 2 \pi / q \to \infty$. However, taking into account that in this case the absorption length $\lambda$ is about 0.1~m~\cite{ref014}, we have $\Gamma L_{ST} \to \infty$. This means that according to Eq.~(\ref{eq006}), the intensity of expected conversion of axions into $\gamma$-quanta is practically equal to zero in this case.

\begin{figure}
    \begin{center}
        \includegraphics[width=10cm]{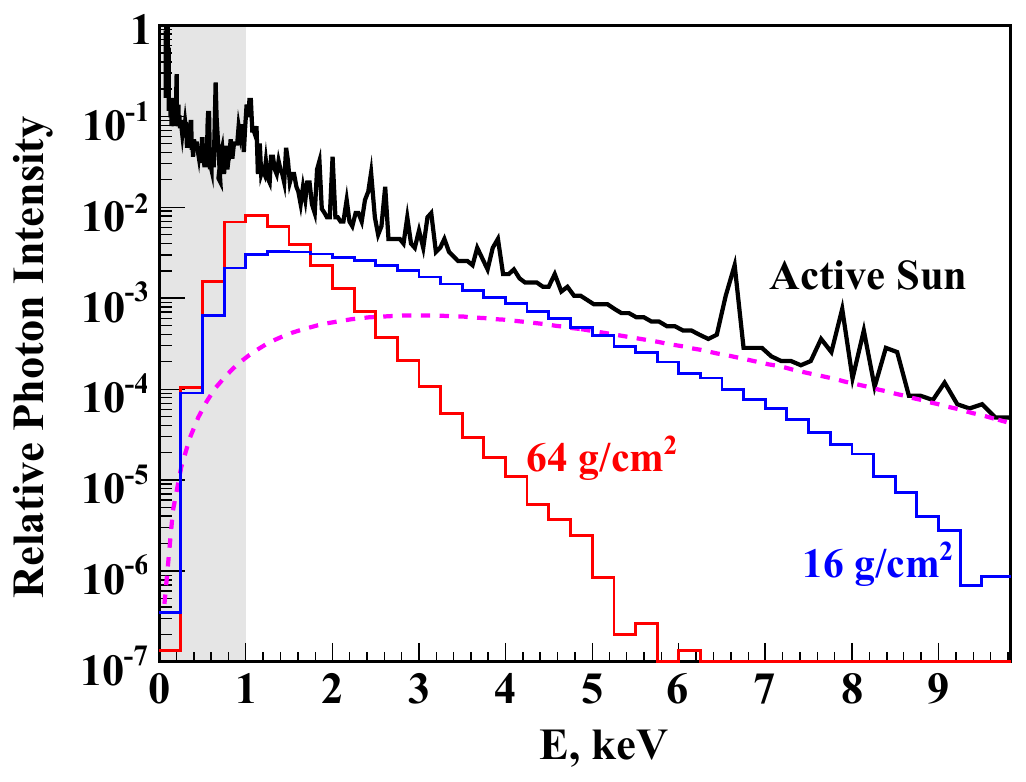}
    \end{center}
\caption{Reconstructed solar photon spectrum below 10~keV from the Active (flaring) Sun (the black line) from accumulated observations~\cite{ref039} (adapted from~\cite{ref014}). The dashed line is the converted solar axion spectrum. Two degraded spectra due to multiple Compton scattering are also shown for column densities above the initial conversion place of $64 ~g / cm^2$, $16 ~g / cm^2$. The pink dotted line represents the initial Primakoff axion spectrum. Note that the Geant4 code photon threshold is at 1~keV and therefore the turndown around $\sim$1~keV is an artifact.}
\label{fig07}
\end{figure}

At the same time, there is a reason to believe (see~\cite{ref014} and Refs. therein) that the conversion of axions into $\gamma$-quanta indeed takes place, and strangely enough, this process goes on quite effectively. For example, the reconstructed solar photon spectrum below 10~keV from the Active Sun (Fig.~\ref{fig07}b) is well described by the sum of secondary Compton spectra obtained e.g. by the simulation of $\gamma$-quanta passage (regenerated from the solar axion spectrum in the tachocline zone of the Sun (Fig.~\ref{fig07}b)) through the areas of the solar photosphere of different thickness but equal density, layers with the thickness of $64 ~g / cm^2$ and $16 ~g / cm^2$. 

In other words, despite the fact that the coherent axion-photon conversion by the Primakoff effect is impossible due to the small absorption length for $\gamma$-quanta ($l \gg 1$) in the medium (see Eq.~(\ref{eq006}) $\Gamma = 1 / \lambda \to \infty$), there is a good agreement between the relative theoretical $\gamma$-quantum spectra generated by solar axions and experimental photon energy spectra detected close to the Sun surface in the period of its active phase (see Fig.~\ref{fig07}). The additional account taken of the bremsstrahlung in the photosphere will surely enhance the quality of the theoretical description of the experimenal solar photon spectrum substantially.

At the same time it is necessary to note that attempts to match the absolute values of these spectra did not succeed so far~\cite{ref014}. It was mainly associated with the absence of a wish to make efforts, since it was absolutely unknown how can the $\gamma$-quanta spectra generated by solar axions in the tachocline be transported in a "virgin", i.e. unchanged, form through the convective zone up to the Solar photosphere.

To overcome the problem of the small absorption length for $\gamma$-quanta and to reach a resonance in Eq.~(\ref{eq006}) it is necessary for the refractive gas, in which the axion-photon oscillation is studied, to have a zero refractive index~\cite{ref040}. It appears that in order to satisfy this condition it is not necessary to use the so-called metamaterials~\cite{ref041} with the negative permittivity ($\varepsilon$) and magnetic permeability ($\mu$) or results of the Pendry superlens theory~\cite{ref042}, which are not practically realized in nature\footnote{Though it should be noted that the metamaterial technology is frequently used nowadays for laboratory simulations of some celestial mechanics and cosmology phenomena~\cite{ref043,ref044,ref045,ref046}. And the "..."artificial  atoms" used as building blocks in metamaterial design offer much more freedom in constructing analogues of various exotic spacetime metrics, such as black holes, wormholes, spinning cosmic strings, and even the metric of Big Bang itself. Explosive development of this field promises new insights into the fabric of spacetime, which cannot be gleaned from any other terrestrial experiments"(\cite{ref047} and Refs. therein).}. Taking into account the known difficulties~\cite{ref048} induced by the so-called problem of electromagnetically induced transparency for X-rays and the recent significant advances in this field~\cite{ref049,ref050}, let us consider two (possibly related) alternative ways of solving the problem of "unperturbed" $\gamma$-quanta spectra transfer through the Solar convective zone.

\subsection{Channeling of $\gamma$-quanta in periodical structure}

We can use the results from papers~\cite{ref051,ref052}, where the possibility of the electromagnetic X-radiation in a microwave range channeling in a multi-layered metal-dielectric structure is theoretically and experimentally shown.

As it is stated in Appendix~\ref{A1} in detail, the essence of the electromagnetic X-ray channeling in long-period media lies in a fact that the rays are reflected from the layers of higher electron density when propagating at small angles to these layers. It leads to a non-uniform intensity distribution over the cross-sectional plane because of the rays concentration within the "channels" -- the layers with lower electron density. It decreases the absorption substantially and makes it possible for the rays to penetrate much deeper into the sample \textit{along} the layers than in the case of an arbitrary angle of arrival.

According to~\cite{ref051}, the intensity $J(x)$ of the photons (see Fig.~\ref{fig-a1.2} and (\ref{eq-a1.26})) passed through a sample of a thickness $x$, may be written in the form

\begin{equation}
J (x) = J_0 \exp (- \sigma x ) = J_0 \exp \left( - \frac{\chi _0}{\cos \alpha} x \right) \cdot Q(\alpha, y_0, x),
\label{eq013a}
\end{equation}

\noindent where

\begin{equation}
Q(\alpha, y_0, x) = 
\begin{cases}
\exp \left[ - \frac{\chi _0}{\cos \alpha} \beta ^2 x \left( 1 - \frac{E(q^{-1})}{K (q^{-1})} \right) \right] & at ~~ q > 1, \\
\exp \left[ - \frac{\chi _0}{q^2 \cos \alpha} \beta ^2 x \left( 1 - \frac{E(q)}{K (q)} \right) \right]             & at ~~ q < 1.
\end{cases}
\label{eq013b}
\end{equation}

\noindent with the same notation used in expressions (\ref{eq-a1.26})-(\ref{eq-a1.27}).

Although we give a complete analysis of the Eqs.~(\ref{eq013a}) and~(\ref{eq013b}) in \ref{A1}, let us make a short remark regarding the physical nature of these equations. Here the multiplier $J_0 \exp (- \chi x / \cos \alpha$) in (\ref{eq013a}) corresponds to the case of $\gamma$-quanta propagation through a homogeneous medium with the electron density $N_e$ and the absorption coefficient $\chi _0$. The additional multiplier $Q(\alpha,y_0,x)$ characterizes the influence of the medium layering.

As is shown in~\ref{A1}, the condition $Q(\alpha, y_0, x) \to 1$ is theoretically feasible for a majority of the multilayer metal-dielectric structures~\cite{ref051,ref053}, which are an effective emulator of a plasma medium~(Fig.~\ref{fig06}). This condition is obviously necessary, but not sufficient. The layers with ultralow, if not with "quasi-zero" density, are also required for the ideal photon channeling. Such layers suppress the photon absorption processes almost completely, i.e. minimize the effect of the multiplier $J_0 \exp (-\chi_0 x / \cos \alpha$) in~(\ref{eq013a}).

Surprisingly enough, it turns out that such long-period (in terms of density) media with one of the two alternating media having almost zero density can take place, and not only in plasmas in general, but straight in the convective zone of the Sun. Here we generally mean the so-called magnetic flux tubes, the properties of which are examined below (see Appendix~\ref{A2} for details).

\subsection{Channeling of $\gamma$-quanta along the magnetic flux tubes (waveguides) in Solar convective zone}

The idea of the energy flow channeled along a fanning magnetic field has been suggested for the first time by Hoyle~\cite{ref054} as an explanation for darkness of umbra of sunspots. It was incorporated in a simple sunspot model by Chitre~\cite{ref055}. Zwaan~\cite{ref056} extended this suggestion to smaller flux tubes to explain the dark pores and the bright faculae as well. Summarizing the research of the convective zone magnetic fields in the form of the isolated flux tubes, Spruit and Roberts~\cite{ref057} suggested a simple mathematical model for the behavior of thin magnetic flux tubes, dealing with the nature of the solar cycle, the sunspot structure, the origin of spicules and the source of mechanical heating in the solar atmosphere. In this model, the so-called thin tube approximation is used (see~\cite{ref057} and Refs. therein), i.e. the field is conceived to exist in the form of slender bundles of field lines (flux tubes) embedded in a field-free fluid. Mechanical equilibrium between the tube and its surrounding is ensured by the reduction of the gas pressure inside the tube, which compensates the force exerted by the magnetic field. In our opinion, this is exactly the kind of mechanism Parker~\cite{ref058} was thinking about when he wrote about the problem of flux emergence: "Once the field has been amplified by the dynamo, it needs to be released into the convection zone by some mechanism, where it can be transported to the surface by magnetic buoyancy"~\cite{ref059}.

In order to understand magnetic buoyancy, let us consider an isolated horizontal flux tube in pressure equilibrium with its non-magnetic surroundings, so that

\begin{equation}
p_{int} + \frac{B^2}{2 \mu _0} = p_{ext}, 
\label{eq013}
\end{equation}

\noindent where $p_{int}$ and $p_{ext}$ are the internal and external gas pressures respectively and $\mu _0$ is the magnetic permeability of the medium, $B$ denotes the uniform field strength in the flux tube. If the internal and external temperatures are equal so that $T_{int} = T_{ext}$ (thermal equilibrium), then since $p_{int} < p_{ext}$, the gas in the tube is less dense than its surrounding ($\rho _{int} < \rho _{ext}$), implying that the tube will rise under the influence of gravity.

In spite of the obvious, though turned out to be surmountable, difficulties of expression~(\ref{eq014}) application to the real problems, it was shown (see~\cite{ref057} and Refs. therein) that strong buoyancy forces act in magnetic flux tubes of the required field strength (10$^4$-10$^5$~G~\cite{ref060}). Under their influence tubes either float to the surface as a whole (e.g. Fig.1 in~\cite{ref061}) or they form loops of which the tops break through the surface (e.g. Fig.1 in~\cite{ref056}) and lower parts descend to the bottom of the convective zone, i.e. to the overshoot tachocline zone. The convective zone, being unstable, enhances this process~\cite{ref062,ref063}. Small tubes take longer to erupt through the surface because they feel stronger drag forces. It is interesting to note here that the phenomenon of the drag force which raises the magnetic flux tubes to the convective surface with the speeds about 0.3-0.6~km/s was discovered in direct experiments using the method of time-distance helioseismology~\cite{ref064}. Detailed calculations of the process~\cite{ref065} show that even a tube with the size of a very small spot, if located within the convective zone, will erupt in less than two years. Yet, according to ~\cite{ref065}, the horizontal fields are needed in the overshoot tachocline zone, which survive for about 11 yr, in order to produce an activity cycle.

\begin{figure}[tb!]
    \begin{center}
          \includegraphics[width=15cm]{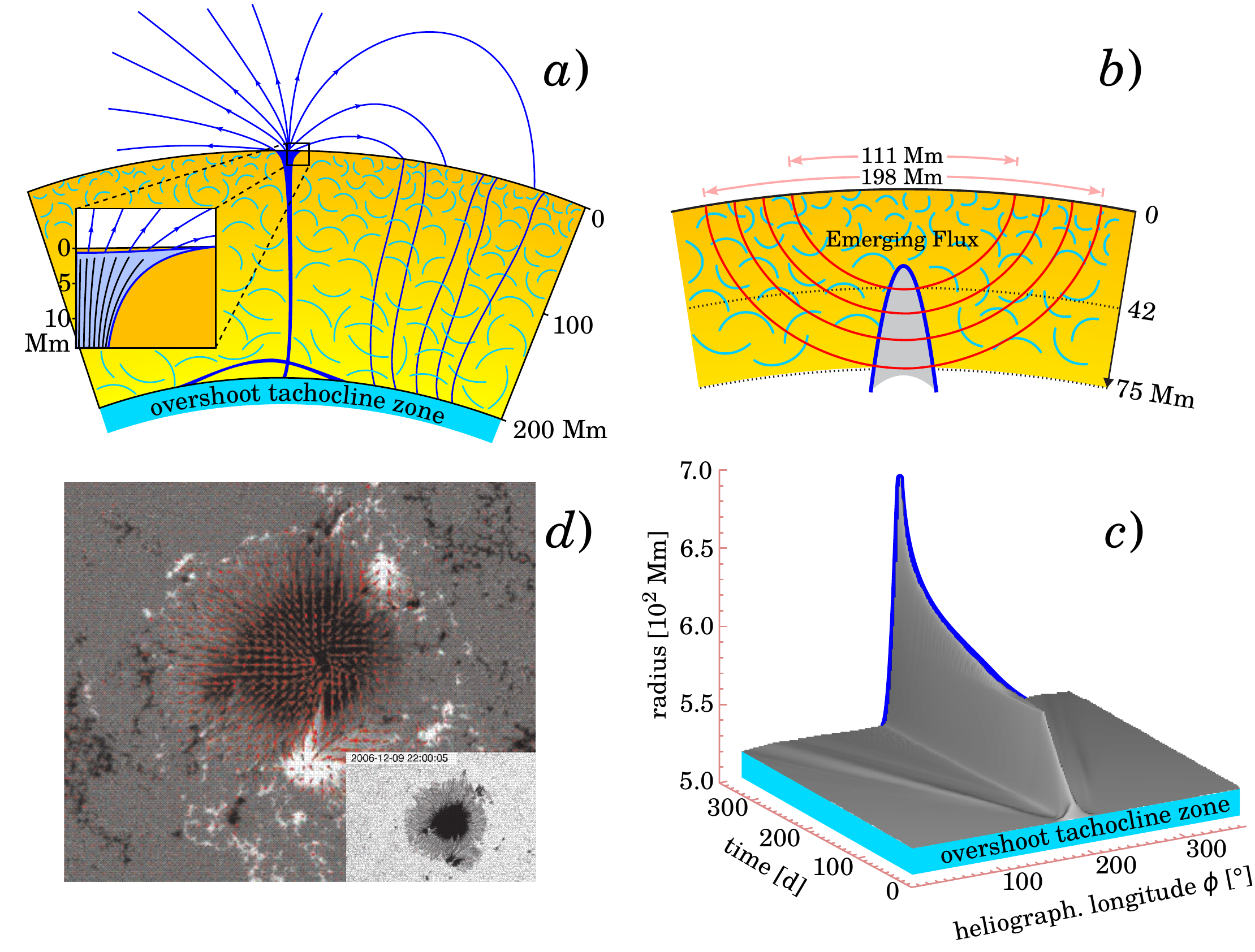}
    \end{center}
\caption[caption]{\textbf{(a)} Vertical cut through an active region illustrating the connection between a sunspot at the surface and its origins in the toroidal field layer at the base of the convection zone. Horizontal fields are stored at the base of the convection zone (the overshoot tachocline zone) during the cycle. Active regions form from sections brought up by buoyancy (one is shown in the process of rising). After the eruption through the solar surface a nearly potential field is set up in the atmosphere (broken lines), connecting to the base of the convective zone via almost vertical flux tube. Hypothetical small scale structure of a sunspot is shown in the inset (adopted from Spruit~\cite{ref066} and Spruit and Roberts~\cite{ref057}).\\
\textbf{(b)} Detection of emerging sunspot regions in the solar interior~\cite{ref064}. Acoustic ray paths with lower turning points between 42 and 75 Mm (1 Mm=1000 km) are crossing the region of the emerging flux. For simplicity, only four out of a total of 31 ray paths used in this study (the time-distance helioseismology experiment) are shown here. Adopted from~\cite{ref064}.\\
\textbf{(c)} Emerging and anchoring of stable flux tubes in the overshoot tachocline zone, and its time-evolution in the convective zone. Adopted from~\cite{ref067}.\\
\textbf{(d)} Vector magnetogram of the white light image of a sunspot (taken with SOT on a board of the Hinode satellite -- see inset) showing the direction of the magnetic field and its strength (the length of the bar) in red. The movie shows the evolution in the photospheric fields that has led to an X class flare in the lower part of the active region. Adopted from~\cite{ref068}.}
\label{fig08}
\end{figure}

A simplified scenario of magnetic flux tubes (MFT) birth and space-time evolution (Fig.~\ref{fig08}a) may be presented as follows. MFT is born in the overshoot tachocline zone (Fig.~\ref{fig08}d) and rises up to the convective zone surface without separation from the tachocline (the anchoring effect), where it forms the sunspot (Fig.~\ref{fig08}b) or other kinds of active solar regions when intersecting the photosphere. There are more fine details of MFT physics expounded in overviews by Hassan~\cite{ref059} and Fisher~\cite{ref061}, where certain fundamental questions, which need to be addressed to understand the basic nature of magnetic activity, are discussed in detail: \textit{How is the magnetic field generated, maintained and dispersed? What are its properties such as structure, strength, geometry? What are the dynamical processes associated with magnetic fields? \textbf{What role do magnetic fields play in energy transport?}}

Dwelling on the last extremely important question associated with the energy transport, let us note that it is known that the thin magnetic flux tubes can support longitudinal (also called sausage), transverse (also called kink), torsional (also called torsional Alfv\'{e}n), and fluting modes (e.g.~\cite{ref069,ref070,ref071,ref072,ref073}); for the tube modes supported by wide magnetic flux tubes see Roberts and Ulmschneider~\cite{ref072}.  Focusing on the longitudinal tube waves known to be an important heating agent of solar magnetic regions, it is necessary to mention the recent papers by Fawzy~\cite{ref075}, which showed that the longitudinal flux tube waves are identified as insufficient to heat the solar transition region and corona in agreement with previous studies~\cite{ref076}.

In other words, the problem of generation (the source) and transport of energy by magnetic flux tubes remains unsolved in spite of its key role in physics of various types of solar active regions.

Interestingly, this problem may be solved in the natural way in the framework of the "axion" model of the Sun. As it is shown in Appendix \ref{A2}, the inner pressure, temperature and matter density decrease rapidly in a magnetic tube "growing" between the tachocline and the photosphere. The analysis of these parameters evolution within the equation of the growing magnetic flux tube medium state not only gives the ultralow values for them, but also leads to the so-called hydrostatic condition of an ideal (without absorption) $\gamma$-quanta channeling inside the thin magnetic flux tubes

\begin{equation}
p_{ext} \simeq \frac{\vert \vec{B} \vert ^2}{2 \mu_0}, 
\label{eq013c}
\end{equation}

\noindent which is well satisfied for the "axion" model of the Sun, according to estimations in Appendix \ref{A2}.  It means that such thin magnetic flux tubes are the ideal $\gamma$-quanta waveguides, which reveal the essence of the unique energy transport mechanism between the tachocline and the photosphere.

As a matter of fact, the phenomenon of $\gamma$-quanta channeling along the magnetic flux tubes not only makes it possible to solve a problem of the energy transport to the photosphere, but may also be a basis for solving other important and critical problems in solar physics. If we assume that the vertically oriented thin magnetic flux tubes play the role of waveguides for $\gamma$-quanta produced in the tachocline via the axion mechanism of Sun luminosity, virtually all known anomalies of experimental data interpretation in physics of active solar regions, helioseismology and solar neutrino are withdrawn. Since this assumption needs to be substantiated, let us describe our phenomenology, consequences and experimental proofs of this hypothesis below in short.

\begin{figure}
    \begin{center}
        \includegraphics[width=18cm]{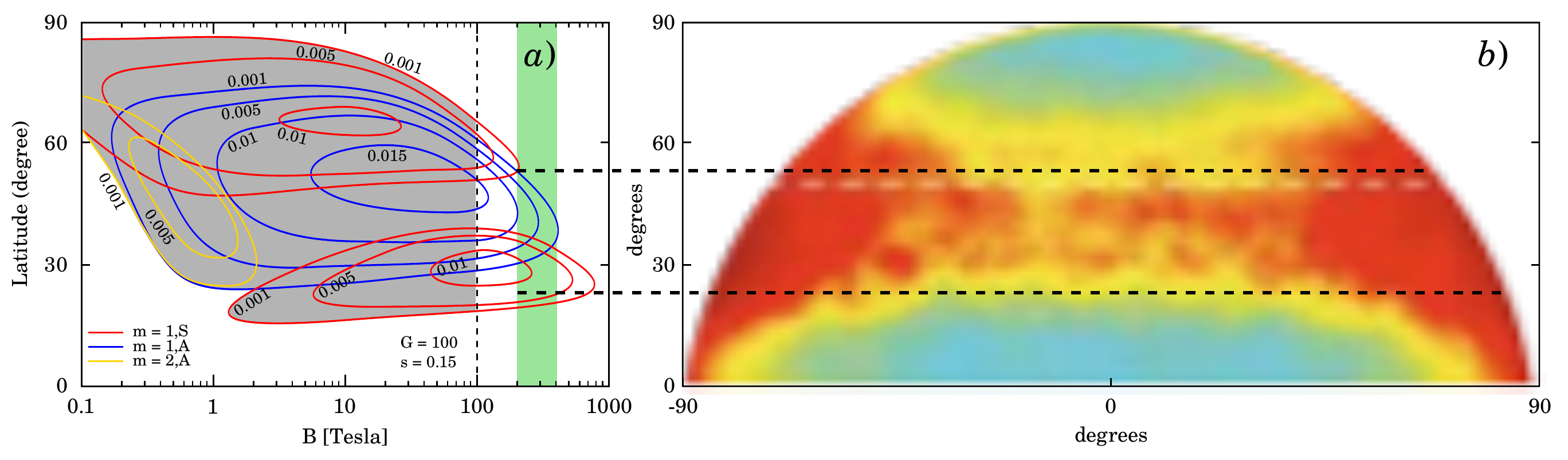}
    \end{center}
\caption[caption]{\textbf{(a)} Growth rates for magnetic shear instabilities are plotted as a function of the initial latitude (vertical axes) and the field strength (horizontal axes) of a toroidal band. Shaded areas indicate instability in 0.1-100T band (gray) and 200-400T band (green). Contour lines represent $m = 1$ and $m = 2$ symmetric (S) and antisymmetric (A) modes as indicated. The non-dimensional model is normalized in such a way that the growth rate of 0.01 corresponds to an e-folding growth time of 1 year. The parameter $s$ is the fractional angular velocity contrast between equator and pole and the reduced gravity $G$ (adopted from~\cite{ref077}). In addition, a hidden part of the "latitude – magnetic field in overshoot tachocline zone" dependence, which was missing on the original plot (Fig.11 in~\cite{ref077}), is plotted to the right of the dashed line.\\
\textbf{(b)} Solar images at photon energies from 250~eV up to a few keV from the Japanese X-ray telescope Yohkoh (1991-2001) (adapted from~\cite{ref014}). The following shows solar X-ray activity during the last maximum of the 11-year solar cycle.}
\label{fig09}
\end{figure}

\underline{First} of all, if one takes into account the sufficiently strong magnetic field in the balance equation of (\ref{eq013}) type, it becomes clear that the vertically oriented thin magnetic flux tubes may serve as the X-ray waveguides for the radiation originating from the overshoot tachocline zone because of the high magnetic pressure. Naturally, in this case the X-ray spectrum coincides with the observed Solar X-ray spectrum. All these facts, i.e. the strong magnetic field $\sim$200-400T (Fig.~\ref{fig09}), X-rays (Fig.~\ref{fig08}b) and their spectrum (e.g. Fig.~7 in~\cite{ref068}) in active solar regions, are in good agreement with observational data.

\begin{figure}
    \begin{center}
        \includegraphics[width=15cm]{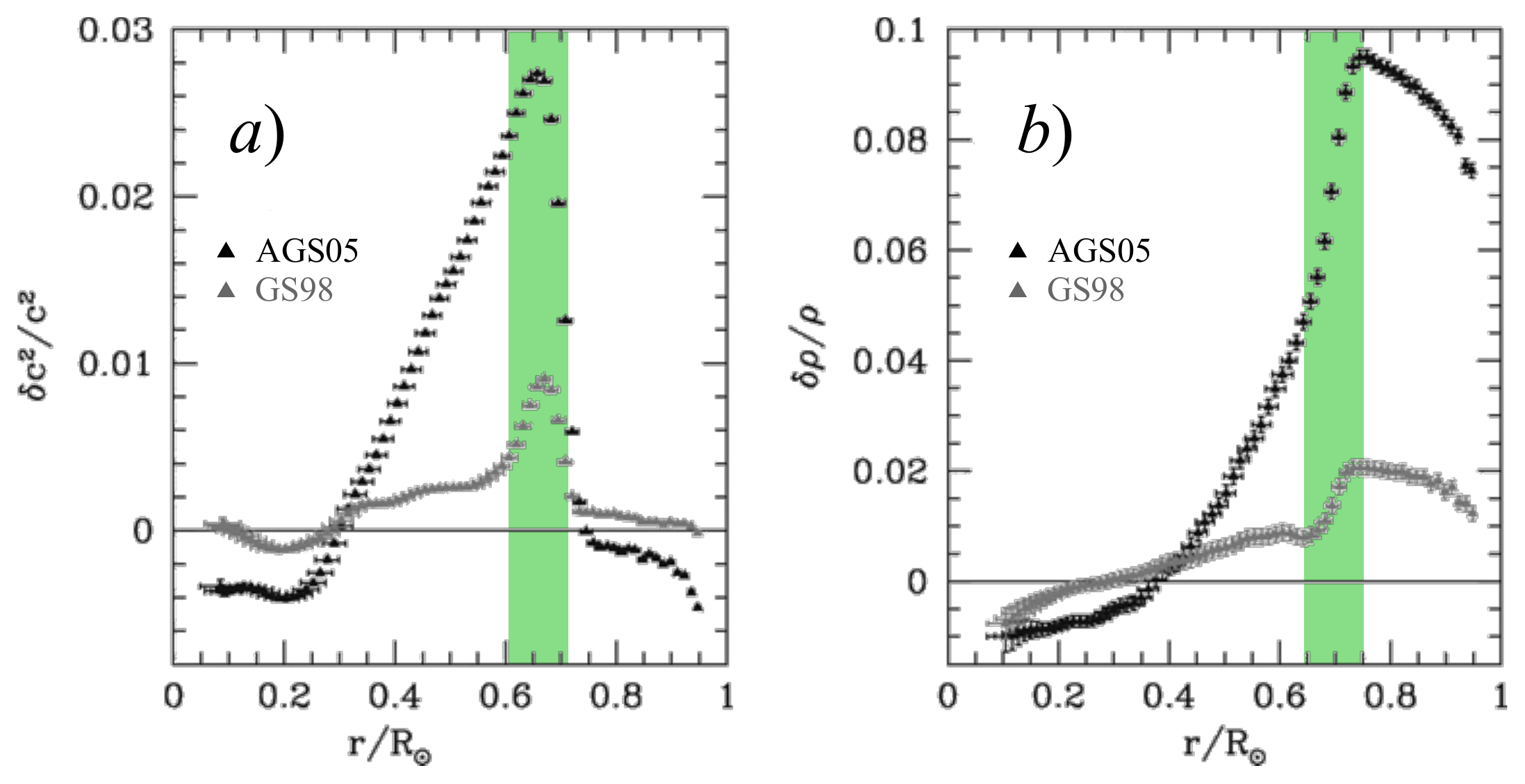}
    \end{center}
\caption{The relative sound-speed (the panel a) and density differences (the panel b) between the Sun and the model constructed with the AGS05 abundances~\cite{ref-bahcall2005}. For comparison we also show the results for the model constructed with the GS98 abundances. MDI 360 day data have been used for the inversions. The overshoot tachocline anomaly is highlighted with green. Note: GS98 is a solar model with the solar heavy-element mixture $Z / X = 0.0245$; AGS05 is a solar model with low heavy-element mixture $Z / X = 0.0122$. Adopted from~\cite{ref078}.}
\label{fig10}
\end{figure}

\underline{Second}, it clears up a way to the solution of the known problem associated with the over-shoot tachocline anomaly (Fig.~\ref{fig10}) which arises when interpreting the helioseismology and solar abundances data. And here is why.

It is known~\cite{ref078} that the problem comes from the attempts to improve agreement between solar models with low heavy-element abundances and seismic inference. The low-metallicity models that have the least disagreement with seismic data require changing all input physics to stellar models beyond their acceptable ranges. Let us consider the way it happens in the framework of a solar model built upon the axion mechanism of Sun luminosity.

Since solar luminosity is determined by the $\gamma$-quanta born in the tachocline in the framework of the axion mechanism, it is clear that an old heat flux transport mechanism (from the radiative zone to the overshoot) \textit{by radiation}, used in the standard model of the Sun, should be highly depressed because the major part of the radiation is converted into axions in the core of the Sun~\cite{ref025} and does not get into the radiative interior. It is easy to see from the traditional statement of the problem involving helioseismology and solar abundances as described by Basu~\cite{ref078} that this is one of the main and fundamental differences from the standard model of the Sun: "The most easily detectable effect of the reduction of heavy-element abundances is a change in the position of the base of the convection zone. The temperature gradient in the radiative interior is determined by opacity, and hence, its structure is affected by the heavy-element abundances. The base of the convection zone occurs at a point \textit{where opacity is just small enough to allow the entire heat flux to be transported by radiation}, and thus the \textbf{\textit{location of this point depends on the abundance of those heavy elements}} that are the predominant sources of opacity in that region. If these abundances are reduced, opacity reduces, and the depth of the convection zone also reduces. Since the depth of the convection zone has been measured very accurately, it is the most sensitive indicator of opacity or heavy-element abundances".

The axion mechanism of Sun luminosity implies that because of a virtually complete transparency of the magnetic flux tubes for $\gamma$-radiation there are no reasons for moving the location of the center of the tachocline "by hand", since the radiative opacity determined by the effect of heavy-element abundance loses its impact on the location of this point and, consequently, its significance, because of almost total suppression of the radiative heat flux transport mechanism itself in this case. In other words, the effect of absolute magnetic flux tubes transparency for the $\gamma$-radiation is dominant in the overshoot tachocline zone and levels the influence of radiative opacity. As a consequence, the value of the temperature gradient in the radiative interior is almost entirely free from the strict limit introduced by opacity which is still affected by the heavy element abundances, but not as dramatically as it is in several other solar models -- standard and nonstandard -- that have been published recently~\cite{ref078}. The latter opens up a possibility to build a new standard solar model on the basis of the axion mechanism of Sun luminosity which may become a key to the solar abundance problem solution.

\underline{Third}, let us consider the axion mechanism of Sun luminosity compatibility with the nuclear energy generation pathways in the solar core and solar neutrino fluxes generation.

The axion mechanism of Sun luminosity is compatible with the standard nuclear energy generation pathways scheme and does not disturb the known values for solar neutrino fluxes, since the "invisible" axion losses almost do not change the Sun energy balance in our model (see Section~\ref{sec-02-5} below), and therefore do not introduce any problems related to energy-producing regions (i.e. the solar core).

It actually means that introducing the axion mechanism of Sun luminosity in the framework of the standard model of the Sun leads to such value of axion losses which does not contradict the Gondolo-Raffelt limit on the "invisible" axion and Sun luminosities ratio, $\Lambda_a ^{invis} / \Lambda _{Sun} \leqslant 0.1$~\cite{ref080}, for which a good coincidence between the theoretical values and experimental data of modern helioseismological and solar neutrino experiments is still observed~\cite{ref080,ref081}.

And \underline{forth}, if the vertically oriented thin magnetic flux tubes in the convective zone play a role of the waveguides for the $\gamma$-quanta born in the tachocline with total luminosity equal to that of the Sun, what are the nature and the power of the energy source maintaining the convective processes on the Sun?

In order to find it out, let us assume that this source is the radiative zone and perform an estimation of its power basing on the magnetic field $B_{OT}$ in the overshoot tachocline zone dependence on the total ohmic dissipation $D_{CZ}^{ohmic}$ in the convective zone~\cite{ref082}.

\begin{equation}
D_{CZ}^{ohmic} = \int \frac{\eta}{\mu} \left( \nabla \times \vec{B}_{OT} \right)^2 dV \propto \frac{2 \eta}{H_{p}^2} E_{mag},
\label{eq014}
\end{equation}

\noindent where $E_{mag}$ is the magnetic energy of the field which could be possibly maintained by the currents that produce the ohmic dissipation of the solar dynamo~\cite{ref083}.

It is easy to show~\cite{ref082} that the expression (\ref{eq014}) may be written down in the following form:

\begin{equation}
D_{CZ}^{ohmic} = \frac{\eta \cdot V_{CZ}}{\mu \cdot H_{p}^2} B_{in}^2 \approx 0.015 \Lambda_{Sun},
\label{eq015}
\end{equation}

\noindent where $D_{CZ}^{ohmic} = 0.015 \Lambda_{Sun}$ is the heat power of the radiative zone near the border of the overshoot tachocline zone, equal to the uncertainty of the known Solar luminosity~\cite{ref073,ref084,ref090}; $\eta \sim 10^4 ~cm^2 / s$ is the magnetic diffusivity~\cite{ref085,ref086}, $V_{CZ}$ is the volume of the Sun convective zone; $\mu \sim 1$ is the permeability; $B_{OT} = 400 ~T$ is the magnetic field in the overshoot tachocline zone; $H_P = 6.5 \cdot 10^3 ~km$ is the pressure scale height\footnote{A larger value of the pressure scale height is a consequence of the fact that the rigidity~\cite{ref087} of the interior can be provided only by the large-scale magnetic field (cf. Mestel \& Weiss~\cite{ref088}; Gough \& McIntyre~\cite{ref089}) that the tachocline provides the interface in which radial field lines might connect the convection zone with the radiative interior only near the latitudes at which there is essentially no radial shear.}~\cite{ref056,ref090}

The approximate equality implies that not only the total solar energy balance is preserved in the framework of the axion mechanism of Sun luminosity, but also that the temperature transport changes substantially with respect to the standard model of the Sun. This is because of the fact that the old heat flux transport mechanism (from the radiative zone to the overshoot) by radiation, used in the standard model of the Sun, is highly depressed because the majority of the radiation is converted into axions in the core of the Sun and therefore does not reach the radiative interior. At the same time, this change may not seem so dramatic, since almost all known anomalies of the experimental data interpretation on the active solar regions, helioseismology and solar neutrino may be leveled as it was noted above.

And, finally, turning back to the possible mechanisms of $\gamma$-quanta channeling in a periodical structure and along the magnetic flux tubes in the Sun convective zone, one may suppose with confidence that they are not only physically compatible, but may turn out to be just two different versions of the same mechanism, which naturally manifests itself, for example, in the so-called hexagonal magnetoconvection kinetics (see e.g.~\cite{ref091}).

This means, in its turn, that the absorption length $\lambda$ for photons in such a medium (see~(\ref{eq006})) will become considerably greater than the thickness of the overshoot tachocline zone, i.e., $\lambda \gg L_{OTZ}$. At the same time, it is obvious that $\Gamma = \lambda ^{-1} \to 0$, whence a necessary condition $\Gamma L_{OTZ} \to 0$ follows. In the particular (non-coherent) case in which the magnetic field where axions are converted into photons is under vacuum ($\Gamma \to 0$, $m_{\gamma} \to 0$), equation~(\ref{eq006}) becomes 

\begin{equation}
P_{a \to \gamma} = \left( \frac{g_{a\gamma} B_{OT} L_{OT}}{2} \right)^2 \sin ^2 \left( \frac{q L_{OT}}{2} \right) / \left( \frac{q L_{OT}}{2} \right)^2
\label{eq16a}
\end{equation}

\noindent where $q = m_{a}^2 / 2 E_a$  (see~(\ref{eq007})).

Obviously, in coherent case $q \to 0$, regardless of the $\gamma$-quanta channeling mechanism type, the probability~(\ref{eq16a}) for an axion to be converted back to an "observable" photon inside the magnetic field may be expressed in the following simple form

\begin{equation}
P_{a \gamma} \simeq \left( \frac{g_{a \gamma} \bar{B}_{OT} \bar{L}_{OT}}{2} \right)^2,
\label{eq016}
\end{equation}

\noindent where $\bar{B}_{OT}$ is the mean value of the magnetic field in the overshoot tachocline zone with the effective thickness $\bar{L}_{OT}$. A value for $\bar{L}_{OT}$ was chosen basing on the analysis of the following data set. The most well known results obtained by Charbonneau et~al.~\cite{ref092} yield a tachocline thickness of $\Delta _t / R_S = 0.039 \pm 0.013$ at the equator and $\Delta _t / R_S = 0.042 \pm 0.013$ at the latitude of 60$^{\circ}$, suggesting that the tachocline may get somewhat wider at high latitudes but that the result is not statistically significant. On the other hand, Basu and Antia~\cite{ref093} argue for the statistically significant increase in the tachocline thickness with the latitude, from $\Delta _t / R_S \sim 0.016$ at the equator to $\Delta _t / R_S \sim 0.038$ at latitudes of 60$^{\circ}$ (when the width is defined as in~\cite{ref093}). Furthermore, they suggest that the variation may not be smooth; there may be a sharp transition from a narrow tachocline at low latitudes to a wider tachocline at high latitudes, possibly associated with the sign of the radial angular velocity gradient which reverses at the latitude of $\sim 35 ^{\circ}$. Other estimates for the width of the tachocline range from $0.01 R_S$ to $0.09 R_S$ (Kosovichev~\cite{ref094}, Basu~\cite{ref095}, Corbard et al.~\cite{ref096}, Elliott and Gough~\cite{ref097}, Basu and Antia~\cite{ref098}).

Taking into account that, first, the tachocline is a transition layer between two distinct rotational regimes (the differrentially rotating solar envelope and the radiative interior) where the rotation is uniform, second, the maximum estimate of the tachocline thickness reaches $0.09 R_S$ and third, the thickness of the overshoot tachocline zone is somewhat larger than that of the tachocline, we took the value of $\bar{L}_{OT}$ equal to $0.1 R_S$.

Then using Eq.~(\ref{eq016}) and the parameters of the magnetic field, it is possible to write down the expression for the solar axion flux\footnote{Hereinafter we use rationalized natural units to convert the magnetic field units from Tesla to $eV^2$, where the conversion is 1 T = 195 $eV^2$~\cite{ref040}.} probability at the Earth as 

\begin{equation}
P_{a \rightarrow \gamma} = \frac{1}{4} \left( \frac{g_{a \gamma}}{7.07 \cdot 10^{-11} ~GeV^{-1}} \right)^2 \left(\frac{\bar{B}_{OT}}{400 ~T} \right)^2 \left( \frac{\bar{L}_{OT}}{7.25 \cdot 10^7 ~m} \right)^2 = 1.
\label{eq017}
\end{equation}

\noindent where the value of the magnetic field $B_{OT} = 400 ~T$ (cf.~\cite{ref034,ref035,ref036}) was chosen so that it satisfied the "experimental" estimates of 200-400~T (see. Fig.~\ref{fig09}) and induced via~(\ref{eq017}) such a value of the axion-photon coupling constant ($g_{a \gamma} = 7.07 \cdot 10^{-11}$~GeV$^{-1}$) that would in its turn be strictly consistent with the known and very important limits~(\ref{eq094})-(\ref{eq095}) taking into account~(\ref{eq012}). More detailed justification of such self-consistent choice will be given in Section~\ref{sec-05}.

It is necessary to make a deviation concerning some important features of the oscillation length ($l = 2 \pi / q$) here. It is known that in order to maintain the maximum conversion probability, i.e. zero momentum transfer ($q \to 0$), the axion and photon fields, put into some medium ($m_{\gamma} \equiv m_{a}$), need to remain in phase over the length of the magnetic field. This coherence condition is met when $q L \leqslant \pi$, and along with~(\ref{eq007}) lets one obtain the following remarkable relation~\cite{ref014} between the medium density variations and axion mass variations for the coherent case of $q \to 0$, i.e. $m_{\gamma} \equiv m_a$

\begin{equation}
\frac{\Delta \rho}{\rho} = 2 \frac{\Delta m_a}{m_a} = \frac{4 \pi E_a}{m_a^2 L_{OT}}
\label{eq016b}
\end{equation}

\begin{figure}[tbp!]
  \begin{center}
    \includegraphics[width=12cm]{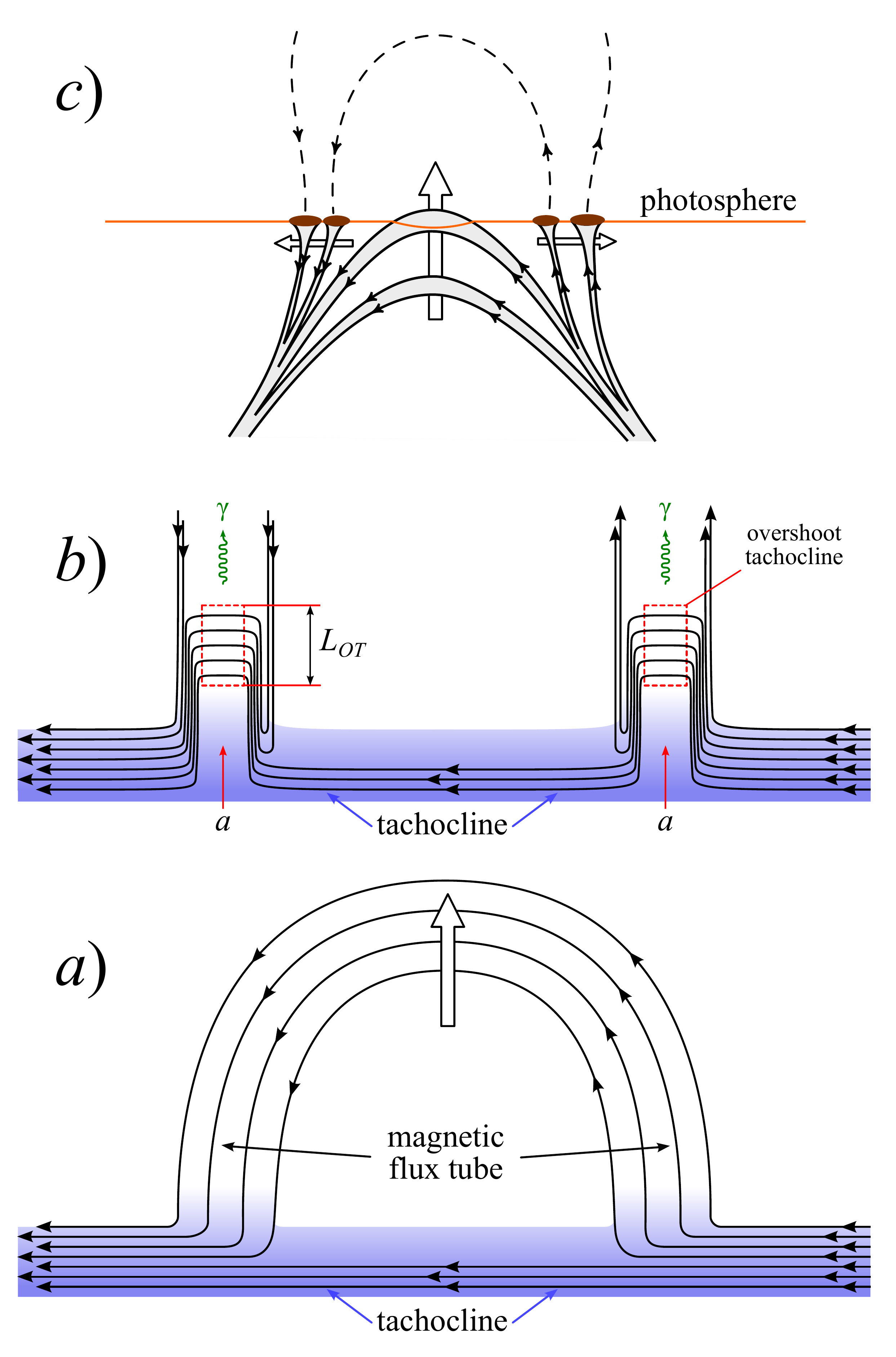}
  \end{center}
\caption{(a) Magnetic loop tubes formation in the tachocline through the shear flows instability development; (b) "Capillary" effect in magnetic tubes and the sketch of the axions (red arrows) conversion into $\gamma$-quanta inside the magnetic flux tubes containing the magnetic steps. Here $L_{OT}$ is the height of the magnetic shear steps. The tubes' rotation is not shown here for the sake of simplicity; (c) Emergence of magnetic flux bundle and coalescence of spots to explain the phenomenology of active region emergence (adopted from \cite{ref056},~\cite{ref066}).}
\label{fig10a} 
\end{figure}

It is easy to show that for the mean energy $E_a = 4.2 ~keV$, axion mass $m_a = 17 ~eV$ and the thickness of the overshoot tachocline zone $L_{OT} = 7.25 \cdot 10^7 ~m$ the density variations in~(\ref{eq016b}) are $\sim$10$^{-13}$. It means that the inverse coherent Primakoff effect takes place only when the variations of the medium density inside a cylindric volume of the "height" $L_{OT}$ (see Fig.~\ref{fig10a}b) do not exceed the value of $\sim$10$^{-13}$. This is a very strong restriction, since it is hard to imagine any kind of a physical process in the magnetic flux tube (see Fig.~\ref{fig10a}b) which would "freeze" the plasma (low-Z gas) in this magnetic volume so much so that this restriction on the density variations is fulfilled.

In other words, such mechanism that would validate the possibility of such locally "frozen" plasma existence is not known to us. At the same time, there are some arguments suggesting that such limitation is possible.

First of them is related to the experimentally obtained solar images (Fig.~\ref{fig12}, adapted from~\cite{ref014}) from the Japanese X-ray telescope Yohkoh (1991-2001) which illustrate the solar X-ray activity during the last maximum of the 11-year solar cycle (Fig.~\ref{fig12}b). There is currently no model alternative to the axion mechanism of sun luminosity which would have described the anomalous distribution of the X-ray radiation over the active Sun surface.

The second argument is related to finding of the axion with mass $m_a = 17 ~eV$ during the study of the EBL spectral intensity (see Section~\ref{sec-05}). On the one hand, this experimentally established fact suggests that the solar axions with mass $m_a = 17 ~eV$ are not only a theoretical prediction, but they really exist; and on the other hand, it is crucial for substantiation of the axion mechanism of Sun luminosity  and the solar dynamo -- geodynamo connection.

Finally, the third argument is related to the lack of understanding the link between the magnetic tubes formation and lifetime in the tachocline, and the length of the solar cycle. It means that we do not understand the mechanisms of solar activity as well as the processes of generation, accumulation and release of the magnetic energy responsible for the 11-year solar cycle as yet. This is applies especially to the current level of understanding of the causes and effects of  the differential rotation and meridional circulation in the tachocline. Let us remind that the tachocline is a thin transitional zone with the width of only $0.05 R_S$, where the latitudinal differential rotation of the convective zone turns into almost solid-body rotation of the radiative zone (e.g.~\cite{ref245}). It is generally believed that the main process of magnetic field generation -- solar dynamo -- responsible for the 11-year cycle takes place in this zone~\cite{ref246}. However, there are no evidence of the 11-year variations in the tachocline so far.Instead, mysterious variations of the rotation velocity with period of 1.3 year are observed here~\cite{ref247}, and curiously enough, they coincide with en estimate of the magnetic flux tube lifetime in the convective zone~\cite{ref065}.

Although there is no deep and detailed understanding of the magnetic tubes formation in the tachocline, one may assume the following picture of this process in the framework of the axion mechanism of Sun luminosity.First of all, numerous magnetic tubes appear (Fig.~\ref{fig10a}a) as a consequence of the shear flows instability development in the tachocline. As is shown in Appendix~\ref{A2}, the pressure inside such tubes is ultralow, which directly leads to formation and "floating-up" of the magnetic steps in these tubes (Fig.~\ref{fig10a}b). In other words, a kind of "capillary" effect is observed in this case. The whole picture of the magnetic tube spatio-temporal evolution in the convective zone depicted at Fig.~\ref{fig10a} leaves the question about the locally "frozen" plasma existence open, but at the same time it illustrates the process of axions conversion into $\gamma$-quanta in the overshoot tachocline and thus the mechanism of sun luminosity and active solar regions formation in the photosphere (Fig.~\ref{fig10a}c).

By normalizing the expression~(\ref{eq017}) the probability of the total conversion of axions into photons is assumed to be equal to a unit at given parameters of the magnetic field, i.e. $P_{a \to \gamma} = 1$. The primary criterion for such choice of the axion-photon coupling strength is the assumption about the maximum contribution of the luminosity produced by the axion to $\gamma$-quanta conversion in the tachocline zone (see Fig.~\ref{fig06}c,d) into the total Solar luminosity ($\Lambda_{Sun}$) during the active phase.

\begin{align}
\nonumber \Delta _a \cdot \left[ \Phi_{Pr} \cdot \left\langle E_a \right\rangle _{Pr} +  \Phi_{Brems} \cdot \left\langle E_a \right\rangle _{Brems} + \Phi_{Compt} \cdot \left\langle E_a \right\rangle _{Compt} + \Phi_{M1} \cdot \left\langle E_a \right\rangle _{M1}  \right] \times \\ 
\times (4 \pi R_{SE}^2) \times P_{a \rightarrow \gamma} = \Lambda_{Sun},
\label{eq018}
\end{align}

\noindent where\footnote{In the balance Eq.~(\ref{eq018}) we considered M1 transition of the $^{57}$Fe nuclei only and ignored the $^{55}$Mn and $^{23}$Na nuclei, because the favorable Boltzman factor of $^{57}$Fe produces the largest cooling rate near $T_8 \sim 1$~\cite{ref099}.} $\Delta _a = 0.90$ is a portion of the axion flux that transforms into $\gamma$-quanta in the tachocline by the inverse Primakoff effect\footnote{The choice of the $\Delta _a$ value is dictated by the spatial geometry of the solar tachocline zone (see Fig.~\ref{fig06}c,d) which is described in more detail below (see Fig.~\ref{fig09}). Let us note that the portion of the "invisible" axions equal to $(1 - \Delta _a$) satisfies the neutrino limit on axions, i.e. the Gondolo-Raffelt criterion~\cite{ref080,ref100} at the same time.}; $P_{a \to \gamma} = 1$; $\Lambda _{Sun} = 3.84 \cdot 10^{26} \pm 1.5\% ~W$~\cite{ref084,ref101}; $R_{SE} = 1.496 \cdot 10^{13} ~cm$ is the distance from the Earth to the Sun; $\langle E_a \rangle _{Pr} = 4.2~keV$, $\langle E_a \rangle _{Brems} = 1.6 ~keV$, $\langle E_a \rangle _{Compt} = 5.1 ~keV$, $\langle E_a \rangle _{M1} = 14.4 ~keV$ are the average energies of the solar axions spectra; $\Phi_{Pr}$, $\Phi_{Brems}$, $\Phi_{Compt}$, $\Phi_{M1}$ are the integral solar axions fluxes generated by the Primakoff effect, the bremsstrahlung, the Compton process and the M1 transition in the $^{57}$Fe nuclei on the Sun respectively, which were obtained using the known differential spectra $d \Phi_{Pr} / dE$~\cite{ref025}, $d \Phi_{Brems} / dE$~\cite{ref006}, $d \Phi_{Compt} / dE$~\cite{ref006}, $d \Phi_{M1} / dE$~\cite{ref002}:

\begin{equation}
\Phi _{Pr} = 3.75 \cdot 10^{31} g_{a \gamma}^2 ~~ cm^{-2}s^{-1},
\label{eq019}
\end{equation}

\begin{equation}
\Phi _{Brems} = 1.43 \cdot 10^{35} g_{a e}^2 ~~ cm^{-2}s^{-1},
\label{eq020}
\end{equation}

\begin{equation}
\Phi _{Compt} = 2.16 \cdot 10^{34} g_{a e}^2 ~~ cm^{-2}s^{-1},
\label{eq021}
\end{equation}

\begin{equation}
\Phi _{M1} = 1.66 \cdot 10^{23} g_{a n}^2 ~~ cm^{-2}s^{-1};
\label{eq022}
\end{equation}

\noindent where $g_{a \gamma}$, $g_{ae}$, $g_{an}$ are the axion coupling constants to photons ($g_{a \gamma}$), electrons ($g_{ae}$) and nucleons ($g_{an}$) respectively\footnote{Although Eqs.~(\ref{eq019})-(\ref{eq022}) were obtained in the framework of the standard model of the Sun, they may be applied to the axion model as well, since the basic thermodynamic parameters (temperature, pressure, plasma density etc.) of the solar core are roughly the same in both models, and almost all the axions are produced in the solar core, regardless of the exact way of their birth.}. 

We obtained only the axion-photon coupling constant $g_{a \gamma}$ (see~(\ref{eq017})) out of the three axion coupling constants so far. Next it is not too hard to estimate the axion-nucleon coupling constant basing on the hadronic axion models~\cite{ref102,ref103}.

\begin{equation}
g _{an} = \vert g_0 \beta + g_3 \vert,
\label{eq023}
\end{equation}

\noindent where

\begin{equation}
g_0 = - \frac{m_N}{6 f_a} \left[ 2S + (3F - D) \frac{1 + z - 2w}{1 + z +w} \right],
\label{eq024}
\end{equation}

\begin{equation}
g_3 = -\frac{m_N}{2 f_a} \left[ (D+F) \frac{1-z}{1+z+w} \right].
\label{eq025}
\end{equation}

Here the value $\beta = -1.19$ for the M1 transition in the $^{57}$Fe nucleus was calculated in~\cite{ref102,ref103}; spontaneous breaking of the PQ-symmetry (in view of~(\ref{eq010})-(\ref{eq011})) takes place at the energy $f_a \approx 0.119 \cdot 10^6$ GeV; $m_N = 939$~MeV is the nucleon mass. The exact values of $D$ and $F$ parameters determined from semileptonic hyperon decays are equal to $D = 0.808 \pm 0.006$ and $F = 0.462 \pm 0.011$~\cite{ref104}. Parameters $z = m_u / m_d \approx 0.56$ and $w = m_u / m_s \approx 0.029$ are quark mass ratios~\cite{ref032,ref033}.

The parameter $S$ characterizing the flavor singlet coupling still remains a poorly constrained one. Its value varies from $S = 0.68$ in the naive quark model down to $S = −0.09$ which is given on the basis of the EMC collaboration measurements~\cite{ref105}. The more stringent boundaries ($0.37 \leqslant S \leqslant 0.53$) and ($0.15 \leqslant S \leqslant 0.5$) were found in~\cite{ref106} and~\cite{ref107}, accordingly. As a result the value of the sum~(\ref{eq023}) may significantly decrease and, due to negativity of the parameter $\beta$, actually vanish. Taking into account that the usually accepted value of $u$- and $d$-quark mass ratio $z = 0.56$ can vary in $0.35 \div 0.6$ range~\cite{ref108}, the exact interpretation of experimental results is significantly restricted.

Calculations performed using the expressions~(\ref{eq023})-(\ref{eq025}) and~(\ref{eq012}) let one derive the value of $g_{an} = 2.46 \cdot 10^{-6}$ for the axion-nucleon coupling constant at $S = 0.55$. However, it should be noted that such a model-dependent value does not suit our needs, and here is why.

It is easy to show using (\ref{eq022}) that the luminosity $\Lambda _{M1}$ of the axions produced by M1 transition of $^{57}$Fe nuclei depends on the solar photon luminosity $\Lambda _{Sun}$ in the following way\footnote{Here we follow the calculations by Derbin (e.g.~\cite{ref003}), that is why the expression~(\ref{eq026}) differs slightly from the analogous expression (2.13) in~\cite{ref110}. The difference mainly comes from the different values of the Doppler spectrum broadening used during the calculation of the monoenergetic solar axions flux produced by the $^{57}Fe$  nuclear de-excitations (e.g.~\cite{ref003,ref110}).}:

\begin{equation}
\Lambda _{M1} \cong 2.8 \cdot 10^9 g_{an}^2 \Lambda_{Sun},
\label{eq026}
\end{equation}

Substituting the value of $g_{an} = 2.46 \cdot 10^{-6}$ into~(\ref{eq026}) we derive that the relative axion luminosity $\Lambda _{M1}$ is about $\sim 2\%$. This value is inadmissible for the axion mechanism of Sun luminosity, because otherwise the resulting solar photon spectrum (Fig.~\ref{fig07}) would contain a rather high peak near $E_a = 14.4$~keV. As a matter of fact, if it is there, then it is very weak considering the spectrum uncertainty in this band (see Fig.~9 in~\cite{ref109}).

In this connection let us from now on assume that the relative axion luminosity $\Lambda _{M1}$ is

\begin{equation}
\Lambda_{M1} / \Lambda_{Sun} = 0.003.
\label{eq027}
\end{equation}

The choice of such relation was made so that the relative axion luminosity $\Lambda _{M1}$ allowed the 14.4~keV peak existence within the experimental Sun photon spectrum uncertainty and conformed to the axion-mucleon coupling constant from the theoretically allowed limitations known as the SN1987A limit ($3 \cdot 10^{-7} \leqslant g_{an} \leqslant 10^{-6}$~\cite{ref111,ref112}).

Substituting~(\ref{eq027}) into~(\ref{eq026}) we obtain a consistent value of the axion-nucleon coupling constant\footnote{Hereinafter by $g_{an}$ and $g_{ae}$ we always mean $\vert g_{an} \vert$ and $\vert g_{ae} \vert$ respectively.}

\begin{equation}
g_{an} \cong 3.2 \cdot 10^{-7}.
\label{eq028}
\end{equation}

Now let us turn again to Eq.~(\ref{eq017}) which describes the axion mechanism of Sun luminosity hypothesis. Apparently, the found axion coupling constants to photons ($g_{a \gamma} = 7.07 \cdot 10^{-11} ~GeV^{-1}$) and nucleons ($g_{an} = 3.2 \cdot 10^{-7}$) let one calculate the axion-electron coupling constant ($g_{ae}$) by means of Eqs.~(\ref{eq017})-(\ref{eq022}). It turned out to be

\begin{equation}
g_{ae} \cong 5.28 \cdot 10^{-11}.
\label{eq029}
\end{equation}

It is interesting to note that within such approach, bremsstrahlung ($\sim 2.59 \cdot 10^{26}$ W) has the major part in the Solar luminosity, which is $\sim 67.48 \%$, while the  Compton process ($\sim 1.24 \cdot 10^{26}$ W), the Primakoff effect ($\sim 3.19 \cdot 10^{23}$ W) and the M1 transition of $^{57}$Fe nuclei ($\sim 9.93 \cdot 10^{22}$ W) make 32.41\%, 0.08\% and 0.03\% respectively. Differential solar axions spectra as observed at the Earth originating from bremsstrahlung, the Compton process, the Primakoff effect and M1 ground-state nuclear transition in solar $^{57}$Fe are shown on Fig.~\ref{fig02}b.

Thus, it is obvious that the mechanism of luminosity and the X-ray spectrum shape for the active and quiet Sun (Fig.~\ref{fig07}) can be easily explained by our model, in which axions are converted into $\gamma$-quanta in the solar tachocline under certain conditions. First of all, the $\gamma$-quanta energy spectrum generated by axions in the solar tachocline zone due to the channeling effect practically does not change to the boundary of the photosphere of the Sun, where it transforms because of the Compton scattering, as is shown in Figs.~\ref{fig06}c,d and Fig.~\ref{fig07}. And, finally, it is obvious that the integral of this spectrum (see Fig.~\ref{fig07}) by virtue of the equality~(\ref{eq018}) coincides by the order of magnitude with the estimation of the Sun luminosity.

\subsection{Invisible axions and Solar Equator effect}
\label{sec-02-5}

\begin{figure}
    \begin{center}
        \includegraphics[width=10cm]{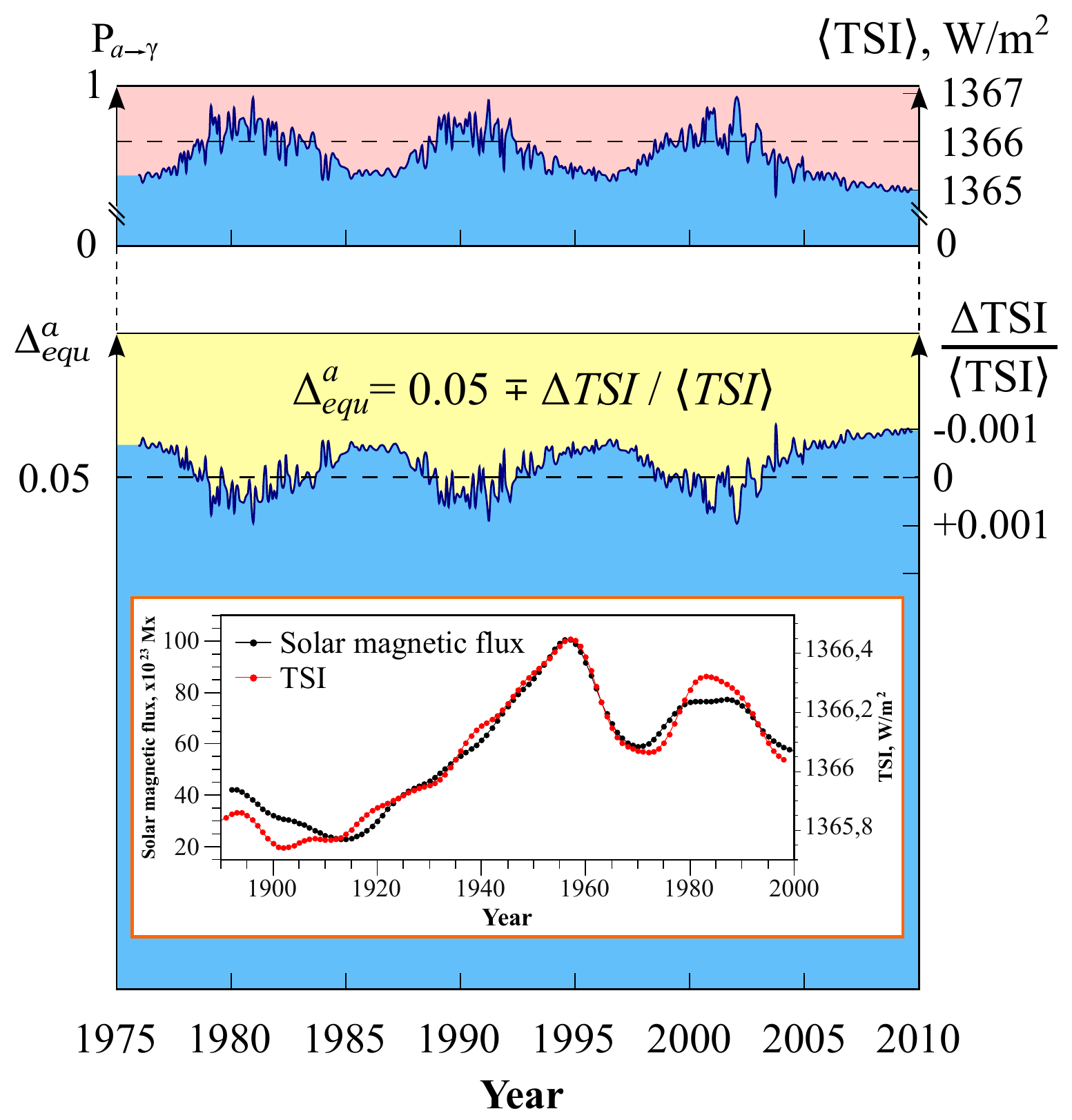}
    \end{center}
\caption{Time evolution of $P_{a \to \gamma}$ and TSI during 1975-2010. Inset: time evolution of \textbf{(a)} the variations of the magnetic flux in the tachocline zone of the Sun~\cite{ref012}), \textbf{(b)} TSI annual variations~\cite{ref101}. Curves are smoothed by the sliding intervals in 5 and 11 years.}
\label{fig11}
\end{figure}

As it was already stated before, the solar magnetic field variations (Fig.~\ref{fig03}) must drive (with the help of the solar axions and the inverse Primakoff effect) the total solar irradiance (TSI) and 14.4~keV axions variations at the same time (see the inset in Fig.~\ref{fig11}). It is important to note here that, however strange it may seem, TSI variations are not the modulator of the Earth climatic system (ECS) global temperature, because the strong inverse\footnote{A lot of climatologists still believe that it is the TSI variations that are responsible for the Earth global temperature variations obstinately disregarding the facts of a very small contribution of TSI variations into the energy balance in the Earth atmosphere and the inverse correlation between TSI and global temperature variations (see Fig.1 in~\cite{ref009}, where the ocean level is a proxy for the global temperature).} correlation with the 22-year lag is observed between them~\cite{ref009}. And vice versa, the variations of the solar axion flux manifest the strong positive correlation with the global temperature variations with the same time lag. This fact plays a key role in the new global climate theory~\cite{ref113,ref114,ref115}, which considers the variations of the 14.4~keV solar axions (which are resonantly absorbed in the Earth core) a trigger-like modulator of all the thermal processes in ECS and, particularly, in the atmosphere. However, this problem requires a special discussion and therefore will be examined in a separate paper.

On the other hand, as follows from Fig.~\ref{fig11}, the TSI variations during the active phase of the Sun are so small ($\sim 1 ~W / m^2$~\cite{ref101}), that the relative portion of 14.4~keV axions must also be small at the Earth.

\begin{equation}
p_{TSI} = \frac{ \Delta R_a}{R_a} = \frac{(TSI) \cdot 4 \pi r_{SE}^2}{L_{Sun}} \sim 10^{-3}
\label{eq030}
\end{equation}

Therefore their heat power in the Earth core

\begin{equation}
\mathcal{R} = p_{TSI} \cdot R_a \cdot N_{Fe}^{57} \cdot E_a \sim 20 ~~W,
\label{eq031}
\end{equation}

\noindent is not enough not only for the geomagnetic field generation (which requires at least $\geqslant$0.1~TW~\cite{ref018}), but for the geomagnetic field variations also ($\sim$0.01~TW). Here $R_a = 5.16 \cdot 10^{-3} g_{an}^4$, $N_{Fe}^{57} \sim 3 \cdot 10^{47}$ is the number of $^{57}$Fe nuclei in the Earth core, $E_a = 14.4$~keV is the $^{57}$Fe solar axions energy.

At the same time it is not difficult to see that the relative part of the axions $\Delta_a$ that are almost not affected by the Primakoff effect in the polar ($\Delta_{pol}$) and equatorial ($\Delta_{equ}$) sectors of the tachocline zone~(Fig.~\ref{fig12}) is a considerable quantity:

\begin{figure}[tb!]
    \begin{center}
        \includegraphics[width=15cm]{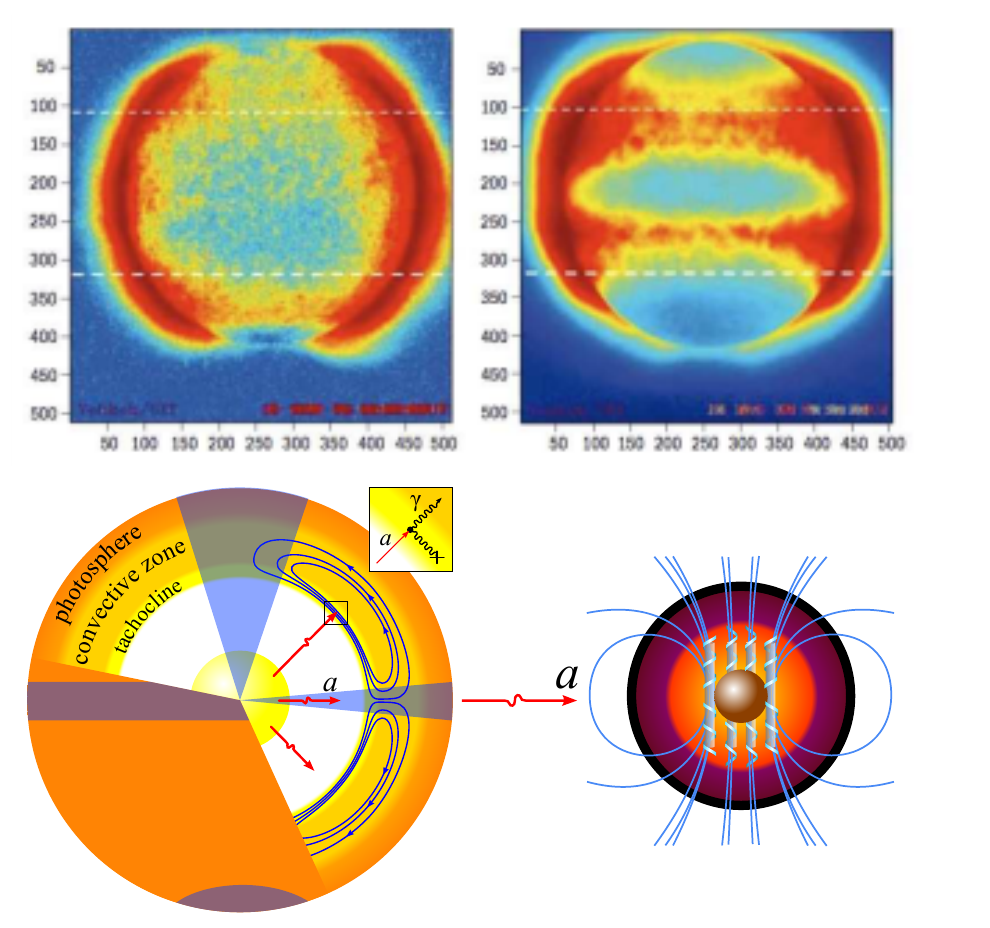}
    \end{center}
\caption[caption]{\textbf{Top:} Solar images at photon energies from 250~eV up to a few keV from the Japanese X-ray telescope Yohkoh (1991-2001) (adapted from~\cite{ref014}). The following is shown: on the left, a composite of 49 of the quietest solar periods during the solar minimum in 1996. On the right, the solar X-ray activity during the last maximum of the 11-year solar cycle. According to Fig.~\ref{fig06} (and Fig.~\ref{fig12}, bottom) most of the X-ray solar activity (right) occurs at a wide bandwidth of $\pm 45^{\circ}$ in latitude, being homogeneous in longitude. Note that $\sim 95\%$ of the solar magnetic activity covers this bandwidth~\cite{ref116} (see also a similar topology for microflares measured with RHESSI~\cite{ref117}). \\
\textbf{Bottom:} Schematic picture of the solar tachocline zone, the Earth's liquid outer (the red region) and inner (the brown region) core. Blue lines on the Sun designate the magnetic field. In the tachocline axions are converted into $\gamma$-quanta,  which form the experimentally observed solar photon spectrum (Fig.~\ref{fig07}) after passing through the photosphere. Solar axions moving towards the poles (blue cones) and in the equatorial plane (blue bandwidth) are not transformed by the Primakoff effect, since the magnetic field vector is almost collinear to their momentum vector in these regions. Solar axions are then resonantly absorbed by iron in the Earth core transforming into $\gamma$-quanta, which are the supplementary energy source in the Earth core (see the text).}
\label{fig12}
\end{figure}

\begin{equation}
\Delta _a = \Delta_{equ}^a + \Delta_{pol}^a,
\label{eq032}
\end{equation}

If we assume that the equatorial surface ($S_{equ}$) formed by two cones going from the center of the Sun, and a part of the sphere with the radius of the Sun ($R{Sun}$) has the dihedral angle of $\sim 5^{\circ}$, it is easy to show that\footnote{This estimate was made on the basis of the numerous computational experiments on magnetic field evolution in the convective zone of the Sun (e.g.~\cite{ref012} and Refs. therein).}

\begin{equation}
\Delta_{equ}^a = \frac{S_{equ}}{S_{Sun}} \sim 0.05.
\label{eq033}
\end{equation}

The expression~(\ref{eq033}) means that because of the quasi-collinearity of the Earth and Sun rotation axes, the main axion flux directed towards the Earth originates from the equatorial sector of the Sun (Fig.~\ref{fig12}). In this connection a short remark should be made regarding the anti-correlation between the solar magnetic field and geomagnetic field variations. Weak variations of TSI are obviously produced by the solar magnetic field variations. The same solar magnetic field variations are the cause of the "equatorial" axion flux variations. In other words, the "equatorial" effect not only generates the "invisible" axions, but also modulates their intensity inversely proportional to the solar magnetic field changes, producing the observed inverse correlation between them (Fig.~\ref{eq011}). The latter is supposed to be the main cause of the anticorrelation between the solar magnetic field variations and the geomagnetic field variations.

The assumptions used to derive the expressions~(\ref{eq013}) and~(\ref{eq033}) are called forth by the necessity of justification both the axion mechanism of Sun luminosity and the axion mechanism of solar dynamo -- geodynamo connection, and will be additionally substantiated below.

\subsection{Power required to maintain the Earth magnetic field and nuclear georeactor}

It is not hard to show that the resonant absorption rate of 14.4~keV solar axions in the Earth core, which contains the $N_{Fe}^{57}$ nuclei of $^{57}$Fe isotope, is about~\cite{ref003}

\begin{equation}
R_a \cong 5.16 \cdot 10^{-3} g_{an}^4 \cdot N_{Fe}^{57} \cdot \Delta_{equ}^a,
\label{eq034}
\end{equation}

\noindent where $\Delta _{equ}$ is the portion of axions reaching the Earth via the solar equator effect (see~(\ref{eq033})).

It is known, that the number of $^{57}$Fe nuclei in the Earth core is $N_{Fe}^{57} \sim 3 \cdot 10^{47}$ and the average energy of $^{57}$Fe solar axions is $E_a = 14.4$~keV. Then with an allowance for Eq.~(\ref{eq034}) and the value of the axion-nucleon coupling constant~(\ref{eq029}) the maximum energy release rate $\Delta D_{ohmic}^a$ in the Earth core is equal to

\begin{equation}
\Delta D_{ohmic}^a = R_a \cdot E_a \simeq 1.2 ~~ kW.
\label{eq035}
\end{equation}

This estimate of the heat power supplied to the Earth core by the absorbed axions~(\ref{eq035}) apparently is much less then the value necessary to generate the magnetic field of the Earth ($\geqslant 0.1$~TW~\cite{ref018}). Moreover, it is small even in comparison with the heat power fluctuations ($\sim$0.01~TW) responsible for the geomagnetic field variations in the Earth core (see the inset in Fig.~\ref{fig11}).

It is easy to illustrate this using the known dependence of the core magnetic field $B_{in}$ on ohmic dissipation $D_{ohmic}$ in the Earth core~\cite{ref018,ref082,ref118} in the form:

\begin{equation}
D_{ohmic} = \frac{\eta \cdot V}{\mu \cdot l_B ^2} B_{in}^2 \sim 0.1 ~~ TW,
\label{eq036}
\end{equation}

\noindent where $\eta \sim 2 ~m^2 / s$ is magnetic diffusivity~\cite{ref017}, $V = (4/3) \pi R_C^3$ is the volume of the Earth core, $R_C = 3480 ~km$ is the radius of the Earth core, $\mu \sim 1$ is permeability, $l_B \sim 0.8 \cdot 10^5 ~m$ is the characteristic length scale on which the field vector changes~\cite{ref017}, $B_{in} \sim 4~mT$ is the core magnetic field~\cite{ref119}.

In order to estimate the heat power fluctuations $\Delta D _{ohmic}$ responsible for the magnetic field variations in the Earth core let us represent~(\ref{eq036}) in differential form:

\begin{equation}
dD_{ohmic} = \frac{2 \eta \cdot V}{\mu \cdot l_B ^2} B_{in} dB_{in},
\label{eq037}
\end{equation}

The right-hand side of~(\ref{eq037}) contains the known estimates except for the magnetic field fluctuations $\Delta B_{in} \sim d B_{in}$. On the other hand, there are well known long-term records of magnetic field variations measured on the Earth surface (e.g.~\cite{ref013}). This lets us estimate $d D_{ohmic}$ by writing down the expression~(\ref{eq034}) in the following form:

\begin{equation}
dD_{ohmic} = \frac{2 V}{\mu} B_{in} \frac{dB_{in}}{dt},
\label{eq038}
\end{equation}

\noindent where the magnetic field variations ($dB_{in} / dt$) in the Earth core appear after taking into account the magnetic diffusion

\begin{equation}
dB_{in} \cong \frac{dB_{in}}{dt} \delta t = \frac{dB_{in}}{dt} \cdot \frac{l_B^2}{\eta}.
\label{eq039}
\end{equation}

If one also takes into account the known relation between the inner ($dB_{in} / dt$) and outer ($dB_{out} / dt$) geomagnetic field variations~\cite{ref120},

\begin{equation}
\frac{dB_{out}}{dt} = \left( \frac{\gamma R_C}{R_E} \right)^2 N \cdot  \frac{dB_{in}}{dt},
\label{eq040}
\end{equation}

\noindent then it is possible to make an estimate of the ohmic dissipation fluctuations ($\Delta D_{ohmic} \sim d D_{ohmic}$) basing on (\ref{eq037})-(\ref{eq039}) necessary for inducing a certain number $N$ of Taylor cells in the core~\cite{ref120}.

\begin{equation}
\Delta D_{ohmic} = \frac{2 V}{\mu} B_{in} \left( \frac{R_E}{\gamma R_C} \right)^2  \frac{dB_{out}}{dt} \simeq 0.02 ~~ TW,
\label{eq041}
\end{equation}

Here $R_E = 6357 ~km$ is the radius of the Earth, $\gamma = 0.1$~\cite{ref120}, $(dB_{out} / dt) \approx 20 ~ nT / yr$ is the annual variation of the external magnetic field of the Earth core~\cite{ref121,ref020}.

A natural question arises from the stated above about the way that 14.4~keV solar axions may provide an effective mechanism of solar dynamo -- geodynamo connection while supplying a rather low heat power. In other words, how does this problem reduce to the mechanism of small heat perturbations critical influence on the convective process in the Earth liquid core. The problem is stated this way because if there is an effective mechanism of convective instabilities generation by weak heat perturbations in the Earth liquid core, then this effect may simultaneously cause substantial weakening of the convective heat removal from the Earth solid core. The intense weakening of the heat removal from the Earth solid core surface, in its turn, causes the corresponding temperature increase in the solid core boundary layer. This is very important because it promotes the subsequent effective convection stability recovery in the liquid core.

\begin{figure}
    \begin{center}
        \includegraphics[width=18cm]{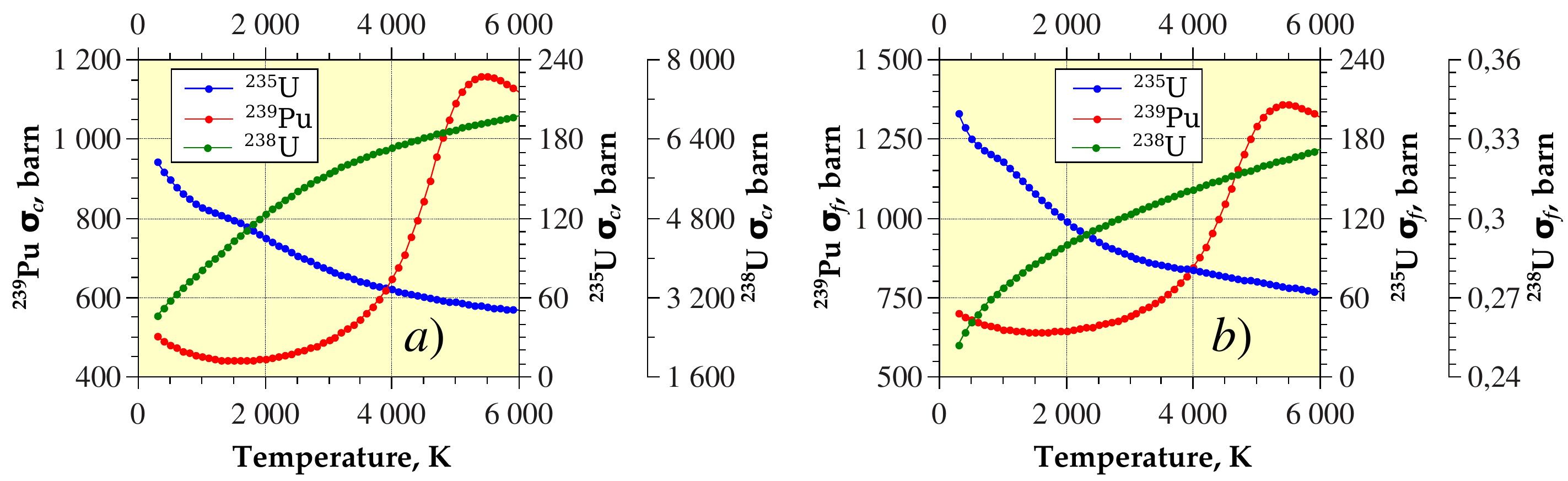}
    \end{center}
\caption{Dependence of \textbf{(a)} capture cross-sections and \textbf{(b)} fission cross-sections for $^{235}$U (blue), $^{238}$U (green), and $^{239}$Pu (red) averaged over the neutron spectrum on the fuel medium temperature for the limiting  energy (3kT) of the Fermi and Maxwell spectra~\cite{ref124}. The neutron spectra averaging procedure was applied for the concentrational fuel composition of the nuclear georeactor discussed in~\cite{ref122}.}
\label{fig13}
\end{figure}

There are strong grounds to believe that there is a natural nuclear georeactor operating at the boundary between the Earth solid core and liquid core. The analysis of the KamLAND experiment neutrino spectra for 2002-2008 shows that the heat power of this non-stationary traveling wave reactor (TWR) is about 30~TW~\cite{ref122,ref123,ref124}. The heat power of such TWR depends on the nuclear fuel composition and the medium temperature. The latter is because of the fact that according to~\cite{ref123,ref124}, $^{238}$U and $^{239}$Pu capture and fission cross-sections depend quasi-linearly on the temperature of the neutron-multiplicating medium in the 3000-5500~K range (Fig.~\ref{fig13}).

These peculiarities of TWR are responsible for the positive feedback that leads to the reactor heat power increase after the corresponding boundary layer temperature increase (see Fig.~\ref{fig13}). The georeactor heat power growth lasts until the steady heat removal from the TWR is reestablished, which implies restoration and stabilization of the convection in the Earth liquid core.

Now let us turn back to the physical essence of the axion mechanism of the weak convective instability thermal perturbations in the liquid core. In this connection it should be noted that the convection in the Earth core is compositional. It means that there are some light elements originating, in particular, from the $^{239}$Pu nuclei fission which take part in the convection along with the "iron" component~\cite{ref122,ref123,ref124}. It turns out that the convective instability may appear in such media even under hydrostatically stable density stratification, i.e. when the density decreases with height~\cite{ref125,ref126,ref127}.

It is known that the phenomenon of the convective instability caused by the double (differential) diffusion was discovered rather long ago and has been described in numerous overviews and monographs in detail (e.g.~\cite{ref125,ref126,ref127}). The principal role in this case usually belongs to the difference between the two hydrodynamic components of heat and admixture~\cite{ref125}. The convection caused by double diffusion is generally believed to appear when the thermal medium stratification is stable, while the weakly diffusing admixture (e.g. light elements) introduces a destabilizing contribution into the density stratification. Although this contribution may be relatively small, it may be enough for destabilization of a stably stratified (in terms of density) system owing to the mentioned effects~\cite{ref127}. 

However, we are interested in the conditions of the convective instability formation in a qualitatively different situation, particularly, when a weakly diffusing admixture, on the contrary, introduces a stabilizing contribution into the density stratification. This contribution may even exceed the thermal instability in absolute value. Such possibility may seem paradoxical at first glance since due to double diffusion effects the slowly diffusing admixture usually has the much greater impact on the convective instability than the quickly transportable heat, all other factors being equal. Let us show that this is not always the case by analyzing the situation when there is a slow background motion along the gravity force described in~\cite{ref127} in detail.

\begin{figure}
    \begin{center}
        \includegraphics[width=15cm]{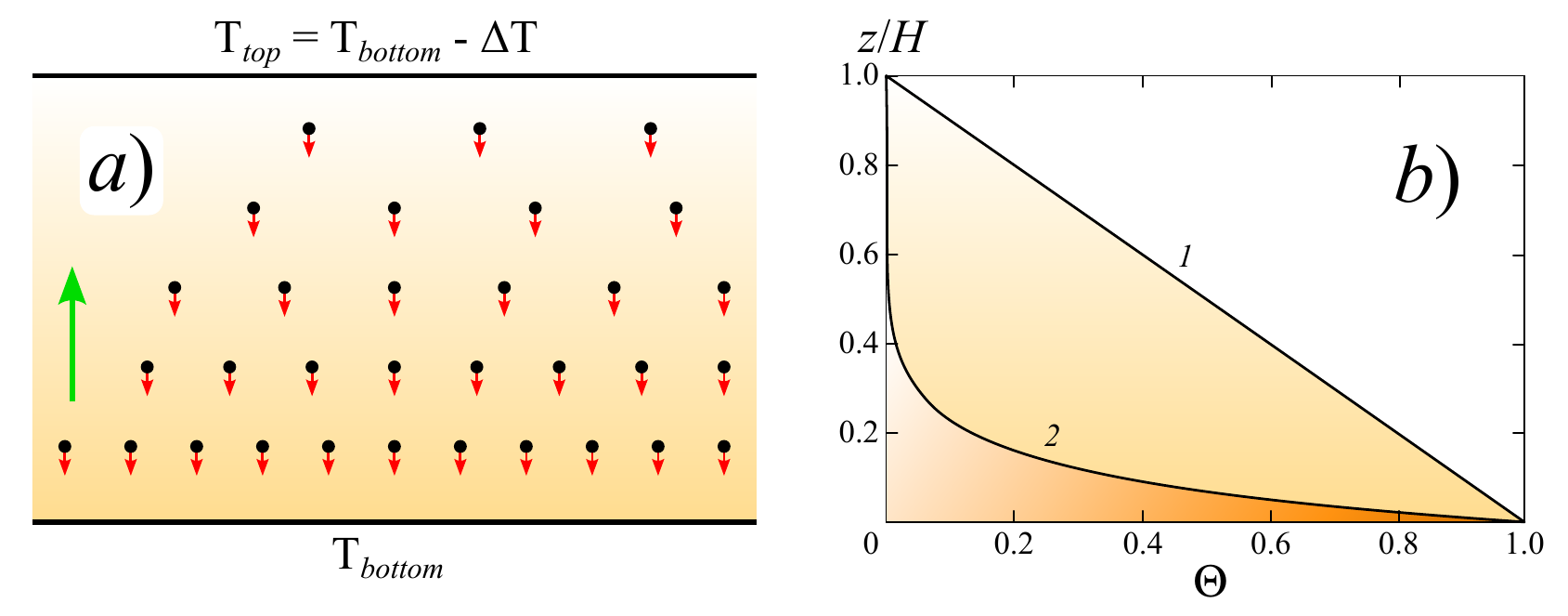}
    \end{center}
\caption{\textbf{(a)} A sketch of the thermal convection with the resonant absortion of 14.4~keV solar axions by iron nuclei producing the 14.4~keV $\gamma$-quanta flux (red arrows). Here $\gamma$-quanta flux emulates a slow background descending motion in the convecting medium. The green arrow denotes the convection direction. \textbf{(b)} Distortion of the vertical background temperature distribution by the downward motion: (1) the zero background vertical velocity, $w = H / h = 0$; (2) $w = 10$ (adapted from~\cite{ref127}).}
\label{fig14}
\end{figure}

A problem on convection on the background of the slow (relative to the characteristic speed of the studied convective motions) vertical motion was first considered in~\cite{ref128}. It is of a considerable interest for us, since the resonant absorption of 14.4~keV solar axions in the iron nuclei may be considered as a process that induces a slow descending background motion (Fig.~\ref{fig14}a) in the convective medium of the liquid core.

Following~\cite{ref127}, let us consider a single-component medium with its density depending on the temperature $T$ only (neglecting the admixture stratification effects). In other words, we are considering a modification of the classical Rayleigh-B\'{e}nard problem on the convective stability of a liquid between two horizontal plates~\cite{ref125}. A slow vertical motion along the gravity force is assumed to be present in the background state. For the sake of simplicity let us consider a motion with the velocity $W < 0$ independent of the vertical coordinate $z$ counted from the top boundary. Let us also consider the bottom and top boundaries temperatures $T_{bot}$ and $T_{top}$ fixed and denote the difference between them by $\Delta T$. Heat transfer in the background flow is described by the equation

\begin{equation}
- W \frac{dT}{dz} = \kappa \frac{d^2 T}{dz^2},
\label{eq042}
\end{equation}

\noindent where $\kappa$ is the thermal diffusivity. A solution for the two boundary conditions mentioned above may be written down in the form

\begin{equation}
\Theta (z) = \frac{\exp (- \xi) - \exp (-w)}{1 - \exp (-w)}.
\label{eq043}
\end{equation}

Here $\Theta (z) = [T(z) - T_{top}] / \Delta T$ is the dimensionless temperature deviation, $\xi = h / \kappa$ is the dimensionless vertical coordinate, and $h = \kappa / W$ is the reference height associated with the vertical motion (infinity in the quiescent fluid). The key dimensional parameter is

\begin{equation}
w = \frac{H}{h} = W(H / \kappa),
\label{eq044}
\end{equation}

\noindent where $H$ is the fluid layer thickness. In the absence of the background vertical motion (in the limit $W \to 0$, $h \to \infty$, and $w \to 0$), the result is the expected linear profile

\begin{equation}
\Theta = 1 - z / H,
\label{eq045}
\end{equation}

\noindent i.e. the solution whose stability is analyzed in the classical Rayleigh-B\'{e}nard problem. Fig.~\ref{fig14}b shows the vertical profiles of (\ref{eq045}) for $w = 0$ and $w = 10$~\cite{ref127,ref128}. Apparently, the background medium sinking "pushes" almost all temperature difference $\Delta T$ to the bottom boundary where it concentrates within a layer with the thickness $h = \kappa / W$.

Generally speaking, a rigorous stability study of the stationary state with a curvilinear temperature profile and the background sinking is a rather complex problem. An estimate of the effective Rayleigh number for the bottom sublayer, which incorporates virtually all the vertical temperature difference $\Delta T$ (Fig.~\ref{fig14}b), was made on the basis of the simple physical reasoning in the paper~\cite{ref128}.

\begin{equation}
Ra \sim \frac{\alpha \cdot g \cdot \Delta T \cdot h^3}{\kappa \nu} \sim \frac{\alpha \cdot g \cdot \Delta T \cdot \kappa^2}{\nu \cdot W^3},
\label{eq046}
\end{equation}

Here $\alpha$ is the thermal expansion coefficient of the fluid, $\nu$ is the kinematic viscosity, and $g$ is the gravitational acceleration.

Let us denote the value of the effective Rayleigh number, which corresponds to  the loss of stability, by $Ra_{cr}$. The sinking rate sufficient for stability loss prevention in this case is expressed by the equation

\begin{equation}
W_{cr} \sim \left( \frac{\alpha \cdot g \cdot \Delta T \cdot \kappa ^2}{\nu \cdot Ra_{cr}} \right)^{1/3} \sim 10^{-4} ~~ m \cdot s^{-1},
\label{eq047}
\end{equation}

\noindent where $Ra_{cr}$  is a complex function of $\varepsilon$~\cite{ref129}

\begin{equation}
Ra_{cr} \sim E^{1.16} \left[ 0.21 \varepsilon^{-2} + 22.4 (1 - \varepsilon^2)^{1/2} \right] \sim 10^{-6}, ~~ \varepsilon = R_{in} / R_{out},
\label{eq048}
\end{equation}

For example, setting $\kappa \sim 0.1 ~m^2 \cdot s$, $\nu \sim 10^3 ~m^2 \cdot s$ (effective turbulent transport coefficients characteristic for the liquid core~\cite{ref130}), $\alpha \sim 10^{-5} ~K^{-1}$~\cite{ref131}, $\Delta T \sim 1000 ~K$~\cite{ref131}, $g = 11 ~m \cdot s^{-2}$~\cite{ref131}, $E \sim 10^6$ is the Ekman number for a given $\kappa$~\cite{ref130}, one obtains $W_{cr} \leqslant 10^{-4} ~m \cdot s^{-1}$.

It is interesting to note that the background sinking leads to an effective Rayleigh number decrease, i.e. $Ra_{cr} < Ra$, where $Ra > 10^6$ is Rayleigh number in Earth core~\cite{ref131}, and consequently, to a decrease in convective instability formation probability. This value of downward velocity is also in good agreement with the value of the characteristic velocity of the Earth core convection ($\sim 4 \cdot 10^{-4} ~m \cdot s^{-1}$~\cite{ref130}), while the results obtained in both field experiments and numerical simulations (see Fig.~\ref{fig14}b) demonstrate that downward flow with this velocity suppresses convection~\cite{ref128}.

Now let us pass on to a two-component medium which has its unstable thermal stratification under the absence of the vertical motion overcompensated by a stable admixture stratification. As is shown in~\cite{ref128}, if there is a slowly diffusing admixture stratification along with the temperature stratification (light elements in our case), admixture is suppressed by the background movement more effectively than the heat in such two-component medium. In other words, even the presence of a very slow downward movement may pull the admixture down as opposed to the heat and thus cancel its stabilizing effect, and the system becomes unstable. It is also known that the temperature profile may be deformed as well under more intense background motions. It is important to keep in mind that the mentioned effects are possible under very small vertical velocities of the background motions. As the authors of~\cite{ref128} point out, we are dealing with a new kind of instability. Note, however, that the times required for a system to evolve into the unstable steady states considered above may be very large~\cite{ref128}.

It may be concluded that the vertical background motions may prevent the convective instability formation in the single-component media while leading to a destabilization of a two-component medium layer. It happens because the background motions are  an immediate cause of the effective vertical drift of a slowly diffusing admixture. An important fact to remember is that the above-mentioned effects are possible even under very small vertical background motion velocities\footnote{It is interesting to note here that a problem on convection in presence of the slow background motions is extremely urgent for the known geophysical applications related to e.g. cloud patterns and atmospheric circulation~\cite{ref128,ref132,ref133,ref134}. For example, the atmospheric and oceanic convection often takes place on the background of the processes with much larger horizontal scales (cyclones and anticyclones) which are characterized by the average vertical motions several orders of magnitude slower than those that appear during a convective instability formation. According to the natural experiments~\cite{ref134}, even a slow background sinking of the medium can effectively suppress the convection in the atmosphere.}.

The problem on convective instability caused by the background motion is examined in its simplest form so far. The rotation effects, magnetic field influence, nonlinear background motion velocity etc. were not taken into account here, but all of them are actually present in a traditional composite media magnetohydrodynamics in the Earth core. A detailed consideration of these effects is beyond the scope of the present paper.  The purpose of the current section is to demonstrate a possibility of a nontrivial impact of background motions, which may be produced by the resonant absorption of the "iron" solar axions along with the convection in the Earth liquid core, within a simple model.

Thus, the essence of the axion mechanism of solar dynamo -- geodynamo connection lies in the following. The resonant absorption of 14.4~keV solar axions by the iron of the Earth core induces a vertical background motion along the gravity force (Fig.~\ref{fig14}a), which in its turn "pulls" almost all the temperature difference $\Delta T$ down to the bottom of the liquid core (Fig.~\ref{fig14}b) and concentrates it within a layer of the thickness $h = \kappa / W$. As it was noted above, this effect takes place both in single-component and in two-component media.

An important result of these processes is a substantial attenuation of a heat removal from the Earth solid core surface which leads to a corresponding temperature increase in the boundary layer between the liquid core and the solid core where the nuclear georeactor (TWR) resides. As it was shown in~\cite{ref124}, one of the peculiarities of such TWR is that its heat power output depends both on the fuel composition and the medium temperature. It means that increase of the temperature in the boundary layer at the solid core and liquid core border leads to a corresponding increase of the nuclear georeactor power output (see Fig.~\ref{fig13}). As a result, the georeactor heat power output grows until a steady heat removal is re-established, i.e. the convection is re-established and stabilized in the liquid core (up to the "next" variation of the thermal perturbations by axions!).

Therefore if such axion mechanism of solar dynamo -- geodynamo connection exists, then the ohmic dissipation caused by a resonant 14.4~keV solar axions absorption in the Earth core should be connected with the heat power perturbations $\Delta D_{ohmic}$, responsible for the magnetic field variations in the earth core, by the following relations:

\begin{equation}
B_{in} = \xi \cdot B_{in}^a,
\label{eq049}
\end{equation}

\begin{equation}
dD_{ohmic} = \xi^2 dD_{ohmic}^a,
\label{eq050}
\end{equation}

\noindent where the trigger gain $\xi$ in our case (see (\ref{eq032}) and (\ref{eq038})) is equal to

\begin{equation}
\xi \sim 3.4 \cdot 10^3.
\label{eq051}
\end{equation}

Here $B_{in}$ and $B_{in}^a$ are the magnetic fields in the Earth liquid core produced by the nuclear georeactor and the solar axions respectively. The physical sense of the expressions (\ref{eq049})-(\ref{eq050}) reveals the reason why all of the known candidates for an energy source of the Earth magnetic field~\cite{ref015} cannot in principle explain one of the most remarkable phenomena in solar-terrestrial physics -- a strong (inverse) correlation between the temporal variations of magnetic flux in the overshoot tachocline zone~\cite{ref012} and the Earth magnetic field (Y-component)~\cite{ref013} (Fig.~\ref{fig03}).

\section{Axion mechanism of Sun luminosity and CUORE experiment}
\label{sec-03}

Recently a CUORE-experimental search was performed for axions from the solar core from 14.4~keV M1 ground-state nuclear transition in $^{57}$Fe~\cite{ref001}. The detection technique employed a search for a peak in the energy spectrum at 14.4~keV when the axion is absorbed by an electron via the axio-electric effect. The cross section for this process is proportional to the photo-electric absorption cross section for photons~\cite{ref135}:

\begin{equation}
\sigma _{ae} = \frac{\sigma _{pe}}{8 \pi \alpha _{EM} } \left( \frac{2 x_{e}' m_e c^2}{f_a} \right)^2 \left( \frac{\hbar \omega}{m_e c^2} \right)^2,
\label{eq052}
\end{equation}

\noindent where $x_e ' \approx 1$ , $m_e c^2$ is the electron mass in GeV, $\alpha _{EM} = 1 / 137$, and $\sigma _{pe} = 55.336 ~cm^2 \cdot gm^{-1}$ is the photoelectric cross sections for TeO$_2$ (taken from~\cite{ref136}).

Substituting the values of these constants one may rewrite Eq.~(\ref{eq052}) in the following form:

\begin{equation}
\sigma _{ae} = \frac{2.18 \cdot 10^{-11} ~~ GeV^2}{f_a^2} \cdot \left( \frac{E_a}{keV} \right)^2 \sigma_{pe},
\label{eq053}
\end{equation}

\noindent where $E_a$ is in keV and $f_a$ is the Peccei-Quinn scale in GeV.

Hence the estimated absorption rate of 14.4~keV solar axions $N_{CUORE}$ detected by the axio-electric effect (\ref{eq052}) in TeO$_2$-detector ($m = 3 ~kg$) of the CUORE experiment is

\begin{equation}
\Phi_a \cdot \pi \sigma_{ae} \cdot \varepsilon_D \leqslant N_{CUORE} = 0.63 ~~ count \cdot kg^{-1} d^{-1},
\label{eq054}
\end{equation}

\noindent where $\varepsilon _D$ is the detection efficiencies; the flux $\Phi_a$ at the Earth (\ref{eq022}) is 

\begin{equation}
\Phi_a = 1.66 \cdot 10^{23} (g_{an})^2  ~~ cm^{-2} s^{-1},
\label{eq055}
\end{equation}

\begin{figure}
    \begin{center}
        \includegraphics[width=18cm]{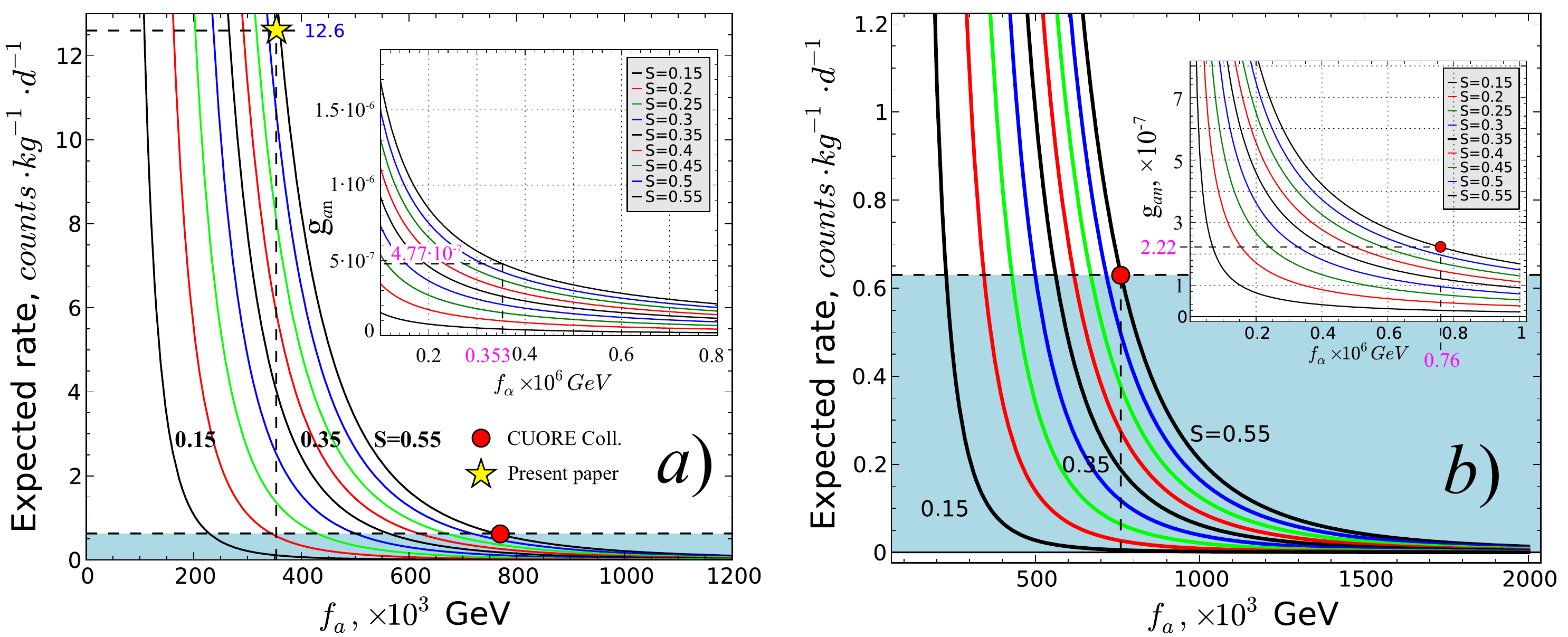}
    \end{center}
\caption{Expected rate in the axion region as a function of the $f_a$ axion constant for different values of the nuclear $S$ parameter. The horizontal line indicates the upper limit obtained in the present paper ($f_a \sim 0.353 \cdot 10^6$~GeV for $S = 0.55$) and in CUORE-experiment ($f_a \sim 0.76 \cdot 10^6$~GeV for $S = 0.55$)~\cite{ref001}.The insets show the correspondence between the axion-nucleon coupling constant and Peccei-Quinn energy scale \textbf{(a)} in our model and \textbf{(b)} in CUORE-experiment. Bounds are given for the interval $0.15 \leqslant S \leqslant 0.55$.}
\label{fig15}
\end{figure}

It is the Eq.~(\ref{eq054}) that made it possible for CUORE collaboration to place a bound (at $S = 0.55$) on the axion coupling constant of $f_a \geqslant 0.76 \cdot 10^6$~GeV at 95\% C.L. (Fig.~\ref{fig15}b). According to Eq.~(\ref{eq010}), the limit on $f_a$ translates into a mass limit $m_a < 8$~eV.

It is necessary to note that if one takes into account the axion mechanism of Sun luminosity and solar dynamo-geodynamo connection, expression (\ref{eq054}) with respect to solar equator ($\Delta _{equ} ^a \sim 0.05$), becomes

\begin{equation}
\Phi_a \cdot \pi \sigma_{ae} \cdot \varepsilon_D \leqslant \frac{N_{CUORE}}{\Delta_{equ}^a} = \frac{0.63}{\Delta_{equ}^a} ~~ counts \cdot kg^{-1} d^{-1},
\label{eq056}
\end{equation}

In other words, within the framework of such mechanism, it is necessary to keep in mind that because of the solar equator effect, only a part of the total axion flux $\Phi_a$ (\ref{eq055}) equal to $\Delta _{equ} \Phi_a$ arrives to the Earth. The solution~(\ref{eq056}) in this case lets us use the expressions\footnote{Let us point out that that the expressions (\ref{eq057}) and (\ref{eq058}) used in CUORE experiment data processing are slightly different from the corresponding expressions (\ref{eq024})-(\ref{eq025}) derived in~\cite{ref102,ref103}.}~\cite{ref001}

\begin{equation}
g_0 = -7.8 \cdot 10^{-8} \left( \frac{6.2 \cdot 10^6 ~GeV}{f_a} \right) \left( \frac{3F - D + 2S}{3} \right),
\label{eq057}
\end{equation}

\begin{equation}
g_3 = -7.8 \cdot 10^{-8} \left( \frac{6.2 \cdot 10^6 ~GeV}{f_a} \right) \left( (D+F) \frac{1-z}{1+z} \right),
\label{eq058}
\end{equation}

\noindent to determine (see Fig.~\ref{fig15}a) the $f_a^{*}$ for a value of $S = 0.55$\footnote{It is appropriate to mention here that E143-experiment provided the value for the parameter $S$ equal to $0.30 \pm 0.06$~\cite{ref137}.} 

\begin{equation}
f_a^{*} \geqslant 0.353 \cdot 10^6 ~~ GeV,
\label{eq059}
\end{equation}

\noindent calculate (see Fig.~\ref{fig15}a) the value of axion-nucleon coupling constant (at $S = 0.55$) 

\begin{equation}
g_{an}^{*} \leqslant 4.8 \cdot 10^{-7},
\label{eq060}
\end{equation}

\noindent and estimate the axion mass by means of (\ref{eq010}) and (\ref{eq059}):

\begin{equation}
m_{a}^{*} \leqslant 17 ~~ eV.
\label{eq061}
\end{equation}

The estimates (\ref{eq060}) and (\ref{eq061}) demonstrate and excellent agreement with the axion-nucleon coupling constant~(\ref{eq028}) and axion mass (\ref{eq011}) obtained in the framework of axion mechanism of Sun luminosity and solar dynamo -- geodynamo connection.

These results make it possible to estimate the limit on the axion-electron coupling constant $g_{ae}$ as well. Derevianko et~al.~\cite{ref138} and Derbin et~al.~\cite{ref007} showed that the cross section $\sigma _{ae} (E_a)$ for the axio-electric effect is proportional to the photoelectric cross section $\sigma _{pe} (E)$, and is given by the formula~\cite{ref007}:

\begin{equation}
\sigma _{ae} (E) = \sigma_{pe} (E) \frac{g_{ae}^2}{\beta} \frac{3}{16 \pi \alpha} \left( \frac{E_a}{m_e c^2} \right)^2 \left( 1 - \frac{\beta}{3} \right),
\label{eq062}
\end{equation}

\noindent where $E_a$ is the axion total energy, $\beta$ is the axion velocity divided by the velocity of light, and $g_{ae}$ is the dimensionless axion-electron coupling constant. At $\beta \to 1$ and $\beta \to 0$, this formula coincides with the cross sections for relativistic and nonrelativistic axions  obtained in~\cite{ref139}.

Since the axion mass is small (see~(\ref{eq061})), let us consider a relativistic form of Eq.~(\ref{eq062}) (i.e. the case of $\beta \to 1$) and compare it to the analogous expression~(\ref{eq052}). As a result of such comparison we derive the upper limit on axion-electron coupling constant with respect to (\ref{eq059}):

\begin{equation}
g _{ae}^{*} = \frac{2 x_e ' m_e}{f_a^{*}} \leqslant 2.89 \cdot 10^{-9}.
\label{eq063}
\end{equation}

The limit~(\ref{eq063}) obviously does not contradict the value of axion-electron coupling constant~(\ref{eq029}) obtained in the framework of axion mechanism of Sun luminosity.

So it may be concluded that the new estimations for the strength of the axion-photon coupling ($g_{a \gamma} \sim 7.07 \cdot 10^{-11} ~GeV^{-1}$), the axion-nucleon coupling ($g_{an} \sim 3.2 \cdot 10^{-7}$), the axion-electron coupling ($g_{a e} \sim 5.28 \cdot 10^{-11}$) and the axion mass ($m_a \sim 17$~eV) obtained in the framework of axion mechanism of Sun luminosity and solar dynamo -- geodynamo connection are in good agreement with the CUORE experiment data. It brings hope that this hypothesis will be justified by the future CUORE experiment with the expected exposure of $1.4 \cdot 10^6 ~kg \cdot day$ which is significantly larger than the current one ($43.65 ~kg \cdot day$)~\cite{ref001}.

\section{Axion mechanism of Sun luminosity and other important experiments}
\label{sec-04}

In view of the axion mechanism of Sun luminosity, let us analyze the data of known experiments on measuring the axion coupling to photon, nucleon and electron for different axion mass ranges.

\subsection{Axion coupling to a photon}

\begin{figure}
    \begin{center}
        \includegraphics[width=18cm]{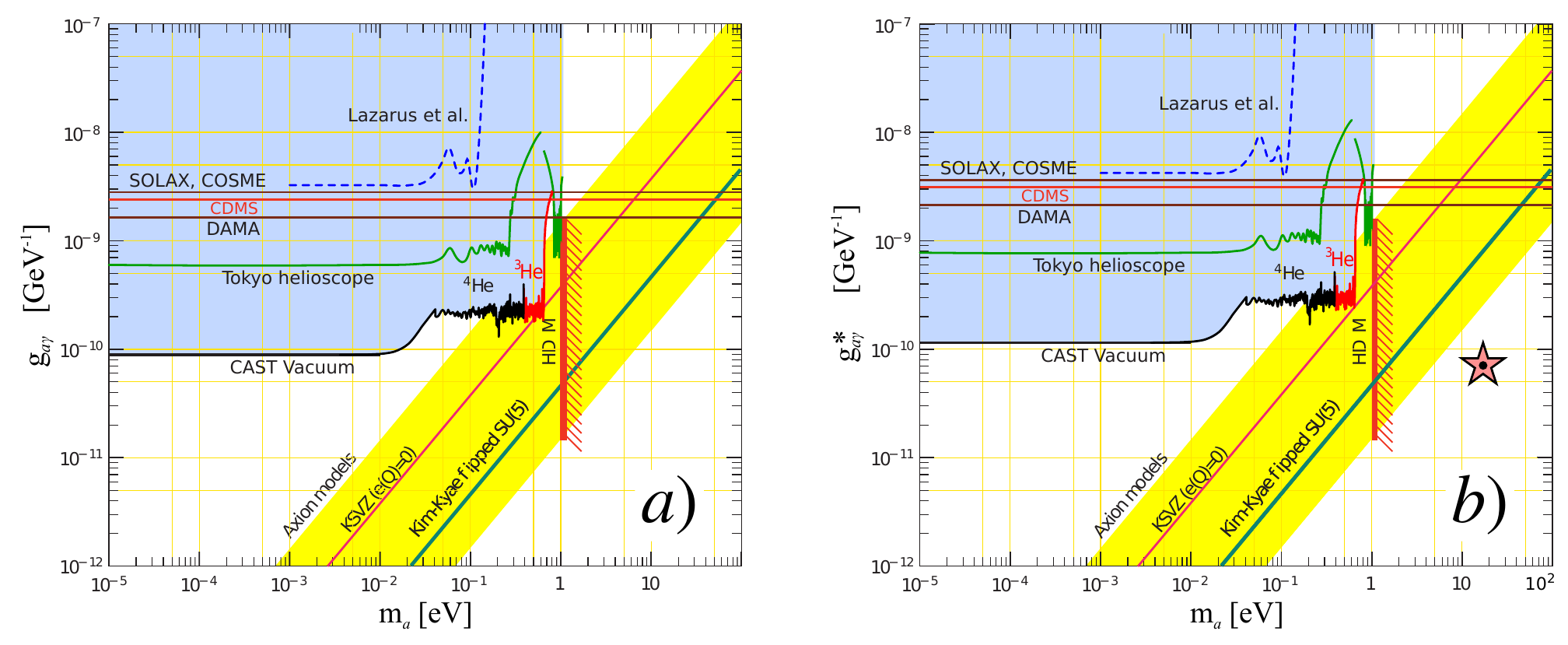}
    \end{center}
\caption{\textbf{(a)} Exclusion regions in the "$m_a$ -- $g_{a \gamma}$"-plane achieved by CAST in the vacuum~\cite{ref140,ref026}, $^4$He and $^3$He phase~\cite{ref141}. We also show constraints from the Tokyo helioscope~\cite{ref142,ref143,ref144}, BNL telescope~\cite{ref145}, SOLAX~\cite{ref146}, COSME~\cite{ref147}, CDMS~\cite{ref148}, DAMA~\cite{ref149} and the hot dark matter limit (HDM) for hadronic axions $m_a < 1.05 ~ eV / c^2$~\cite{ref150} inferred from WMAP observations of the cosmological large-scale structure. The yellow band represents typical theoretical models with $\vert E / N - 1.95 \vert = 0.07 \div 7$. The red solid line corresponds to $E / N = 0$ (the KSVZ model). The field theoretic expectations are shown together with the string theory $Z_{12-I}$  model of Choi, Kim, and Kim (the green line)~\cite{ref151,ref151a}. \textbf{(b)} Same as (a), but all the experimental data are corrected with account for the solar equator effect and a contribution of the Primakoff effect into the total Sun luminosity. A red star denotes the result obtained in the present paper.}
\label{fig16}
\end{figure}

Fig.~\ref{fig16}a shows virtually all the major experiments on estimating the limits of the axion-photon coupling constant. The statement of the problem in all presented experiments included the measurement of the axion flux that could be produced in the Sun by the Primakoff conversion of the thermal photons in the electric and magnetic fields of the solar plasma. The difference between these experiments consisted in the axion flux detection technique only: by the axion-to-photon reconversion (the inverse Primakoff effect) in laboratory transverse magnetic~\cite{ref026,ref110,ref140,ref141,ref142,ref143,ref144,ref145} and electric (the intense Coulomb field of nuclei in a crystal lattice of the detector plus the Bragg scattering technique~\cite{ref146,ref147,ref096,ref097,ref098,ref099}) fields.

The new constants for the axion mechanism of Sun luminosity, obviously, should be calculated with due regard for the solar equator effect and Primakoff effect contribution into the total solar luminosity. It is not hard to show that in this case they should be

\begin{equation}
(g _{a \gamma}^{*})^2 = \left( \frac{\Phi_{Brems} + \Phi_{Compt} +\Phi_{Pr} +\Phi_{M1}}{\Phi_{Pr}} \cdot \Delta_{equ}^a \right)^{-1} \cdot g_{a \gamma}^2
\label{eq064}
\end{equation}

Hence, the values of the partial components of Sun luminosity ($\Phi_{Brems} \sim 67.48\%$, $\Phi_{Compt} \sim 32.41\%$, $\Phi_{Pr} \sim 0.08\%$,  and $\Phi_{M1} \sim 0.03\%$) and the solar equator effect probability ($\Delta _{equ} \sim 0.05$) lead to a new value of $g_{a \gamma}^{*}$ constant:

\begin{equation}
g _{a \gamma}^{*} = 1.26 \cdot g_{a \gamma}.
\label{eq065}
\end{equation}

Fig.~\ref{fig16}b shows the new limits obtained for the axion mechanism of Sun luminosity taking into account (\ref{eq065}) for the corresponding experiments. The change relative to the initial picture (Fig.~\ref{fig16}a) is rather small, since the small contribution of the Primakoff effect into total Sun luminosity "effectively" compensates the influence of the solar equator effect. At the same time it is interesting to note that our values for the axion-photon coupling constant ($g_{a \gamma}^{*} \sim 7.07 \cdot 10^{-11} ~GeV^{-1}$) and the axion mass ($\sim$17~eV) fit well into the existing limits (a red star in Fig.\ref{fig16}).

\subsection{Axion coupling to an electron}

Let us now consider the paper by Derbin~et~al.~\cite{ref007}. In this paper the axio-electric effect in silicon atoms is sought for solar axions appearing owing to bremsstrahlung and the Compton process. Axions are detected using a Si(Li) detector placed in a low-background setup. As a result, new model-independent constraints have been obtained for the axion-electron coupling constant and the mass of the axion. For axions with a mass smaller than 1~keV, the resulting bound is $g_{ae} \leqslant 2.2 \cdot 10^{-10}$  (at 90\% C.L.).

Obviously, the new constants for the axion mechanism of Sun luminosity should be derived with due account taken of the solar equator effect and a contribution of bremsstrahlung and the Compton process into the total Sun luminosity. It is not hard to show that in this case they have the following form:

\begin{equation}
(g_{a e})^2 = \left( \frac{\Phi_{Brems} + \Phi_{Compt} + \Phi_{Pr} + \Phi_{M1}}{\Phi_{Brems} + \Phi_{Compt}} \Delta _{equ}^a \right)^{-1} \cdot g_{ae}^2.
\label{eq079}
\end{equation}

This time, the values of the partial components of Sun luminosity ($\Phi_{Brems} \sim 67.48\%$, $\Phi_{Compt} \sim 32.41\%$, $\Phi_{Pr} \sim 0.08\%$,  and $\Phi_{M1} \sim 0.03\%$) and the solar equator effect probability ($\Delta _{equ} \sim 0.05$) lead to a new estimate of the $g_{ae}^{*}$ constant limit for the axions with a mass smaller than 1~keV:

\begin{equation}
g_{a e}^{*}  \leqslant 4.47 \cdot g_{ae} \cong 9.8 \cdot 10^{-10},
\label{eq080}
\end{equation}

\noindent where $g_{ae} \leqslant 2.2 \cdot 10^{-10}$~\cite{ref007}.

\begin{figure}
    \begin{center}
        \includegraphics[width=18cm]{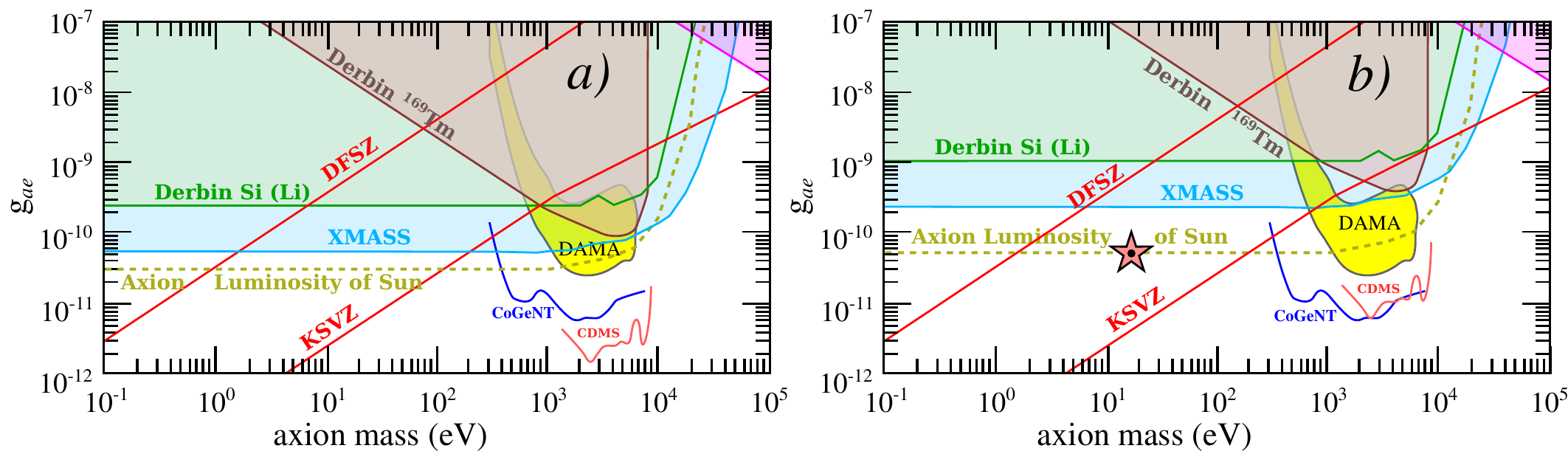}
    \end{center}
\caption{\textbf{(a)} Summary of limits on axion-electron coupling. The limits shown include the astrophysical bound from the solar neutrino flux~\cite{ref080}, dedicated axion experiments by Derbin using $^{169}$Tm~\cite{ref006} and Si(Li)~\cite{ref007}; CDMS, CoGeNT, DAMA and XMASS data obtained from~\cite{ref148,ref153,ref154,ref155}; \textbf{(b)} the data from (a) with a due correction associated with the solar equator effect and a contribution of bremsstrahlung and the Compton process into the total Sun luminosity, except for the astrophysical bound from the solar neutrino flux which is corrected for the contribution of the Compton process into total luminosity only.}
\label{fig18}
\end{figure}

Fig.~\ref{fig18}b shows the new limitations obtained in the context of the axion mechanism of Sun luminosity for the corresponding experiments, according to~(\ref{eq079}). There is a substantial change in Fig.~\ref{fig18}b relative to the initial Fig.~\ref{fig18}a, because the relatively high contribution of bremsstrahlung and the Compton process into the total Sun luminosity cannot compensate the solar equator effect probability largely. At the same time, the determined values of the axion-electron coupling constant ($g_{ae} \sim 5.28 \cdot 10^{-11}$) and the axion mass ($\sim$~17~eV) fit rather well into the known limits (a red star in Fig.~\ref{fig18})

\section{Axion dark matter and extragalactic background light}
\label{sec-05}

Naturally, having the complete axion "portrait", we are entitled to ask a question about the conditions of its detectability.  In other words, provided that axions do exist, once they are produced, where are they likely to be found? Let us turn to the astrophysical observations. Since the current temperature of the cosmic microwave background radiation is $T \cong 2.35 \cdot 10^{-4} ~eV$~\cite{ref156}, axions are nonrelativistic and have been since before decoupling. Therefore axions should, in accord with the equivalence principle, fall with baryons and any other particles into the various potential wells which develop in the Universe~\cite{ref157}. The most likely place to find light relic axions is in clusters of galaxies and the halos of galaxies. It is always possible, however, that the lines of sight to localized regions are partially obscured by previously unrecognized amount of absorbing material, so that the decay photon flux is underestimated~\cite{ref158}. To get around this problem, we shall consider the diffuse extragalactic background (EBL) rather than the flux from any particular region of the sky.

Before we proceed to the discussion of the axions emission and detectability questions, let us consider some important theoretical limitations on the axion parameters. So, within the framework of our mechanism, the new estimations of the strength of the axion coupling to a photon ($g_{a \gamma} \sim 7.07 \cdot 10^{-11} ~GeV^{-1}$), the axion-nucleon coupling ($g_{an} \sim 3.2 \cdot 10^{-7}$) and the axion-electron coupling ($g_{ae} \sim 5.28 \cdot 10^{-11}$) have been obtained. It is necessary to note that obtained estimations cannot be excluded by the existing experimental data (see Fig.~\ref{fig16} and Fig.~\ref{fig18}), because the discussed above effect of solar axion intensity modulation by magnetic field variations in the solar tachocline zone was not taken into account in these observations. The obtained estimates for the strength of the axion-photon coupling and the axion-nucleon coupling also cannot be ruled out by the existing theoretical limitations known as the globular cluster star limit ($g_{a \gamma} < 6 \cdot 10^{-11} ~GeV^{-1}$) and the red giant star limit ($g_{ae} < 3 \cdot 10^{-13}$)~\cite{ref100}, since these values are highly model-dependent\footnote{In this context it is interesting to quote a remark from the paper by Hannestad~S., Mirizzi~A., Raffelt~G.G., and Wong~Y.Y.Y.~\cite{ref150}: "...In principle, $f_a \leqslant 10^9 ~GeV$ is excluded by the supernova SN~1987A neutrino burst duration...  However, the sparse data sample, our poor understanding of the nuclear medium in the supernova interior, and simple prudence suggest that one should not base far-reaching conclusions about the existence of axions in this parameter range on a single argument or experiment alone.  Therefore, it remains important to tap other sources of information, especially if they are easily available".}. That is to say these limitations have a high level of uncertainty because of the absence of the standard theoretical model for globular cluster stars and red giant stars (see, for example the analysis of theoretical models of the hot core in~\cite{ref100} the cross section for axion absorption in~\cite{ref112}, and a note for Fig.~3 in~\cite{ref100}).

There is one more important fact related to the so-called "axion trapping" effect which was not taken into account quantitatively in the paper by Raffelt~\cite{ref100}.  It is known~\cite{ref159,ref159a,ref160,ref161,ref162,ref163}, that the axion flux from the supernova can be suppressed enough in two parameter regions~\cite{ref162,ref163}. If axion-nucleon-nucleon interaction is weak enough, the axion cannot be effectively produced in the core of the supernova. Quantitatively, for $f_a \geqslant 10^9 ~GeV$, the axion flux can be small enough not to affect the cooling process~\cite{ref159,ref159a,ref160,ref161,ref162}. On the contrary, if the axion interacts strongly enough, the mean free path of the axion becomes much shorter than the size of the core, and hence the axions cannot escape from the supernova. In this case, axion is trapped inside the so-called "axion sphere", and the axion emission is also suppressed. In this case, axions are emitted only from the surface of the axion sphere; this type of the axion emission is often called "axion burst". Quantitatively, for $f_a \leqslant 2 \cdot 10^6 ~GeV$ (or equivalently, $m_a \geqslant 3 ~eV$), the axion luminosity from SN1987A is suppressed enough~\cite{ref159,ref159a,ref160,ref161,ref162,ref163}.

For $f_a \leqslant 2 \cdot 10^6 ~GeV$ suggested from the cooling of supernova, following the paper~\cite{ref163} we have another constraint from the detection of axions in water Cherenkov detectors. In this parameter region, the axion  flux  from  the  axion burst is quite sizable for its detection, even though it does not affect the cooling of SN1987A. If the axion-nucleon-nucleon coupling is strong  enough, axions  may  excite  the  oxygen  nuclei  in  the  water Cherenkov detectors ($^{16}\text{O} + a \to {^{16}\text{O}}^{*}$), followed by  radiative decays  of the excited state. If this process had happened, the Kamiokande detector should have  observed  the  photons emitted  from  the decay of  $^{16}\text{O}^{*}$. Due to the non-observation of  this  signal, $f_a \leqslant 3 \cdot 10^5 ~GeV$ is excluded~\cite{ref112}.

Alternatively stated, the hadronic axion with the axion decay constant in the following range  is  still  viable  with  all  the  astrophysical  constraints\footnote{Somewhat more conservative estimates performed by Turner~\cite{ref162} and Ressel~\cite{ref157} give the following limitations for the hadronic axion: $2 ~eV \leqslant m_a \leqslant 5 ~eV$ and $3 ~eV \leqslant m_a \leqslant 8 ~eV$ respectively. However, we are going to use the limitation (\ref{eq081}) since the uncertainty of above-mentioned estimates is determined by the factor of $\geqslant 3$~\cite{ref162}.}~\cite{ref163}

\begin{equation}
3 \leqslant m_a \leqslant 20 ~[eV]
\label{eq081}
\end{equation}

\begin{figure}
    \begin{center}
        \includegraphics[width=8cm]{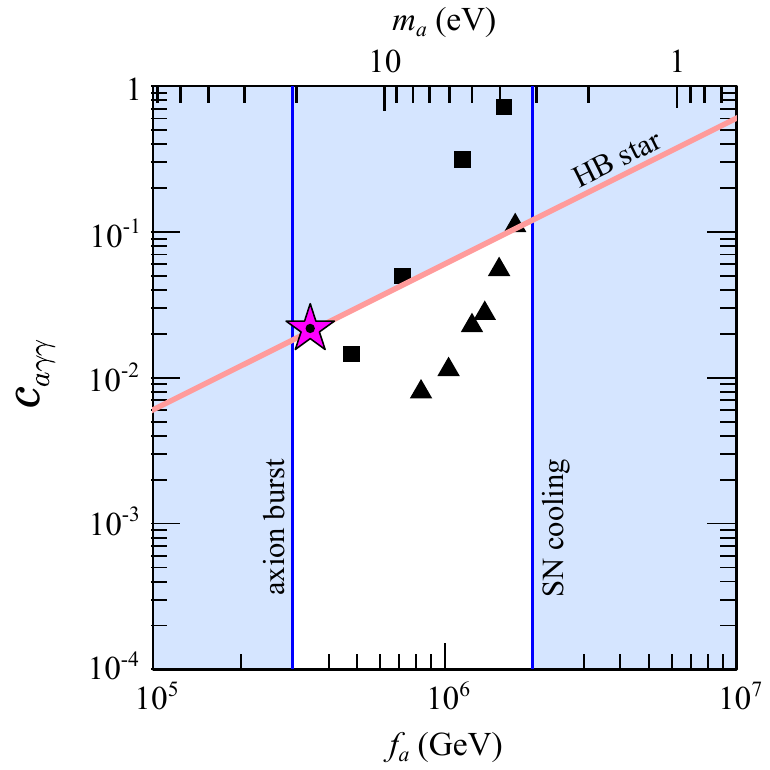}
    \end{center}
\caption{Astrophysical constraints on the axion mass $m_a$ from the cooling of the supernova, axion burst, cooling of the HB stars, the extragalactic background light~\cite{ref158} (squares), and the emission line in clusters of galaxies~\cite{ref157} (triangles). Shaded region is excluded. A purple star marks our result ($c_{a \gamma \gamma} \cong 0.02$, $f_a = 0.353 \cdot 10^6 ~GeV$). The graph is inspired by~\cite{ref163}.}
\label{fig19}
\end{figure}

The mentioned consequences of the "axion trapping" effect may be illustrated as follows (Fig.~\ref{fig19}). Given that the value of the axion coupling to a photon in terms of the axion mechanism of Sun luminosity is $g_{a \gamma} \sim 7.07 \cdot 10^{-11} ~GeV^{-1}$, it is easy to write an equation for a straight line

\begin{equation}
c_{a \gamma \gamma} = \frac{2 \pi f_a}{\alpha} g_{a \gamma} = 0.06 \cdot \frac{f_a}{10^6 ~GeV},
\label{eq082}
\end{equation}

\noindent which marks a sort of a "border" (Fig.~\ref{fig19}) between the allowed and forbidden values of the $c_{a \gamma \gamma}$ constant and the energy scale $f_a$ associated with the break-down of the U(1) PQ symmetry.

Now everything is ready for discussion of the axionic contribution to the EBL. The axion with a mass $m_a = 17 ~eV$ has been termed "invisible" because it interacts very, very weakly. Its lifetime far exceeds the age of  the Universe

\begin{equation}
\tau _{a \rightarrow \gamma \gamma} ^{*} = \frac{64 \pi}{g_{a \gamma}^2 m_a^3} \cong 1439 \cdot t_{Univ},
\label{eq083}
\end{equation}

\noindent where the age of  the Universe $t_{Univ} \approx 4.34 \cdot 10^{17} ~s$~\cite{ref164,ref156}. Notice that the lifetime of the axion is longer than the age of the Universe for $m_a = 17 ~eV$ and $c_{a \gamma \gamma} \cong 0.02$ and hence primordial axions are still in the Universe. However, as we will see later, radiative decay of the axion  may  affect  the  background  UV photons  in spite of the long lifetime.

The  number ($N_a$) of axions  in a cluster (the mass $M$) is~\cite{ref165}

\begin{equation}
N_a \sim 10^{66} \frac{M}{M_{Sun}} \left( \frac{eV}{m_a} \right).
\label{eq084}
\end{equation}

It seems to have gone unnoticed that this number may be so large that the cluster luminosity

\begin{equation}
L_a = \frac{m_a N_a}{\tau_a^{*}}
\label{eq085}
\end{equation}

\noindent is easily measured~\cite{ref165}. In this connection we explore this possibility of observable photon luminosity caused by the cluster's axions decay. Let us remind that from now on we shall consider the \textit{diffuse extragalactic background} (EBL) rather than the galaxy luminosity.

According to the stated above, we shall start with examining the effect of dark matter in the form the light axions on the extragalactic background light. In dong so we follow the theoretical results by Overduin and Wesson~\cite{ref166}, who assumed that the axions are clustered in Galactic halos with nonzero velocity dispersions and derived an expression for the intensity of the axionic contribution which describes the axion halos as a luminous element of a pressureless perfect fluid in the standard Friedman-Robertson-Walker universe. To go further and compare our predictions with observational data, we would like to calculate the intensity of axionic contributions to the EBL as a function of the wavelength $\lambda_0$ after the manner of~\cite{ref166}:

\begin{equation}
I_{\lambda} (\lambda_0) = \frac{\Omega_a \rho_{crit,0}}{\sqrt{32 \pi^3} h H_0 \tau_a} \left( \frac{\lambda_0}{\sigma_{\lambda}} \right) \cdot \int \limits _{0} ^{z_f} \frac{\exp \left\lbrace - \frac{1}{2} \left[ \frac{\lambda_0 / (1+z) - \lambda_a}{\sigma_{\lambda}} \right]^2 \right\rbrace dz}{(1+z)^3 \left[ \Omega_{m,0} (1+z)^3 + 1 - \Omega_{m,0} \right]^{1/2}},
\label{eq086}
\end{equation}

\noindent where $z_f =30$~\cite{ref166};

\begin{equation}
\lambda_a = 24800 \text{\AA} \left( \frac{eV}{m_a} \right)
\label{eq087}
\end{equation}

\noindent is a peak wavelength of the decay photons;

\begin{equation}
\sigma_{\lambda} = 2 \frac{v_c}{c} \lambda_{a} \approx 220 \text{\AA} \left( \frac{eV}{m_a} \right)
\label{eq088}
\end{equation}

\noindent is the standard deviation of the Gaussian spectral energy distribution, for which the velocity dispersion $\upsilon_c$ (for axions bound in galaxy clusters) rises to as much as 1300~km/s~\cite{ref157,ref166};

\begin{equation}
\Omega _a = 5.2 \cdot 10^{-3} h_0 ^{-2} \left( \frac{m_a}{eV} \right)
\label{eq089}
\end{equation}

\noindent is the present density parameter of the thermal axions; $h$ is the Planck constant; $H_0 = 100 h_0 ~km \cdot s^{-1} \cdot
 Mpc^{-1}$ is the Hubble constant, $h_0 = 0.75 \pm 0.15$~\cite{ref166} is the usual value of the Hubble constant expressed in units of $100 ~km \cdot s^{-1} \cdot Mpc^{-1}$; $\rho_{crit,0} = 1.88 \cdot 10^{-29} ~g \cdot cm^{-3}$ is the present critical density; $\Omega_{m,0} = \Omega_{a} + \Omega_{bar} + \Omega_{\nu} = 0.266 \pm 0.029$~\cite{ref167} is the present total density parameter of the axions ($\Omega_{a}$), baryons ($ \Omega_{bar} = 0.028 \pm 0.012$~\cite{ref166}) and neutrinos ($\Omega _{\nu} \leqslant 0.014$~\cite{ref168}); the expression for the decay lifetime of the axion decay into photon pairs was used in the following form:

\begin{equation}
\tau _{a \rightarrow \gamma \gamma}^{*} = \frac{2^8 \pi^3}{c_{a \gamma \gamma}^2 \alpha _{em}^2} \frac{f_a^2}{m_a^3} = (1.54 \cdot 10^7 t_{Univ}) \zeta ^{-2} \left( \frac{m_a}{eV} \right)^{-5}, ~~ \zeta = \frac{c_{a \gamma \gamma}}{0.72};
\label{eq091}
\end{equation}

\begin{figure}
    \begin{center}
        \includegraphics[width=12cm]{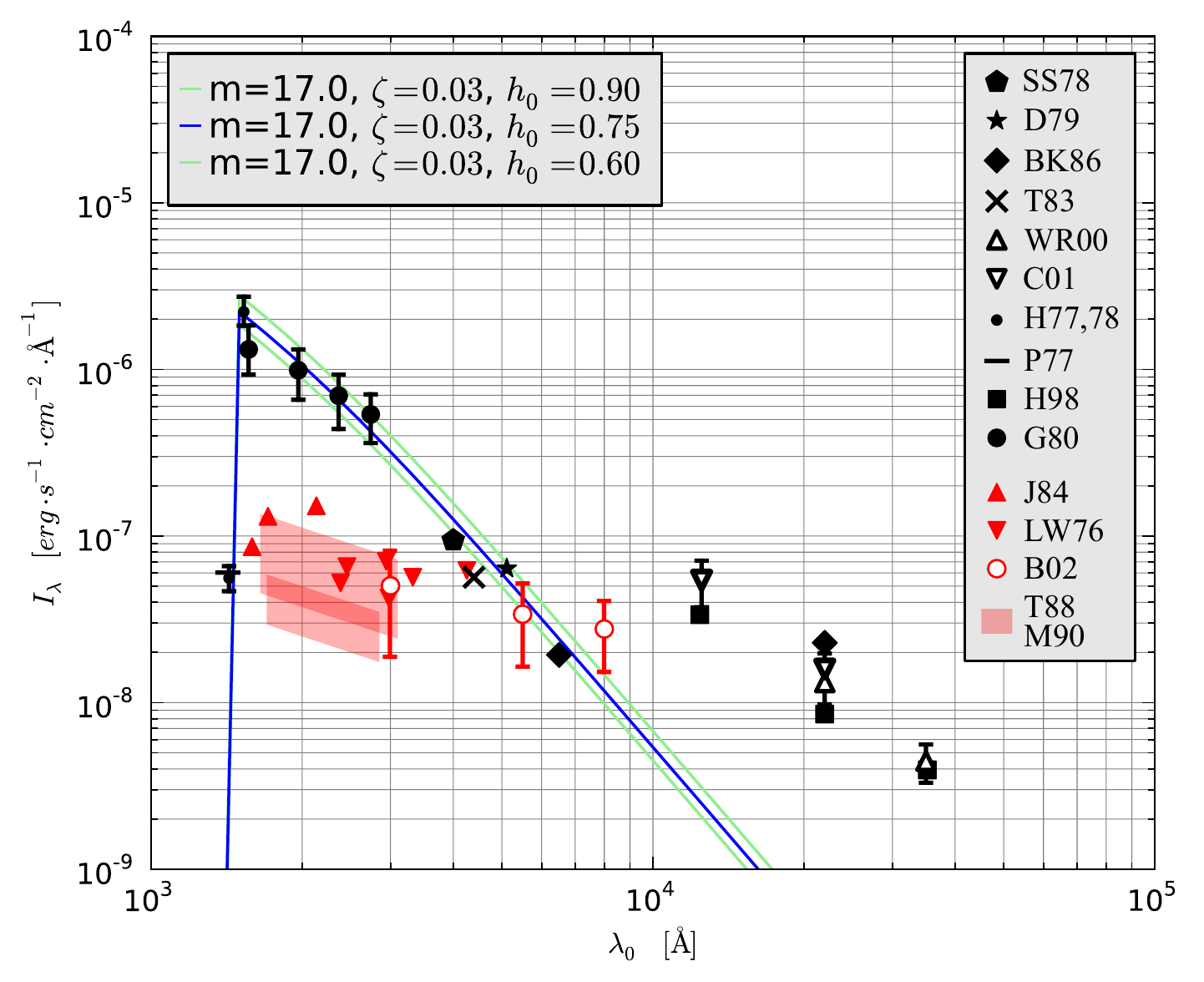}
    \end{center}
\caption{The spectral intensity $I_{\lambda}(\lambda_0)$ of the background radiation from decaying axions as a function of the observed wavelength $\lambda_0$. The curves for the value of $m_a = 17 ~eV$, $\zeta = c_{a \gamma \gamma} / 0.72 = 0.03$ correspond to upper, median and lower limits on $h_0$. Also observational upper limits (solid symbols and a heavy line) and reported detections (empty symbols) over this waveband are shown. Experimental data depicted in red were not taken into account for the theoretical fit of the spectral intensity $I_{\lambda}$ (blue line). See explanations in the text.}
\label{fig20}
\end{figure}

Evaluating Eq.~(\ref{eq086}) over $1500 \text{\AA} \leqslant \lambda _0 \leqslant 20,000 \text{\AA}$  with $\zeta = 0.03$ and $z_f = 30$, we obtain the plots of $I_{\lambda}(\lambda _0)$ shown in Fig.~\ref{fig20}. Apparently, the theoretical fit of spectral intensity of the background radiation~(\ref{eq091}) produced by axion decays describes the known experimental data in the near ultraviolet and optical bands very well. They include data from several ground-based telescope observations (SS78~\cite{ref169}, D79~\cite{ref170}, BK86~\cite{ref171}), sounding rockets (H77~\cite{ref172}, H78~\cite{ref173}), Apollo-Soyuz mission (P77~\cite{ref174}), the Pioneer 10 spacecraft (T83~\cite{ref175}), the DIRBE instrument aboard the COBE satellite (H98~\cite{ref176}, WR00~\cite{ref177}, C01~\cite{ref178}), S2/68 sky-survey telescope aboard TD-1 satellite (G80~\cite{ref179}).

It is important to keep in mind that some of the experimental data were not taken into account at all during the construction of the theoretical fit of the spectral intensity $I_{\lambda}$. For example, OAO-2 satellite (LW76~\cite{ref180}), sounding rockets (J84~\cite{ref181}, T88~\cite{ref182}), the Space Shuttle-borne Hopkins UVX experiment (M90~\cite{ref183}), and combined Hubble Space Telescope -- Las Campanas Telescope observations (B02~\cite{ref184}), shown in red in Fig.~\ref{fig20}.

One of the reasons for not considering the data from LW76~\cite{ref180}, J84~\cite{ref181}, T88~\cite{ref182}, (M90~\cite{ref183} is that these experiments involved measurements of certain parts of the sky only. This is a grave methodological disadvantage which, according to Gondhalekar~\cite{ref179}, may lead to serious distortions of the true EBL value, because "...the individual observation were taken over different regions of the sky and cover different wavelength ranges. It should be noted that intensity of the observed inter-stellar radiation field shows significant variation, not only with galactic latitude but also with galactic longitude and these variations should be taken into account when comparing observation taken over different regions of the sky. Strictly, an accurate determination of the total interstellar radiation density requires integration of observations made over the whole sky". Experimental data from B02~\cite{ref184} were dropped due to the other reason which is related to the fair criticism by Mattila~\cite{ref185}, who casts doubt on the data calibration method and, consequently, on the accuracy of the obtained results.

Let us now make a short comment regarding the quantity $\zeta$. For this purpose we write the effective Lagrangian density which describes the coupling $g_{a \gamma}$ of axions to photons:

\begin{equation}
L_{a \gamma \gamma} = \frac{g_{a \gamma}}{4} F_{\mu \nu} \tilde{F}^{\mu \nu} a = - g_{a \gamma} \vec{E} \cdot \vec{B} a.
\label{eq092}
\end{equation}

\noindent where $a$ is the axionic field, $F$ is the electromagnetic field-strength tensor, $\tilde{F}$ is its dual, and $\vec{E}$ and $\vec{B}$ are the electric and magnetic fields, respectively. The axion-photon coupling constant is

\begin{equation}
g_{a \gamma} = \frac{\alpha}{2 \pi f_a} c_{a \gamma \gamma} = \frac{\alpha}{2 \pi f_a} \left[ \frac{E_{PQ}}{N} - \frac{2 (4+z)}{3 (1+z)} \right],
\label{eq093}
\end{equation}

\noindent where $E_{PQ}$ and $N$, respectively, are the electromagnetic and color anomaly of the axial current associated with the axion field. Here $z = m_u / m_d$ is the $u$- and $d$-quarks masses ratio.

Since the theoretical fit (Fig.~\ref{fig20}) of the spectral intensity $I_{\lambda}$ of the background radiation from decaying axions was performed for the value $\zeta = 0.03$, the constant $c_{a \gamma \gamma}$, according to~(\ref{eq091}), will be

\begin{equation}
c_{a \gamma \gamma} = 0.72 \zeta = \left[ \frac{E_{PQ}}{N} - \frac{2 (4+z)}{3 (1+z)} \right] \simeq 0.02.
\label{eq094}
\end{equation}

This value is extremely important, because according to~\cite{ref186}, it is a strong indicator of (a) the effects of axion emission on the evolution of helium burning low-mass stars~\cite{ref187}, (b)  the effect of decaying relic axions on the diffuse extragalactic background radiation~\cite{ref157,ref158}. That is the effects (a) and (b) provide limit to the axion-photon coupling~\cite{ref186}. It can be shown, that a combined action of effects (a) and (b) with $z = m_u / m_d = 0.56$ and $f_a \cong 10^6 ~GeV$ leads to the following important limitation:

\begin{equation}
c_{a \gamma \gamma}  \leqslant \frac{f_a}{10^7 ~GeV} = 0.02.
\label{eq095}
\end{equation}

The exact match of Eqs.~(\ref{eq094}) and~(\ref{eq095}) reflects a remarkable fact that the properties of the axion studied ($m_a = 17 ~eV$, $f_a = 0.353 \cdot 10^6 ~GeV$) conform to conditions (a) and (b) completely. Moreover, they satisfy the conditions (see~\cite{ref186} and Fig.~\ref{fig19}) of the axion's effect on the neutrino burst from SN~1987A (the so-called "trapping regime" in which the axion emission would not have a significant effect on the neutrino burst~\cite{ref159,ref159a,ref160,ref161,ref162}) and the effect of axions emitted from SN1987A on the Kamiokande II detector, coming from a condition of absence of a large signal at the Kamiokande  detector~\cite{ref112}).

Noteworthy, the axion under question ($m_a = 17 ~eV$, $c_{a \gamma \gamma} \cong 0.02$) may be put into the thermally-produced hadronic axions class (see Fig.~\ref{fig16} and Fig.~\ref{fig18}), and it may also be considered as a real candidate for dark matter at the same time. Here is why.

It is usually assumed that dark matter in standard cosmology models is produced during the radiation-dominated era. If the axion mass $m_a \geqslant 10^{-2} ~eV$, axions were produced thermally, with cosmological abundance

\begin{equation}
\Omega_a h_0^2 = \frac{m_a}{130 ~eV} \left( \frac{10}{g_{*S,F}} \right),
\label{eq096}
\end{equation}

\noindent where $g_{*S,F}$ is the effective number of relativistic degrees of freedom when axions freeze out of equilibrium. Turner showed~\cite{ref159,ref159a} that the vast majority of these would have arisen in the early Universe via thermal mechanisms such as Primakoff scattering and photo-production. The Boltzmann equation can be solved to give their present comoving number density as $n_a = (830 / g_{*S,F}) ~cm^{-3}$~\cite{ref157}, where $g_{*S,F} \approx 15$ counts the number of relativistic degrees of freedom left in the plasma at the time when axions "froze out" of equilibrium. The present density parameter $\Omega _a = n_a m_a / \rho _{crit,0}$ of thermal axions thus leads to the expression~(\ref{eq089}) whence it follows

\begin{equation}
0.11 \leqslant \Omega_a \leqslant 0.25.
\label{eq097}
\end{equation}

Here we have taken $0.6 \leqslant h_0 \leqslant 0.9$ as usual. This is comparable to the density of dark matter.

\begin{figure}
    \begin{center}
        \includegraphics[width=15cm]{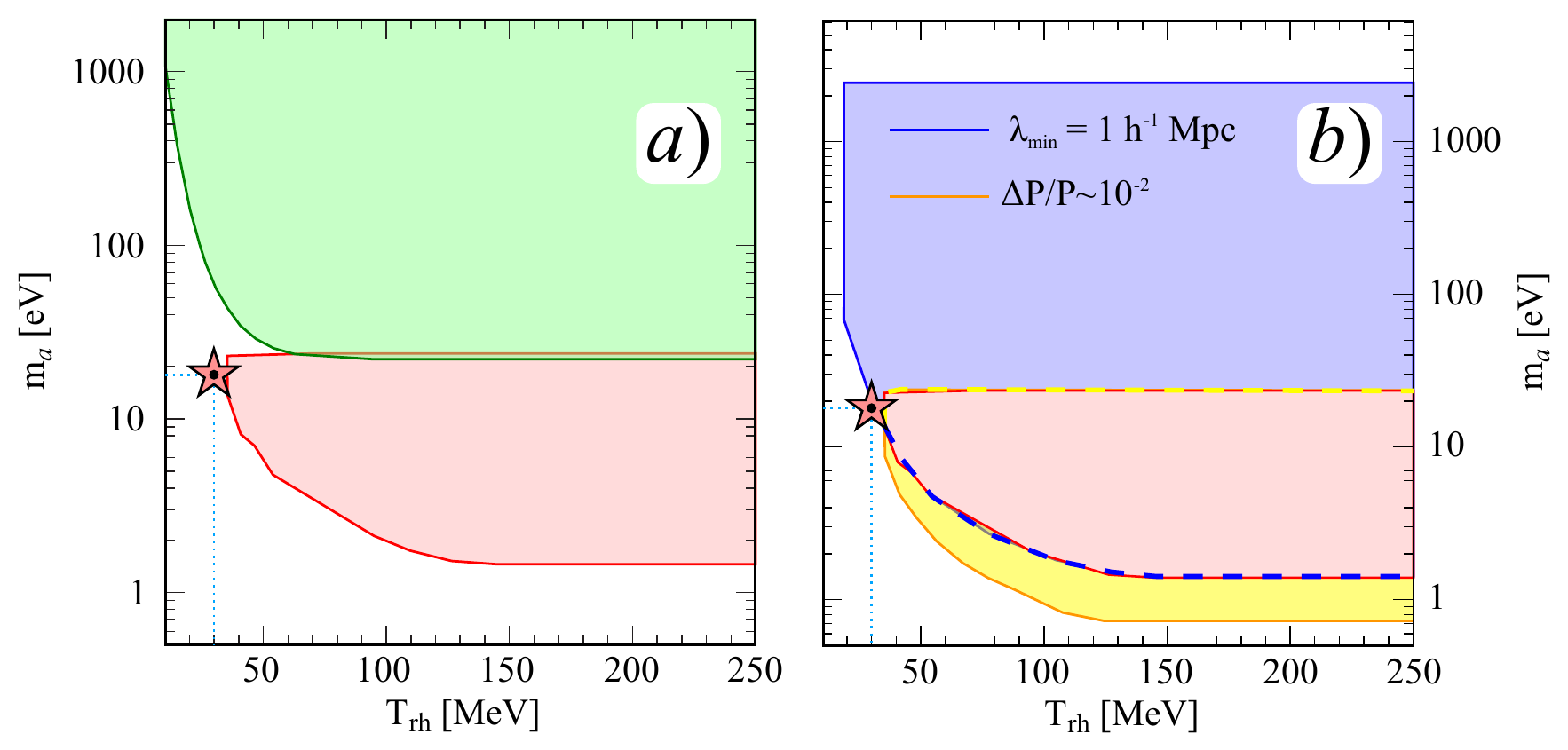}
    \end{center}
\caption[caption]{\textbf{(a)} Upper limits to the hadronic axion mass from cosmology, allowing the possibility of a low-temperature-reheating scenario. The green region shows the region excluded by the constraint $\Omega _a h^2 < 0.135$ as a function of the reheating temperature $T_{rh}$. The red region shows the additional part of axion parameter space excluded by WMAP1/SDSS data. At low reheating temperatures, upper limits to the axion mass are loosened. For $T_{rh} \geqslant 170 ~MeV$, the usual constraints are recovered. Adapted from~\cite{ref188}.\\
\textbf{(b)} Estimated improvement in the accessible axion parameter space from including more precise  measurements of the matter power spectrum (the region bounded by the yellow line), corresponding to LSST~\cite{ref190,ref191}, or from measurements of clustering on smaller length scales, corresponding to Lyman-$\alpha$ forest measurements (the region bounded by the blue line)~\cite{ref192}. The red region indicates the parameter  space excluded by WMAP1/SDSS measurements. Adapted from~\cite{ref188}. Our result (a red star) has the coordinates $\left \lbrace T_{rh} = 30 ~MeV; m_a = 17 ~eV \right \rbrace$.}
\label{fig21}
\end{figure}

At the same time, there are axion constraints in the so-called nonstandard thermal histories. As it was shown in~\cite{ref188}, the most intriguing one among the well-known non-standard thermal histories (\textit{low-temperature reheating} and \textit{kination} cosmologies) is the LTR cosmology which allows for the fact that there is currently no direct evidence for radiation domination prior to big-bang nucleosynthesis~\cite{ref189}.  According to this scenario, radiation domination begins as late as 1~MeV, and is preceded by significant entropy generation. Thermal axion relic abundances are then suppressed, and cosmological limits to axions are loosened. However, for reheating temperatures $T_{rh} \leqslant 35 ~MeV$, the large-scale structure limit to the axion mass is lifted (see Fig.~\ref{fig21}). The remaining constraint from the total density of matter is significantly relaxed in this case. Constraints are also relaxed for higher reheating temperatures. It turns out that axions will be produced thermally, with nonstandard cosmological abundance

\begin{equation}
\Omega_a ^{nth} h_0^2 = \frac{m_a}{130 ~eV} \left( \frac{10}{g_{*S,F}} \right) \cdot \gamma (T_{rh} / T_F ),
\label{eq098}
\end{equation}

\noindent where $T_F$ is the decoupling temperature of axions.

Using the abundance $\Omega _a ^{nth}$ normalized by its standard value $\Omega _a$ as a function of the reheating temperature, obtained by Grin~et~al. (Fig.~4 in~\cite{ref188}) it is easy to derive for the mass $m_a = 17 ~eV$ at $T_{rh} = 30-35 ~MeV$ (see Fig.~\ref{fig21})

\begin{equation}
\frac{\Omega_a ^{nth}}{\Omega_a} =  \gamma (T_{rh} / T_F ) \cong 0.2 ~~ for ~~ T_{rh} = 30 ~~MeV.
\label{eq099}
\end{equation}

Hence,

\begin{equation}
0.022 \leqslant \Omega_a^{nth} \leqslant 0.05.
\label{eq100}
\end{equation}

\noindent which is rather small and comparable to the baryonic density value $0.027 \leqslant \Omega _b \leqslant 0.040$~\cite{ref156} as opposed to the standard case of~(\ref{eq097}).

Which one of them is preferable when it comes to description of real physics -- the standard expression~(\ref{eq097}) for the axion dark matter density or the non-standard one~(\ref{eq100})? Although there is no answer to this question nowadays, one may still hope that, according to~\cite{ref188}, future limits to axions in the standard radiation-dominated and LTR thermal histories may follow from constraints to their contribution to the energy density of relativistic particles at $T \sim$1~MeV. This is due to the fact that a comparison between the abundance of $^4$He and the predicted abundance from  standard big-bang nucleosynthesis (SBBN) places constraints to the radiative content of the Universe at $T \sim$1~MeV~\cite{ref193}. As the paper~\cite{ref188} notes, this can be stated as a constraint to the effective neutrino number ($N_{\nu} ^{eff}$), because at early times, axions contribute to the total relativistic energy density (through $N_{\nu} ^{eff}$), and thus constraints to $^4$He abundances can be turned into constraints on $m_a$ and $T_{rh}$.

It is known that in terms of the baryon-number density $n_b$, the primordial $^4$He abundance is characterized by the expression $Y_P = 4n_{He} / n_b$. In order to translate measurements of primordial $^4$He abundance $Y_P$ to constraints on $m_a$ and $T_{rh}$ we use the scaling relation by Grin~et~al.~\cite{ref188}

\begin{equation}
\Delta N_{\nu}^{eff} (m_a , T_{rh} ) = N_{\nu}^{eff} - 3 = \frac{43}{7} \left\lbrace (6.25 \cdot \Delta Y_p +1 ) - 1 \right\rbrace,
\label{eq101}
\end{equation}

\noindent which describes the relation between the deviation $\Delta N_{\nu}^{eff}$ for the given values of $m_a$ and $T_{rh}$ and the deviation $\Delta Y_P$ of the primordial $^4$He abundance. Here the deviation $\Delta Y_P$ is thought of as a deviation from the value $Y_P = 0.2487 \pm 0.0006$ accepted for SBBN-predicted primordial abundances~\cite{ref194}.

According to the calculations in~\cite{ref188}, for the values $m_a = 17 ~eV$, $T_{rh} = 30 ~MeV$ and the effective neutrino number $N_{\nu}^{eff} \cong 3.44$ (Fig.~7 in~\cite{ref188}), the deviation $\Delta N_{\nu}^{eff}$ is 0.44. Substitution of this value into Eq.~(\ref{eq101}) gives the following value for the $\Delta Y_P$ of the primordial $^4$He abundance:

\begin{equation}
\Delta Y_p = 0.0044,
\label{eq102}
\end{equation}

Let us present some data for the sake of comparison. One careful study gives  the  value $Y_P = 0.2565 \pm 0.0010 ~(stat) \pm 0.0050 ~(syst)$ from 93 H$_{\text{II}}$ regions~\cite{ref195}, leading to the sensitivity limit $N_{\nu}^{eff} \cong 3.61$. Interestingly enough, the deviation $\Delta N_{\nu}^{eff} = 0.61$ may be a  sign of the higher mass axions existence, since according to~\cite{ref186}, for sufficiently high masses, the axionic contribution saturates to $\Delta N_{\nu}^{eff} = 4/7$ at high reheating temperatures.  At the same time the Planck satellite is expected to reach $\Delta Y_P = 0.013$~\cite{ref196}, yielding sensitivity of $N_{\nu}^{eff} \cong 4.04$, while CMBPol (a proposed future CMB polarization experiment) is expected  to approach $\Delta Y_P = 0.0039$, leading to the sensitivity limit $N_{\nu}^{eff} \cong 3.30$~\cite{ref188}.

It makes it clear that the answer to a question about the preference of one expression (standard (\ref{eq097}) or non-standard (\ref{eq100})) for the axion dark matter density over another may be searched for only in direction of a sharp increase in observations precision. According to~\cite{ref197}, measuring CMB temperature and  polarization with cosmic variance accuracy would allow to constrain $Y_P$ to within 1.5\%, or $\Delta Y_P \sim 0.0036$ (assuming flatness). Such an ideal measurement would be able to discriminate between the BBN-guided, deuterium based helium value and the current lowest direct helium observations. In other words, "...if the CMB-determined helium mass fraction turns out to be as high as suggested by SBBN calculations combined with the observed deuterium abundance, this could indicate a systematic error in the present direct astrophysical helium observations. Alternatively, if the CMB could independently determine the helium value with sufficient precision to confirm the present low helium value coming from direct observations, then this would be a smoking gun for new physics"~\cite{ref197}. For example, one could imagine  sterile neutrinos appearing within the nonstandard BBN scenarios, which would agree with present observations of $\eta _{10} = 10^{10} (n_b / n_{\gamma})$, while having a low helium mass fraction.  To put it differently, in our opinion, in spite of the future possible constraints to axions and low-temperature reheating from the helium abundance and next-generation large-scale-structure surveys,  it is the appearance of the sterile neutrinos that may effectively solve the problem of missing dark matter in the framework of LTR cosmology.

And finally, returning to our 17~eV axion "caught" in the extragalactic background, we may say that regardless of the scenario which provides a significant portion of the dark matter, experimentally observed invisible axions (Fig.~\ref{fig20}) must have rest masses in the "semi-visible" range (1500$\text{\AA}$ - 20000$\text{\AA}$) where they do contribute significantly to the light of the night sky. In this sense one may think of our findings as a result of the axion mechanism of Sun luminosity  and the "...nature's most versatile dark-matter detector: the light of the night sky"~\cite{ref166}.

\section{Relic axion-like archion and cosmic infrared background}
\label{sec-06}

It is well known that a direct measurement of the EBL  and, particularly,  the cosmic infrared background (CIB) consists in observing the cumulative emission from various pregalactic objects, protogalaxies, galaxies, cosmic explosions and decaying elementary particles (including dark matter particles) throughout the evolution of the Universe and therefore one can provide important constraints on the integrated cosmological history of star formation~\cite{ref198,ref199} and control the second most important contribution to the cosmic electromagnetic background after the Cosmic Microwave Background generated at the time of recombination at a redshift around 1000~\cite{ref199}.  This background is expected to be composed of three main components~\cite{ref199}:
\begin{itemize}
\item the stellar radiation in galaxies concentrated in the ultraviolet and visible with a redshifted component in the near InfraRed (IR);
\item a fraction of the stellar radiation absorbed by dust either in the galaxies or in the intergalactic medium;
\item the radiation from active galactic nuclei (a fraction of which is also absorbed by dust and reradiated in the far-IR).
\end{itemize}

Its detection is a subject of great scientific interest and the main purpose of the Diffuse Infrared Background Experiment (DIRBE) on the Cosmic Background Explorer (COBE) space-craft~\cite{ref176,ref177,ref178}.  These studies resulted in upper limits on the EBL in the 1.25-100~$\mu m$ region, and in the detection of a positive isotropic signal at 140 and 240~$\mu m$.

However, there are serious problems with interpretation of some DIRBE data at far-IR wavelengths. For example~\cite{ref198}, the energy sources could either be yet undetected dust-enshrouded galaxies, or extremely dusty star-forming regions in observed galaxies, and they may be responsible for the observed iron enrichment in the intracluster medium. Although there is currently no compelling need to invoke non-nuclear energy sources to explain the COBE data, their potential contribution to the observed EBL cannot be ruled out.  It leads to a conclusion that the exact star formation history or scenarios required to produce the EBL at far-IR wavelengths cannot be unambiguously resolved by the COBE observations and must await future observations~\cite{ref198}.

The other type of problems is revealed by the interpretation of the DIRBE data at near-IR wavelengths. For example, in their studies of EBL at near-IR wavelengths Cambr\'{e}sy et~al.~\cite{ref178} obtain a  significantly higher cosmic background than integrated galaxy counts ($3.6 \pm 0.8 ~ kJy \cdot sr^{-1}$ and $5.3 \pm 1.2 ~kJy \cdot sr^{-1}$ for 1.25~$\mu m$ and 2.2~$\mu m$, respectively), suggesting either an increase of the galaxy luminosity function for magnitudes fainter than 30 or the existence of another  contribution to the cosmic background from primeval stars, black holes, or relic particle  decay. However, models predict other possible contributions to the background at these wavelengths~\cite{ref200} such as a burst of star formation either in primeval galaxies or in Population III stars ($z \approx 10$), very massive black holes (accreting from a uniform pre-galactic medium at $z \approx 40$), massive decaying big bang relic particles ($z \approx 300$). In this regard, Cambresy et~al.~\cite{ref178} make a natural conclusion that the new constraints in the near-infrared should encourage revisiting the importance of those contributions to the CIB in cosmological models.

\begin{figure}
    \begin{center}
        \includegraphics[width=12cm]{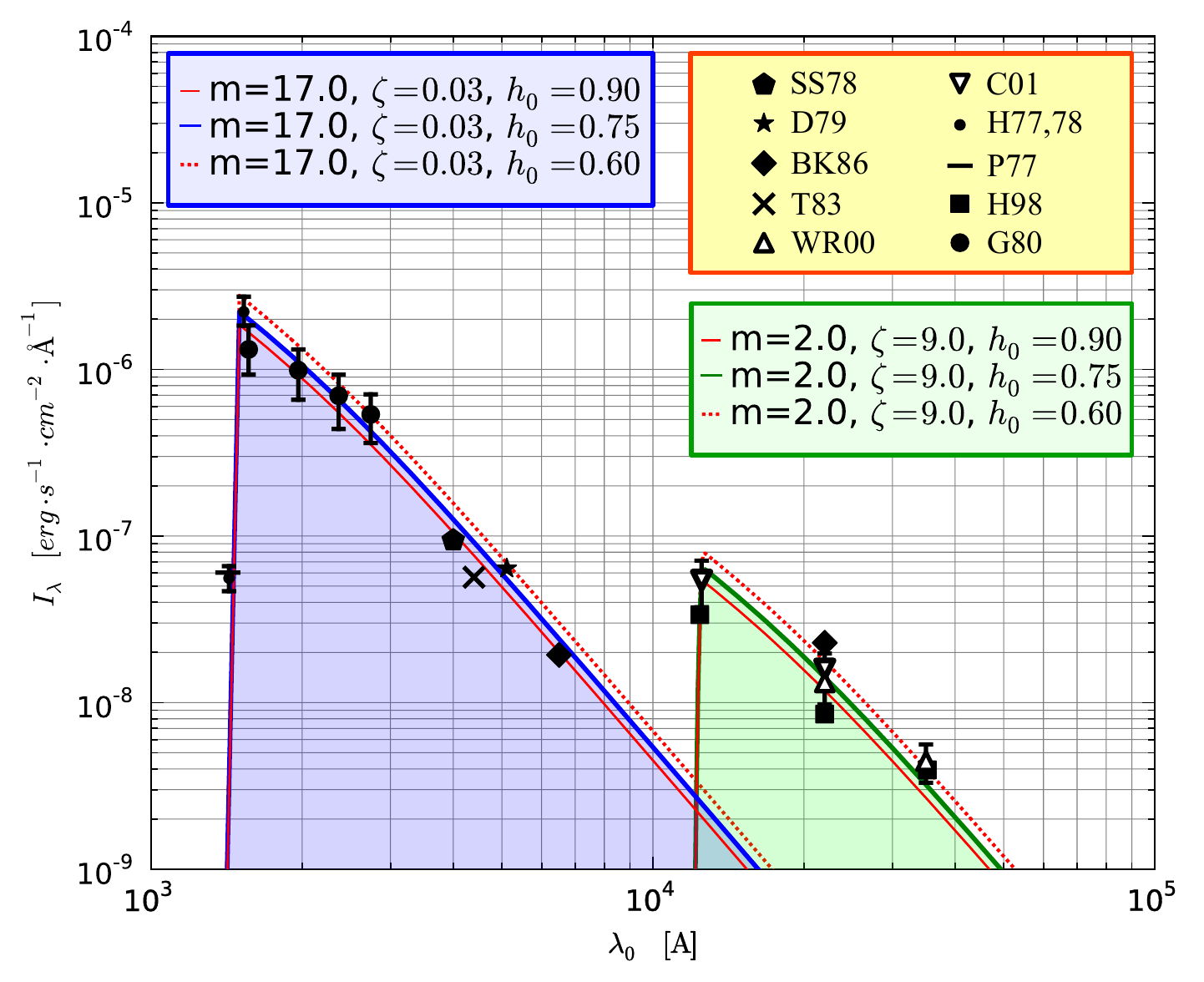}
    \end{center}
\caption{The spectral intensity $I_{\lambda}(\lambda_0)$ of the background radiation from decaying axions as a  function of the observed wavelength $\lambda_0$. The curves for values $m_a = 17.0 ~eV$, $\zeta = c_{a \gamma \gamma} / 0.72 = 0.03$ and $m_a = 2.0 ~eV$, $\zeta = 9.0$ correspond to upper, median and lower limits on $h_0$. Also observational upper limits (solid symbols) over these wavebands are shown.}
\label{fig22}
\end{figure}

In this connection we consider the effect of dark matter in the form of 2~eV relic particle on the extragalactic background light at near-IR region (Fig.~\ref{fig22}). It was obtained on the basis of the intensity~(\ref{eq086}) of axionic contributions to the EBL calculation as a function of the wavelength $\lambda_0$. The theoretical fit (green curves in Fig.~\ref{fig22}) of the spectral intensity of the background radiation~(\ref{eq086}) produced by axion decays, apparently,  describes the known experimental data for the near-infrared band very well.  They include data from ground-based telescope observations (BK86~\cite{ref171}) and the DIRBE instrument aboard the COBE satellite (H98~\cite{ref176}, WR00~\cite{ref177}, C01~\cite{ref178}).

Let us consider some properties of such relic particle with the 2~eV mass. It may be immediately stated that such particle cannot play the role of a sterile neutrino with the mass of the order of a few~eV~\cite{ref201}. And the reason is the following.

It is known that if neutrinos are massive and if the mass eigenstates are not degenerate, then it is possible to have a radiative decay of the form $\nu_s \to \nu + \gamma$. According to~\cite{ref202}, this gives for the decay rate of Majorana neutrinos 

\begin{equation}
\Gamma_{\nu _S \rightarrow \nu \gamma} ^{*} = 5.52 \cdot 10^{-32} \left( \frac{\sin ^2 \theta}{10^{-10}} \right) \left( \frac{m_S}{keV} \right)^5 ~~s^{-1},
\label{eq103}
\end{equation}

\noindent where $m_S$ is the mass eigenstate most closely associated with the sterile neutrino, and $\theta$ is the mixing angle between the sterile and active neutrino. The decay of a nonrelativistic sterile neutrino into two (nearly) massless particles produces a line at the energy $E{\gamma} = m_S / 2$.

Obviously, in case of the decay reaction $\nu_s \to \nu + \gamma$, Eq.~(\ref{eq086}) can be generalized to another relic $\nu_s$ by multiplying it by 

\begin{equation}
I_{\lambda} (\lambda_0) \cdot \frac{1}{2} \cdot \frac{\Omega_{\nu}}{\Omega_a} \cdot \frac{\tau_{a \rightarrow \gamma \gamma}^{*}}{\tau_{\nu_S \rightarrow \nu \gamma}^{*}},
\label{eq104}
\end{equation}

\noindent where the 1/2 is a number of photons produced in each $\nu_s$ decay, $\Omega_{\nu} \leqslant 0.014$ is the present total density parameter of the neutrinos,

\begin{equation}
\tau _{\nu_{\chi} \rightarrow \nu \gamma} ^{*} = ( \Gamma _{\nu_S \rightarrow \nu \gamma}^{*})^{-1} = (1.7 \cdot 10^{15} t_{Univ}) \zeta^{-2} \left( \frac{m_S}{keV} \right)^{-5},
\label{eq105}
\end{equation}

\noindent where

\begin{equation}
\zeta^{-2} = \left( \frac{10^{-10}}{\sin^2 2\theta} \right).
\label{eq106}
\end{equation}

Using the generalization~(\ref{eq104}), it is rather easy to show that a theoretical fit of spectral intensity $I_{\lambda}$ of the background radiation is several orders of magnitude lower than the experimental data in near-infrared band under any reasonable values of the mixing angle $\sin^2 2\theta$.

This relic particle with the 2~eV mass may be supposed to belong to a class of axion-like particles with rather exotic properties. One of them may originate from the fact that in the distant past their birth in the electromagnetic field, or in other words interaction with a photon, was not suppressed for some reason, while at the present time there are simply no conditions suitable for their birth. To put it differently, the mentioned conditions of such particles' birth must be completely suppressed nowadays in order our axion mechanism of Sun luminosity to be possible.

In our opinion, such axion-like particle, provided that it exists, has the properties most similar to those of the so-called archion~\cite{ref203,ref204,ref205,ref206,ref207,ref208,ref209}. An archion may be very similar to a hadronic axion with highly suppressed interaction with leptons under certain conditions. Let us discuss briefly some of its properties below.

As is generally known, in all the models of an invisible axion this particle appears as a Goldstone boson connected with the phase of a complex SU(2)$\times$U(1) singlet Higgs field. The axion coupling to the gauge bosons appears in these models after the U(1)$_{PQ}$ symmetry violation by means of a mechanism, specified by the non-vanishing color anomaly U(1)$_{PQ}$ - SU(3)$_c$ - SU(3)$_c$~\cite{ref210}.

In its most generic form the Lagrangian of axion interaction with fermions (quarks and leptons) and photons is

\begin{equation}
L = c_{\alpha \beta} \alpha \cdot \bar{F}_{\alpha} (\sin \theta_{\alpha \beta} + i \gamma_5 \cos \theta_{\alpha \chi} ) F_{\beta} + g_{a \gamma} a F_{\mu \nu} \bar{F}^{\mu \nu},
\label{eq107}
\end{equation}

\noindent where $\alpha, \beta = 1,2,3$ indices denote the generation of fermions $F$, and the constants $\theta _{\alpha \beta}$,

\begin{equation}
c_{\alpha \beta} \propto f_a ^{-1}
\label{eq108}
\end{equation}

\noindent and

\begin{equation}
g_{a \gamma} \propto f_a ^{-1}
\label{eq109}
\end{equation}

\noindent depend on the axion model chosen.

A model of an archion~\cite{ref203,ref204,ref205,ref206,ref207,ref208,ref209} arose from a model of horizontal unification, the basis of which is expounded in a monograph by Khlopov~\cite{ref210}. This theory includes the global U(1)$_H$ symmetry, the spontaneous breaking of which leads to prediction of a Goldstone boson of the invisible axion type. Such boson called "archion" by authors of~\cite{ref205,ref206,ref207,ref208} has the flavor non-diagonal as well as the flavor diagonal coupling to fermions.

The global U(1)$_H$ symmetry in horizontal unification may be identified with the Peccei–Quinn symmetry U(1)$_{PQ}$~\cite{ref028,ref029,ref030,ref031,ref032,ref032a,ref033,ref033a}, which is due to the fact of triangle anomaly existence in axial currents U(1)$_H$ interaction with gluons~\cite{ref210}.

In the simplest variant of the horizontal unification (the gauge symmetry 

\begin{equation}
S (U)_{c} \otimes SU (2) \otimes U(1) \otimes SU(3)_{H} \otimes U(1)_H
\label{eq110}
\end{equation}

\noindent with a minimal set of heavy fermions) the anomaly is compensated, and the archion remains almost massless. The interaction of the archion with photons is absent because of the parallel compensation which corresponds to the current-photon-photon anomaly.

On the other hand, according to~\cite{ref210}, within any realistic extension of the horizontal unification model up to the grand unification symmetry, for example, within the extension to the SU(5)$_H \otimes $SU(3)$_H$ symmetry, there is no compensation because of the extra heavy fermions, so that the archion appears to be similar to a hadronic axion with strongly suppressed interaction with leptons.

In the framework of the archion model, the archion-photon coupling constant $g_{a \gamma}$ which appears in Eqs.~(\ref{eq107}) and~(\ref{eq109}) has the following form~\cite{ref210}:

\begin{equation}
g_{a \gamma} = \frac{\alpha}{2\pi f_a} c_{a \gamma \gamma} = \frac{\alpha}{4 \pi f_a} \frac{A_c z^{3/2}}{(1+z)^2} \left[ \frac{A_{em}}{A_c} - \frac{2 (4+z)}{3(1+z)} \right]^{-1}.
\label{eq111}
\end{equation}

Here as usual $z = m_u / m_d$ is the $u$- and $d$-quarks mass ratio, $f_a$ is the energy scale associated with the breakdown of the U(1)$_H$ (or equivalently, the U(1)$_{PQ}$ symmetry), $A_c$ and $A_{em}$ are the color and electromagnetic anomalies respectively.

Taking into account the value of the $c_{a \gamma} = 0.72 \zeta$ constant (see~(\ref{eq091}) and Fig.~\ref{fig22}), which is $\sim$6.5 for the archion mass of 2~eV, it is easy to estimate the archion-photon coupling constant from~(\ref{eq111}) and~(\ref{eq010}):

\begin{equation}
g_{a \gamma}^{*} \simeq 2.5 \cdot 10^{-9} ~~ GeV^{-1}.
\label{eq112}
\end{equation}

Obviously, if the archion indeed exists and has such a high archion-photon coupling constant, it must be of the relic origin only in a sense that the conditions of such particles birth must be completely suppressed nowadays. Otherwise, our axion mechanism of Sun luminosity would not be possible, like we pointed out earlier\footnote{It is necessary to note that regardless of whether one assumes the existence of the axion mechanism of Sun luminosity or not, such high value of the archion-photon coupling constant~(\ref{eq112}) is forbidden by the DAMA experiment observations~\cite{ref146} presented in Fig.~\ref{fig16}a and~\ref{fig16}b.}. The value of this constant is also very important because according to~\cite{ref186}, it is a strong indicator of (a) the effects of axion emission on the evolution of helium burning low-mass stars~\cite{ref187}, (b) the effect of decaying relic axions on the diffuse extragalactic background radiation~\cite{ref157,ref166}. In other words, an archion with the 2~eV mass, characterized by the archion-photon coupling constant~(\ref{eq112}), must be relic in order not to violate the known limitation~(\ref{eq095}) which is a consequence of combined action of effects (a) and (b).

For all invisible axion models, including the archion model, Lagrangian of its interaction with nucleons has the same form~\cite{ref209}. Given~(\ref{eq112}), it is not hard to estimate the archion-nucleon coupling constant for the 2~eV mass archion (Fig.~\ref{fig22}) using~(\ref{eq023})-(\ref{eq025}) and~(\ref{eq010}):

\begin{equation}
g_{a n} ^{*} \leqslant 5.6 \cdot 10^{-8}.
\label{eq113}
\end{equation}

If the archion exists with such low value of the archion-nucleon coupling constant, it obviously must have only relic origin in order not to violate the SN1987A limit ($3 \cdot 10^{-7} \leqslant g_{an} \leqslant 10^{-6}$~\cite{ref111,ref112}).

This rises the question of why would Nature need a relic axion-like archion in addition to an ordinary hadronic axion. Curiously enough, the answer suggests itself and is related to the evolutionary formation of the visible large-scale cosmological structure against the background of the invisible "dark" structure.

Apart from the details of this scenario, our axion-like particles may be among the primary participants of this creative action, if they do exist. Let us note however that in spite of the fact that the cosmological structure formation is provided by the single kind of hidden-mass particles -- axions, the "axion liquid" turns out to be two-component. The visible structure in the form of galaxies and superclusters is formed by the shortwave component of the 17~eV axions density perturbations spectrum, while the relic thermal 2~eV archions background plays an important role in sub-shortwave density perturbations spectrum evolution. The latter includes, in particular, the massive halos formation beyond the visible parts of the galaxies. Following~\cite{ref210}, it is worth mentioning that the so-called phase-space argument by Tramaine and Gunn~\cite{ref211}, which gives rise to a known limit on particles mass in the halo, may be substantially weakened or even omitted for the Bose gas~\cite{ref212,ref213,ref214,ref215}.

If all said above is true, it is more than sufficient for a "sensible" existence of axions, and particularly, a relic axion-like archion. On the other hand, except for the indirect indication of axions existence in the form of Fig.~\ref{fig22}, we have no other -- direct -- arguments. In order to fill this gap a little, let us try to substantiate the quantitative relations between the experimental spectral intensities of the background radiation from decaying axions and archions shown in Fig.~\ref{fig22} theoretically on the basis of a simplified cosmological model which takes into account the processes of axion dark matter radiative decay.

\subsection{Decaying axion and relic archion as two components of luminous dark matter}

Let us consider the flat (for simplicity) Universe after the recombination consisting of the following components: relic radiation, usual (light, visible) matter, dark matter (presented by axions and archions), non-relic radiation (resulting from their decay) and dark energy (described by the cosmological constant $\Lambda$), then for the corresponding metrics 

\begin{equation}
ds^2 = c^2 dt^2 - a^2 (t) ( dx^2 + dy^2 + dz^2),
\label{eq114}
\end{equation}

\noindent where $a(t)$ is the scale factor, the first Friedmann equation reads

\begin{equation}
\frac{3 \dot{a}^2}{c^2 a^2} = \kappa T_{00} + \Lambda,
\label{eq115}
\end{equation}

\noindent where a dot denotes the derivative with respect to $t$, $\kappa = 8 \pi G_N / c^4$ ($G_N$ is the Newtonian gravitational constant) and the $00$ covariant component of the total energy-momentum tensor reads

\begin{equation}
T_{00} = \varepsilon_{rr} + \varepsilon_{\nu m} + \varepsilon_{a 2} + \varepsilon_{a17} + \varepsilon_{r2} + \varepsilon_{r17},
\label{eq116}
\end{equation}

\noindent where, in their turn, $\varepsilon_{rr, \nu m, a1, a17, r2, r17}$ denote energy densities of relic radiation (assumed to be independent of all other components), visible matter (also assumed to be independent), archions (with the mass $m_{a2} = 2 ~eV$ and the disintegration constant $\lambda _{a2} = 3.88 \cdot 10^{-22} ~s$, assumed to be completely nonrelativistic), axions (with the mass $m_{a17} = 17 ~eV$ and the disintegration constant $\lambda _{a17} = 1.91 \cdot 10^{-22} s^{-1}$, also assumed to be completely nonrelativistic), radiation resulting from archion decay and radiation resulting from axion decay respectively.

Here some additional comments should be made. First, the disintegration constants are estimated simply as $\lambda = 1 / \tau^{*}$ on basis of the formula~(\ref{eq091}). Second, we assume both dark matter components (archions and axions) completely nonrelativistic, in other words, we completely neglect their velocities (or temperatures). This assumption is simultaneously in agreement and disagreement with~\cite{ref166}, where the formula~(185) (which is equivalent to~(\ref{eq086}) in our text) contains simultaneously the scent of the standard cosmological $\Lambda$CDM-model without matter velocities in the denominator of the integrand and the scent of the velocity dispersion in its numerator. It seems that this fact does not mean that there is a self-contradiction in the formula~(185), because the numerator may be much more sensitive to the local velocity of dark matter than the denominator to the global one. However, in the subsequent analysis we shall be interested in the evolution of the average non-relic radiation energy density without taking into account its frequency distribution and the nonzero dark matter temperature (for simplicity).

Substituting~(\ref{eq116}) into~(\ref{eq115}) and introducing the standard portions  

\begin{equation}
\Omega = \frac{\kappa c^2}{3 H_0 ^2} \varepsilon_{(0)}, ~~ \Omega_{\Lambda} = \frac{c^2}{3 H_0^2} \Lambda,
\label{eq117}
\end{equation}

\noindent where the subscript $(0)$ corresponds to the current moment of time $t = 0$ (this value may be chosen without loss of generality) and $H = \dot{a} / a$ is the Hubble parameter, we obtain

\begin{align}
\nonumber \left( \frac{\dot{a}}{a} \right)^2 = H_0 ^2 & \left( \Omega_{rr} \frac{\varepsilon_{rr}}{\varepsilon_{rr(0)}} + \Omega_{\nu m} \frac{\varepsilon_{\nu m}}{\varepsilon_{\nu m(0)}} + \Omega_{a2} \frac{\varepsilon_{a2}}{\varepsilon_{a2(0)}} + \right.\\
& \left. + \Omega_{a17} \frac{\varepsilon_{a17}}{\varepsilon_{a17(0)}} + \Omega_{r2} \frac{\varepsilon_{r2}}{\varepsilon_{r2(0)}} + \Omega_{r17} \frac{\varepsilon_{r17}}{\varepsilon_{r17(0)}} + \Omega_{\Lambda} \right).
\label{eq118}
\end{align}

Here $\varepsilon_{rr} \sim 1 / a^4$, $\varepsilon_{\nu m} \sim 1 / a^3$, consequently, as usual,

\begin{equation}
\Omega_{rr} \frac{\varepsilon_{rr}}{\varepsilon_{rr(0)}} +  \Omega_{\nu m} \frac{\varepsilon_{\nu m}}{\varepsilon_{\nu m(0)}} = \Omega_{rr} \left( \frac{a_0}{a} \right)^4 +  \Omega_{\nu m} \left( \frac{a_0}{a} \right)^3,
\label{eq119}
\end{equation}

\noindent where $a_0$ is the current value of the scale factor: $a(0) = a_0$. Further,

\begin{equation}
\varepsilon_{a2} : \frac{1}{a^3} \exp (- \lambda_{a2} t ), ~~ \varepsilon_{a17} : \frac{1}{a^3} \exp (- \lambda_{a17} t ) \Longrightarrow \varepsilon_{a2(0)} : \frac{1}{a_0^3}, ~~ \varepsilon_{a17(0)} : \frac{1}{a_0^3},
\label{eq120}
\end{equation}

\noindent whence

\begin{equation}
\Omega_{a2} \frac{\varepsilon_{a2}}{\varepsilon_{a2(0)}} + \Omega_{a17} \frac{\varepsilon_{a17}}{\varepsilon_{a17(0)}} = \Omega_{a2} \left( \frac{a_0}{a} \right)^3 \exp (-\lambda_{a2} t) +  \Omega_{a17} \left( \frac{a_0}{a} \right)^3 \exp (-\lambda_{a17}t).
\label{eq121}
\end{equation}

Finally, combining first and second Friedmann equations, we come to the following equations:

\begin{equation}
d \left[ (\varepsilon_{a2} + \varepsilon_{r2}) a^3 \right] + \frac{1}{3} \varepsilon_{r2} d(a^3) = 0, ~~ d \left[ (\varepsilon_{a17} + \varepsilon_{r17}) a^3 \right] + \frac{1}{3} \varepsilon_{r17} d(a^3) = 0.
\label{eq122}
\end{equation}

From the first one we immediately get

\begin{equation}
\frac{d (\varepsilon_{a2} a^3)}{da} + a^3 \frac{d \varepsilon_{r2}}{da} + 4a^2 \varepsilon_{r2} = 0, ~~ \frac{d (\varepsilon_{a2} a^3)}{da} = \varepsilon_{a2(0)} a_0^3 \exp(-\lambda_{a2}t) \frac{(-\lambda_{a2})}{\dot{a}},
\label{eq123}
\end{equation}

\begin{equation}
\frac{d \varepsilon_{r2}}{da} + \frac{4 \varepsilon_{r2}}{a} = \varepsilon_{a2(0)} \left( \frac{a_0}{a} \right)^3 \exp (-\lambda_{a2}t) \frac{\lambda_{a2}}{\dot{a}},
\label{eq124}
\end{equation}

\noindent whence finally

\begin{align}
\nonumber \varepsilon_{r2} & = \frac{1}{a^4} \left( \varepsilon_{a2(0)} a_0^3 \lambda_{a2} \int \limits _{a_0}^{a} \exp (-\lambda_{a2}t) \frac{a}{\dot{a}} da + \varepsilon_{r2(0)} a_0^4 \right) = \\
& = \frac{1}{a^4} \left( \varepsilon_{a2(0)} a_0^3 \lambda_{a2} \int \limits _{0}^{t} a(t) \exp (-\lambda_{a2}t) dt + \varepsilon_{r2(0)} a_0^4 \right).
\label{eq125}
\end{align}

Similarly,

\begin{equation}
\varepsilon_{r17} = \frac{1}{a^4} \left( \varepsilon_{a17(0)} a_0^3 \lambda_{a17} \int \limits _{0}^{t} a(t) \exp (-\lambda_{a17}t) dt + \varepsilon_{r17(0)} a_0^4 \right).
\label{eq126}
\end{equation}

The substitution of ~(\ref{eq119}),  (\ref{eq121}),  (\ref{eq125}) and~(\ref{eq126}) into~(\ref{eq118}) gives the equation defining the function $a(t)$ satisfying the condition $a(t) = a_0$. Taking into account that the values of both disintegration constants $\lambda_{a2}$ and $\lambda_{a17}$ are extremely small and the radiation energy density decreases faster than the nonrelativistic matter energy density when $a$ increases, one can neglect all (i.e. both relic and non-relic) radiation contributions and get  

\begin{equation}
\left( \frac{\dot{a}}{a} \right)^2 = H_0^2 \left[ \Omega_{\nu m} \left( \frac{a_0}{a} \right)^3 +  \Omega_{a2} \left( \frac{a_0}{a} \right)^3 \exp (-\lambda_{a2}t) + \Omega_{a17} \left( \frac{a_0}{a} \right)^3 \exp (-\lambda_{a17}t) + \Omega_{\Lambda} \right].
\label{eq127}
\end{equation}

Introducing the dimensionless quantities 

\begin{equation}
\tilde{t} = H_0 t, ~~ \tilde{a} = \frac{a}{a_0}, ~~ \tilde{\lambda}_{a2} = \frac{\lambda_{a2}}{H_0}, ~~ \tilde{\lambda}_{a17} = \frac{\lambda_{a17}}{H_0},
\label{eq128}
\end{equation}

\noindent from~(\ref{eq120}) we obtain  

\begin{equation}
\left( \frac{1}{\tilde{a}} \frac{d \tilde{a}}{d \tilde{t}} \right) ^2 = \frac{\Omega_{\nu m}}{\tilde{a}^3} + \frac{\Omega_{a2}}{\tilde{a}^3} \exp (- \tilde{\lambda}_{a2} \tilde{t}) + \frac{\Omega_{a17}}{\tilde{a}^3} \exp (- \tilde{\lambda}_{a17} \tilde{t}) + \Omega_{\Lambda}, ~~ \tilde{a} (0) = 1.
\label{eq129}
\end{equation}

According to the recent observations~\cite{ref216}, we consider the following values: $\Omega _{\nu m} = 0.046$, $\Omega _{a2} + \Omega _{a17} = 0.236$, $\Omega _{\Lambda} = 1 - \Omega _{\nu m} - (\Omega _{a2} + \Omega _{a17}) = 0.718$. For $\Omega _{a2}$ and $\Omega _{a17}$ we use the values $0.016$ and $0.220$ respectively. Again, due to extreme smallness of $\lambda _{a2}$ and $\lambda _{a17}$ (in the presence of nonzero $\Lambda$) they do not affect noticeably the dependence $a(t)$. The following graphs confirm this statement:

\begin{figure}
    \begin{center}
        \includegraphics[width=8cm]{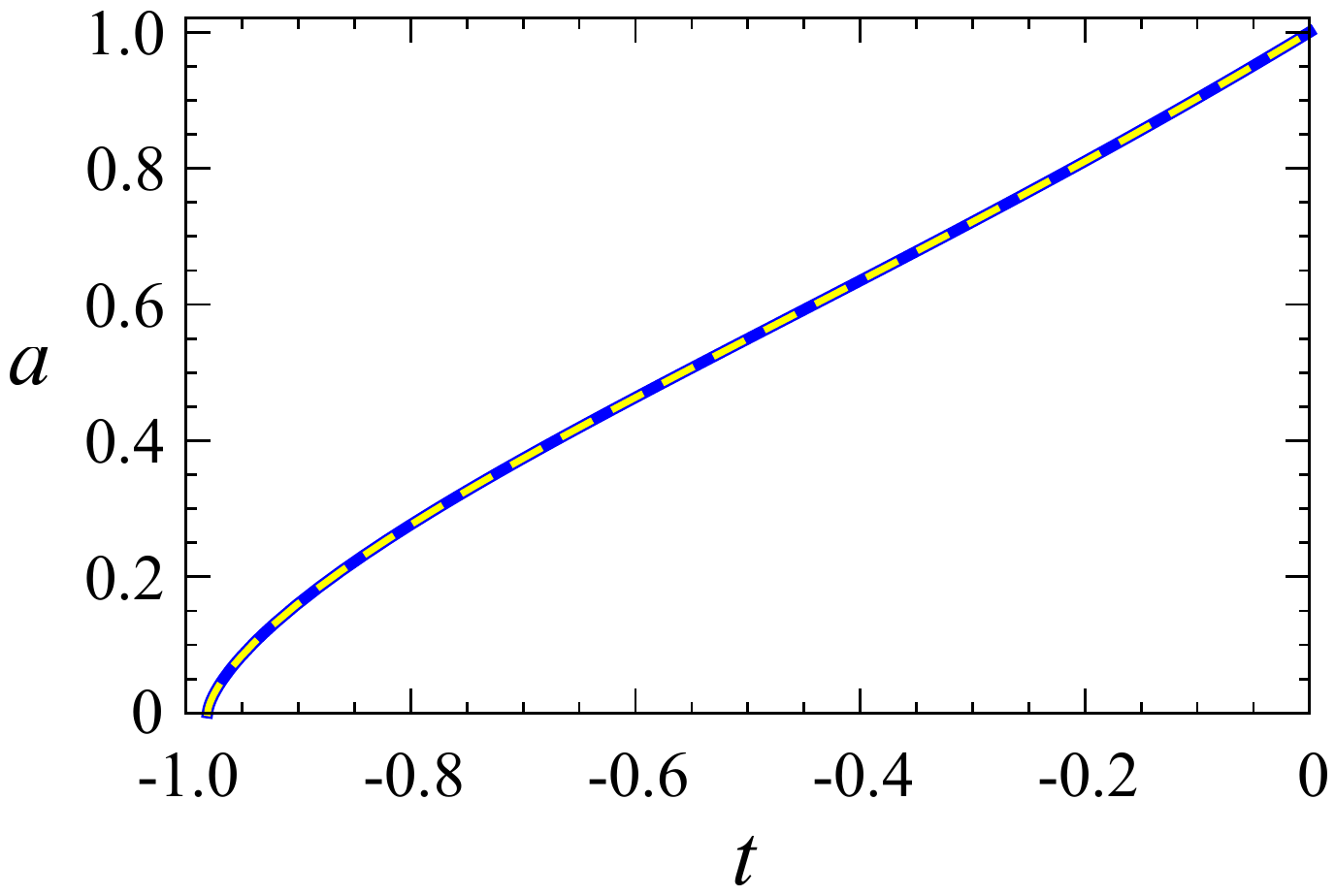}
        \includegraphics[width=8cm]{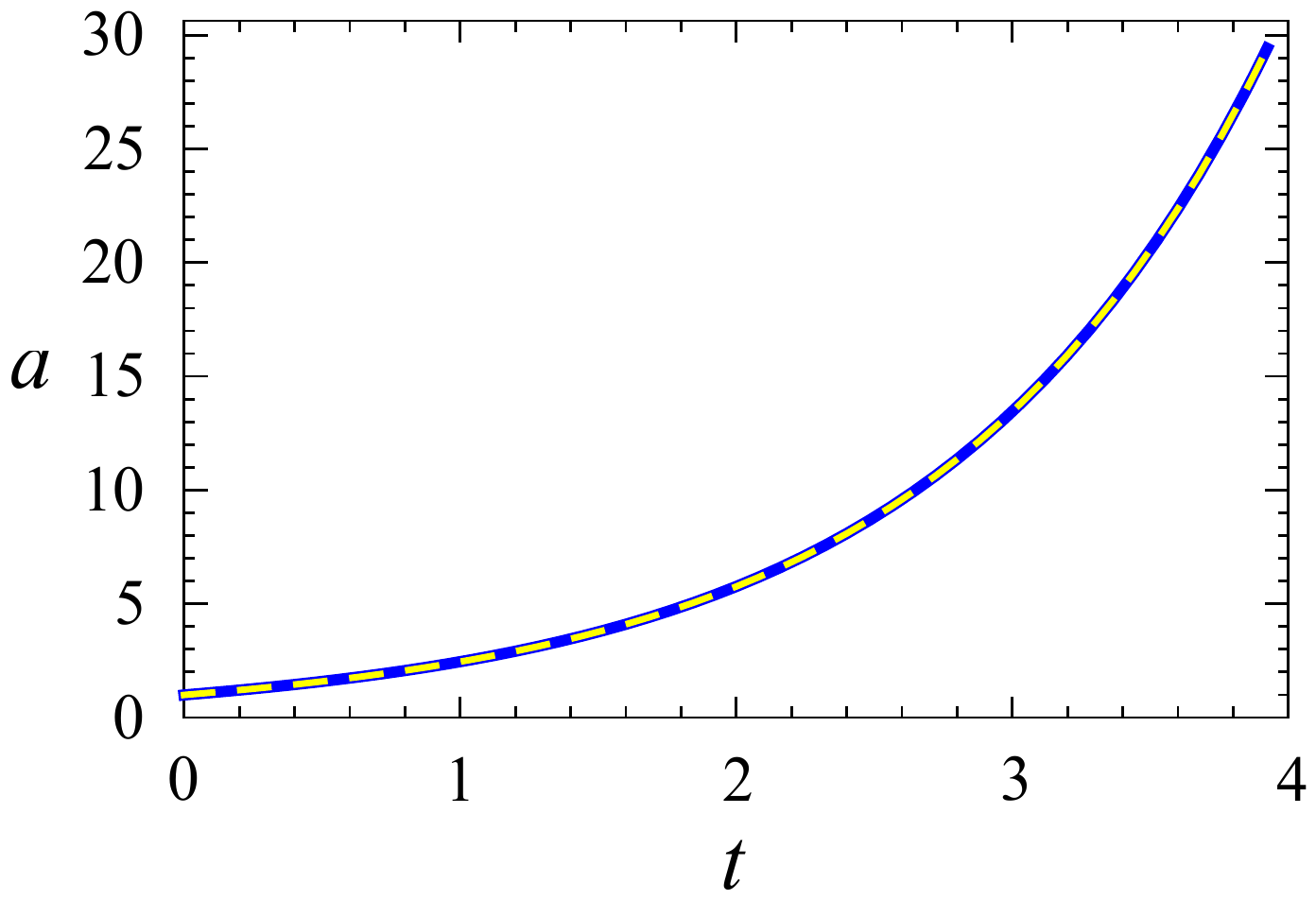}
    \end{center}
\caption{The numerical solution of Eq.~(\ref{eq129}) (the blue firm line) is indistinguishable from its numerical solution for $\lambda _{a2} = 0$ and $\lambda _{a17} = 0$ (the yellow dashed line).}
\label{fig23}
\end{figure}

From~(\ref{eq125}) and~(\ref{eq126}) we obtain respectively  

\begin{equation}
\tilde{t} \tilde{a} (\tilde{t}) \exp (- \tilde{\lambda}_{a2} \tilde{t}) d\tilde{t} + \frac{\varepsilon_{r2(0)}}{\varepsilon_{a2(0)}},
\label{eq130}
\end{equation}

\begin{equation}
\tilde{t} \tilde{a} (\tilde{t}) \exp (- \tilde{\lambda}_{a17} \tilde{t}) d\tilde{t} + \frac{\varepsilon_{r17(0)}}{\varepsilon_{a17(0)}}.
\label{eq131}
\end{equation}

Let us introduce the convenient functions  

\begin{equation}
f_{a2} (\tilde{t}) = \frac{\varepsilon_{r2}}{\varepsilon_{a2(0)}} \tilde{a}^4 - \frac{\varepsilon_{r2(0)}}{\varepsilon_{a2(0)}} = \tilde{\lambda}_{a2} \int \limits _{0} ^{\tilde{t}} \tilde{a} (\tilde{t}) \exp (-\tilde{\lambda}_{a2} \tilde{t} ) d\tilde{t},
\label{eq132}
\end{equation}

\begin{equation}
f_{a17} (\tilde{t}) = \frac{\varepsilon_{r17}}{\varepsilon_{a17(0)}} \tilde{a}^4 - \frac{\varepsilon_{r17(0)}}{\varepsilon_{a17(0)}} = \tilde{\lambda}_{a17} \int \limits _{0} ^{\tilde{t}} \tilde{a} (\tilde{t}) \exp (-\tilde{\lambda}_{a17} \tilde{t} ) d\tilde{t},
\label{eq133}
\end{equation}

For $a = 0$ we get $f_{a2} = -9.26 \cdot 10^{-5}$ and $f_{a17} = -4.56 \cdot 10^{-5}$. These negative values may be completely compensated by the proper choice of $\varepsilon _{r2(0)}$ and $\varepsilon _{r17(0)}$ in order to give $\varepsilon _{r2} = 0$ and $\varepsilon _{r17} = 0$ at the same moment of time (when $a = 0$). This has clear physical sense: we assume that at the recombination moment the non-relic radiation is absent (all radiation, present at that moment of time, may be considered as the relic one). From this requirement we immediately obtain $\varepsilon _{r2(0)} = 9.26 \cdot 10^{-5} \varepsilon _{a2(0)}$ and $\varepsilon _{r17(0)} = 4.56 \cdot 10^{-5} \varepsilon _{a17(0)}$.

\begin{figure}
    \begin{center}
        \includegraphics[height=5cm]{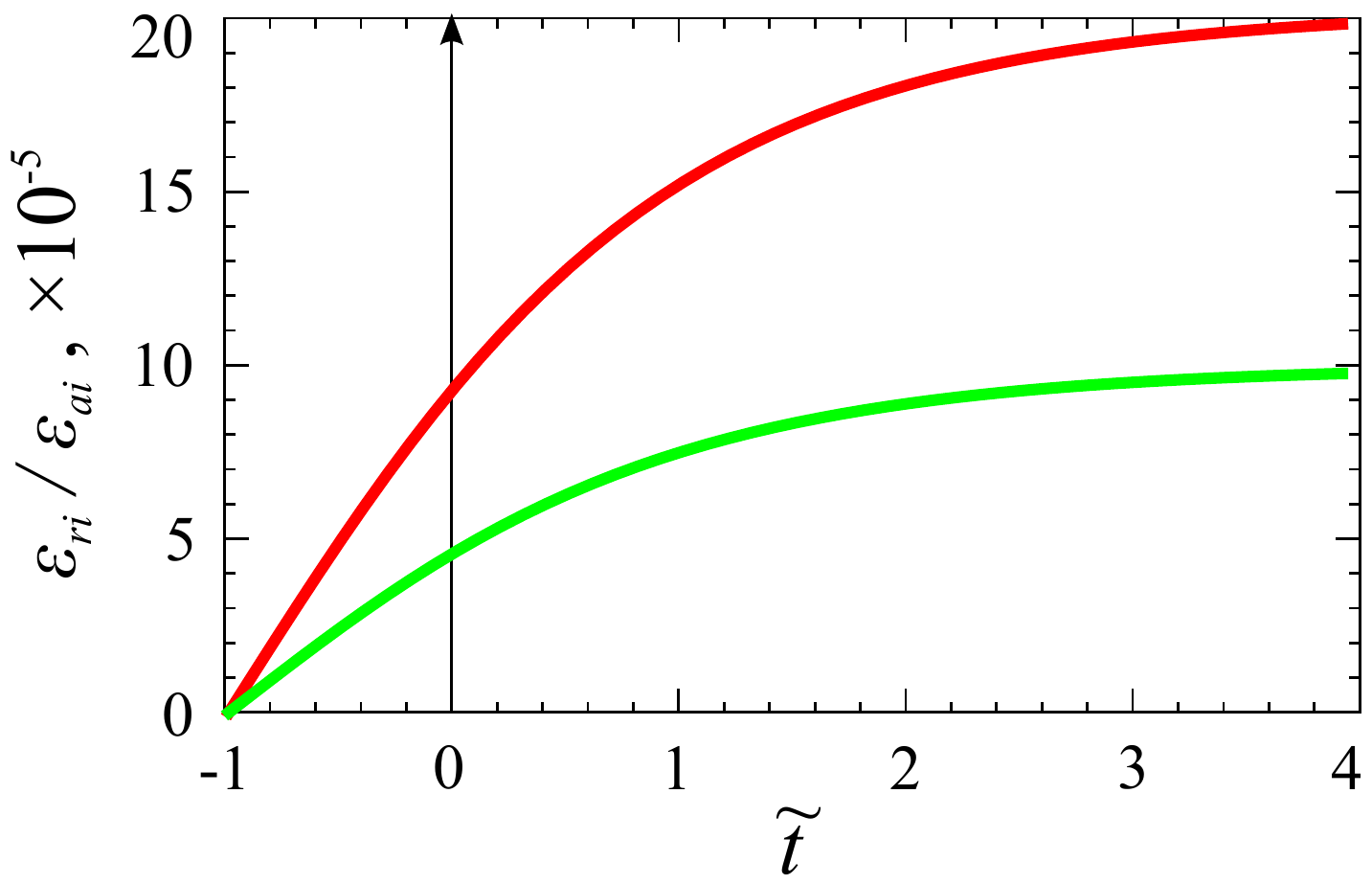}
        \includegraphics[height=5cm]{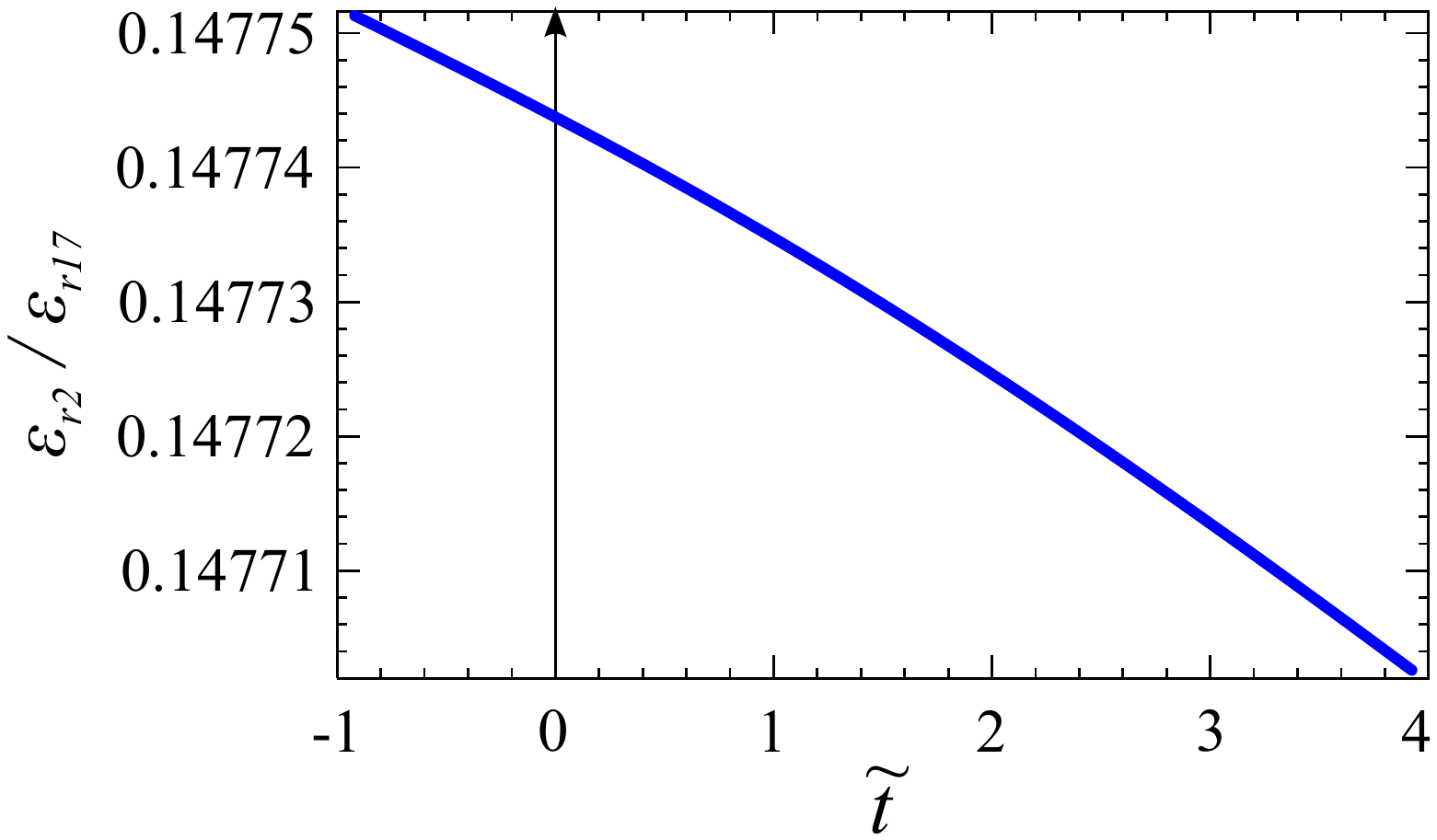}
    \end{center}
\caption{Graphs of the ratios $\varepsilon _{r2} / \varepsilon _{a2}$ (red), $\varepsilon _{r17} / \varepsilon _{a17}$ (green) \textbf{(a)} and $\varepsilon _{r2} / \varepsilon _{r17}$ \textbf{(b)} as functions of time $\tilde{t}$.}
\label{fig24}
\end{figure}

Now let us also depict the helpful graphs of the ratios $\varepsilon _{r2} / \varepsilon _{a2}$ and $\varepsilon _{r17} / \varepsilon _{a17}$  as functions of $\tilde{t}$ (Fig.~\ref{fig24}a, red and green lines respectively):

\begin{equation}
\frac{\varepsilon_{r2}}{\varepsilon_{a2}} = \frac{\exp (\tilde{\lambda}_{a2} \tilde{t})}{\tilde{a}} \left( \tilde{\lambda}_{a2} \int \limits _{0} ^{\tilde{t}} \tilde{a} (\tilde{t}) \exp (-\tilde{\lambda}_{a2} \tilde{t} ) d\tilde{t} + \frac{\varepsilon_{r2(0)}}{\varepsilon_{a2(0)}} \right),
\label{eq134}
\end{equation}

\begin{equation}
\frac{\varepsilon_{r17}}{\varepsilon_{a17}} = \frac{\exp (\tilde{\lambda}_{a17} \tilde{t})}{\tilde{a}} \left( \tilde{\lambda}_{a17} \int \limits _{0} ^{\tilde{t}} \tilde{a} (\tilde{t}) \exp (-\tilde{\lambda}_{a17} \tilde{t} ) d\tilde{t} + \frac{\varepsilon_{r17(0)}}{\varepsilon_{a17(0)}} \right),
\label{eq135}
\end{equation}

In particular, for $t = 0$ we get $\varepsilon _{r2} / \varepsilon _{a2} = 9.26 \cdot 10^{-5}$ and $\varepsilon _{r17} / \varepsilon _{a17} = 4.56 \cdot 10^{-5}$, as it should be (Fig.~\ref{fig24}a).

Let us also depict the helpful graph of the ratio $\varepsilon _{r2} / \varepsilon _{r17}$ as a function of $\tilde{t}$ (Fig.~\ref{fig24}b). 

The theoretical estimate of the ratio $\varepsilon _{r2} / \varepsilon _{r17}$ is, apparently, equal to

\begin{equation}
\left( \frac{\varepsilon_{r2}}{\varepsilon_{r17}} \right) _{theory} \cong 0.15.
\label{eq136}
\end{equation}

On the other hand, it is rather easy to estimate the experimental $\varepsilon _{r2} / \varepsilon _{r17}$ ratio using the data shown in Fig.~\ref{fig22} and the expression for the spectral intensity $I_{\lambda}(\lambda_0)$ of the background radiation from decaying axions:

\begin{equation}
\left( \frac{\varepsilon_{r2}}{\varepsilon_{r17}} \right) _{exper} = \frac{\int \limits_{12400}^{50000} I_{\lambda 2} (\lambda_0) d\lambda_0}{\int \limits_{1500}^{15000} I_{\lambda 17} (\lambda_0) d\lambda_0} \cong 0.23.
\label{eq137}
\end{equation}

Curiously enough, the theoretical estimate~(\ref{eq136}) and the experimental one~(\ref{eq137}) agree fairly well, which increases one's optimism as to the essence of the results obtained in the current section. At the same time, our estimates require additional thorough verification, especially when it comes to the experimental justification of the validity and reliability of the data on the extragalactic background light in near-ultraviolet, optical and near-infrared bands (1500-10000\AA). Indeed, it is a very important step to perform, because otherwise our unusual results may seem more like just the luckily guessed rules of calculation, which do not reflect the actual nature of things.

\section{Summary and Conclusion}
\label{sec-07}

In the present paper we present a self-consistent model of the axion mechanism of Sun luminosity and solar dynamo -- geodynamo connection, in the framework of which we estimate the values of the axion mass ($m_a \sim 17 ~eV$) and the axion coupling constants to photons ($g_{a \gamma} \sim 7.07 \cdot 10^{-11} ~GeV^{-1}$), nucleons ($g_{an} \sim 3.20 \cdot 10^{-7}$) and electrons ($g_{ae} \sim 5.28 \cdot 10^{-11}$). Their verification on the basis of the model results comparison with the known experiments is also provided, including the CAST-, CUORE- and XMASS-experiments.

In order to explain the solar-terrestrial magnetic connection we propose the axion mechanism explaining the Sun luminosity physics and "solar dynamo -- geodynamo" connection, where the total energy of axions, which appear in the Sun core, is initially modulated by the magnetic field of the solar tachocline zone due to the inverse coherent Primakoff effect and after that it makes its way towards the Earth where its "iron" component containing the 14.4~keV solar axions is resonantly absorbed inside the nickel-iron core of the Earth. It results in the fact that the variations of the axion intensity play a role of an energy source and a modulator of the Earth magnetic field. In other words, the solar axion mechanism is not only responsible for formation of a thermal energy source in the liquid core of the Earth necessary for generation and maintenance of the Earth magnetic field, but unlike other alternative mechanisms~\cite{ref015}, naturally explains the cause of the experimentally observed strong negative correlation of the magnetic field in the tachocline zone of the Sun and the magnetic field of the Earth.

It is necessary to note that obtained estimations can't be excluded by the existing experimental data (see Fig.~\ref{fig16}a and Fig.~\ref{fig18}a) because the effect of solar axion intensity modulation by temporal variations of the toroidal magnetic field of the solar tachocline zone discussed above was not taken into account in these observations (see Fig.~\ref{fig16}b and Fig.~\ref{fig18}b). On the other hand, the obtained estimates for the axion-photon coupling and the axion-nucleon coupling cannot be ruled out by the existing theoretical limitations known as the globular cluster star limit ($g_{a \gamma} < 6 \cdot 10^{-11} ~GeV^{-1}$) and the red giant star limit ($g_{ae} < 3 \cdot 10^{-13}$)~\cite{ref100}, since these values are highly model-dependent.  It actually means that the axion parameters obtained in the present paper do not contradict any of the known experimental and theoretical model-independent limitations.

Let us give some major ideas of the paper which formed a basis for the statement of the problem justification, and the corresponding experimental data which comport with these ideas either explicitly or implicitly.

\subsection{Axion mechanism of Sun luminosity}

One of the key ideas behind this mechanism is the effect of $\gamma$-quanta channeling along the magnetic flux tubes (waveguides) in the Sun convective zone (Fig.~\ref{fig08}), which may be represented by the $\gamma$-quanta channeling in the periodical structure (Fig.~\ref{fig06}) in the particular case. A low refraction (i.e. a high transparency) of the thin magnetic flux tubes is achieved due to the ultrahigh magnetic pressure (see~(\ref{eq013})), induced by the magnetic field of about 200-400~T (Fig.~\ref{fig09}a, adapted from~\cite{ref077}). It is noteworthy that although such strong magnetic fields have never been used for the explanation or interpretation of the simulation results such as the stability analysis of tachocline latitudinal differential rotation and coexisting toroidal band using an MHD analog of the shallow-water model~\cite{ref077}, they have always been present implicitly in the very same simulation results, as our analysis shows. Fig.~\ref{fig09}a provides an illustrative example where a hidden part of the "latitude -- magnetic field of the overshoot tachocline zone" dependence (which is absent on the original plot in Fig.11 of~\cite{ref077}) is added. A right to exist and physical validity of this part is confirmed by the bare fact of its direct correspondence (Fig.~\ref{fig09}a,b) with the real X-ray image of the Sun in its active phase (Fig.~\ref{fig09}b) obtained in the experiments performed with the Japanese X-ray telescope Yohkoh (1991-2001) (adapted from~\cite{ref014}).

The direct experiments on monitoring the space-time evolution of the magnetic flux tubes, rising from the deep layers of the convective zone of the Sun (Fig.~\ref{fig08}b, adapted from~\cite{ref064}) proved the existence of the "hollow" ideal magnetic waveguides for $\gamma$-quanta. These tubes cross the photosphere and form the solar active regions such as sunspots which are the sources of X-rays (Fig.~\ref{fig08}d, adapted from~\cite{ref068}). There are also some theoretical results substantiating the effect of anchoring magnetic flux tubes in the tachocline (Fig.~\ref{fig08}c, adapted from~\cite{ref067}).

Finally, the most impressive evidence of the axion mechanism of Sun luminosity are the solar images (Fig.~\ref{fig12}, adapted from~\cite{ref014}) taken at photon energies from 250~eV up to a few keV from the Japanese X-ray telescope Yohkoh (1991-2001), which depict the solar X-ray activity during the last maximum of the 11-year solar cycle (Fig.~\ref{fig12}b). It is hard to imagine another model or considerations which would explain such anomalous X-ray radiation distribution over the active Sun surface just as well. One should also keep in mind that the alternative mechanism of axion transformation into $\gamma$-quanta in the magnetic field of sunspots or other solar active regions is completely ruled out, since there are no signs of an X-ray bright spot at the disk center (see Fig.~\ref{fig12}a and b), which should otherwise be observed according to~\cite{ref217}.

\subsection{Invisible axions and Solar Equator effect}

According to the axion mechanism of Sun luminosity, a part of axions that pass through the tachocline zone near the equator and poles (Fig.~\ref{fig06} and Fig.~\ref{fig12}) is not converted into $\gamma$-quanta by the Primakoff effect because of the magnetic field vector collinearity to the axion momentum. It is the equatorial part~(Fig.~\ref{fig12}) of invisible axions that reaches the Earth, where its "iron" component (i.e. the 14.4~keV solar axions) is resonantly absorbed in the Earth core. The energy of the solar axions supplied to the Earth core in this way plays a role of a trigger for the generation and maintenance of the geomagnetic field, thus giving birth to the effect of the steady anticorrelation between the variations of the Solar magnetic field and geomagnetic field (Fig.~\ref{fig03}).

In this sense, it is appropriate to make a short remark related to the anticorrelation between the solar and terrestrial magnetic fields variations. The small variations of TSI ($\Delta$TSI) are apparently produced by the solar magnetic field variations. At the same time, these variations of the solar magnetic field drive the equatorial sector width variations (Fig.~\ref{fig12}), and consequently, the "equatorial" axion flux (see Fig.~\ref{fig11}) 

\begin{equation}
\Delta _{equ}^a = 0.05 \mp \Delta TSI / \langle TSI \rangle.
\label{eq138}
\end{equation}

In other words, the "equator" effect is not only the source of "invisible" axions, it also modulates their intensity inversely proportional to the solar magnetic field change (see~(\ref{eq138})), thus maintaining the inverse correlation between the solar magnetic field and the "invisible" axions flux (Fig.~\ref{fig11}). The latter is also the cause of the inverse correlation between the solar and terrestrial magnetic fields variations (Fig.~\ref{fig03}).

\subsection{Axion mechanism of the solar dynamo -- geodynamo connection}

This mechanism is responsible for the physical (axionic) nature of the steady anticorrelation between the variations of the solar magnetic field and the geomagnetic field (the Y-component) which is directly proportional to the westward drift of magnetic features (Fig.~\ref{fig04}, Fig.~\ref{fig05}) at the measurement points (Western Europe and Australia). Figuratively speaking, in this case the Y-component of the Earth magnetic field acts as a "measuring instrument" which tracks the influence of the "equatorial" 14.4~keV solar axions on thermal and magnetic processes in the liquid core of the Earth. The physical scenario for this may be following.

The "iron" solar axions that are resonantly absorbed in the Earth core, activate the vertical background motion along the gravity force. This motion, in its turn, "pushes" the temperature gradient to the bottom of the liquid core more or less heavily (depending on the total energy of axions), thereby changing the temperature profile in the convective medium of the Earth core. The change in the temperature profile also changes the thermal conditions near the nuclear georeactor situated at the boundary of the liquid and solid core of the Earth. Since the power output of such reactor is proportional to temperature in the range of 3000-5000C$^{\circ}$ typical for the Earth liquid core, it means that the variations of the Earth core temperature generated by the mechanism of solar dynamo-geodynamo connection induce the corresponding variations of the nuclear georeactor thermal power.  If the georeactor hypothesis is true, the fluctuations of georeactor thermal power, induced by the variations of the absorbed axions energy in the Earth's core, can influence the Earth global climate in the form of anomalous temperature jumps in the following way. Strong fluctuations of the georeactor thermal power can lead to partial blocking of the convection in the liquid core~\cite{ref113,ref114,ref115} and the change of the angular velocity of liquid geosphere rotation, thereby changing the angular velocities of the Earth mantle and crust by virtue of the total angular momentum conservation law. It means that the heat, or more precisely the dissipation energy caused by friction of the Earth surface and lower atmosphere, can make a considerable contribution to the total energy balance of the atmosphere and thereby influence on the Earth global climate evolution significantly~\cite{ref113,ref114,ref115}.

On the other hand, it is clear that understanding of the mechanism of the solar-terrestrial  magnetic correlation can become the clue of the so-called problem of solar power pacemaker related to possible existence of some hidden but crucial mechanism of the Sun's energy influence on the fundamental geophysical processes. It is interesting that the "tracks" of this mechanism have been observed for a long time and manifest themselves in different problems of solar-terrestrial physics, in particular, in climatology where the mechanisms, by which small changes in the Sun's energy output during the solar cycle can cause change in the weather and climate, have been a puzzle and the subject of intense research in recent decades.

If we also add the fact that the dissipation energy caused by friction of the Earth surface and the boundary layer has a considerable value, which is enough to cover the known deficiency of the total energy balance of the atmosphere, it becomes clear that in the framework of the axion mechanism of the solar dynamo -- geodynamo connection it is the solar magnetic field that is a host power pacemaker of the Earth global climate~\cite{ref113}.

\subsection{Axion-like particle and extragalactic background light}

A complete physical "portrait" of the axion, obtained from the axion mechanism of Sun luminosity and the solar dynamo -- geodynamo connection, actually predefined the necessary conditions of its detection in astrophysics. For example, for the search of the axion with the mass $m_a = 17 ~eV$ in radiative decays of the $a \to \gamma \gamma$ type, when the decay photons are emitted at or near a peak wavelength

\begin{equation}
\lambda_a = \frac{2hc}{m_a c^2} \cong 1459 \text{\AA},
\label{eq139}
\end{equation}

\noindent we used the known observational data obtained during the study of the diffuse extragalactic background which, according to~\cite{ref116}, are more suitable than the flux from any particular region of the sky.

Surprisingly, our theoretical fit of the spectral intensity of the background radiation (Fig.~\ref{fig20}) produced by axion decays describes the experimental data in the near ultraviolet and optical bands including the data from several ground- and satellite-based telescope observations~\cite{ref169,ref170,ref171,ref172,ref173,ref174,ref175,ref176,ref177,ref178,ref179} with very good accuracy.

In this sense we may say that our result, obtained within the self-consistent model of Sun luminosity and solar dynamo -- geodynamo connection, has a strong and evident "experimental" support represented by the astrophysical solar-like axions, provided that this is not just a fortuitous coincidence of the data (Fig.~\ref{fig20}), and the axions with the mass $m_a = 17 ~eV$ really exist.

\subsection{Plausible dark matter candidate: hadronic axion or axion-like arhion?}

Along with the identification of the solar-like axion with the 17~eV mass, we also found the possible relic axion-like archion with the 2~eV mass in near-infrared band of the extragalactic background (Fig.~\ref{fig22}), which behaves similarly to the hardronic axion with highly suppressed interaction with leptons under certain conditions.

Discovering both the hadronic axion and the axion-like archion (with similar properties, but different physical nature) made us think about what we actually see and identify. Aren't these two particles just axion-like archions in reality? All that we know about the World, tells us that Mother Nature, although is notable for its Darwinian diversity in the elementary particles zoo, would not "multiply entities beyond necessity". Keeping in mind this principle of Ockham's razor and our intuitive notion about the laws of astroparticle physics, let us give some arguments in favour of the statement that we are really dealing with the axion-like archions and not with hadronic axions.

The main reason of our doubts is the following. It is known~\cite{ref209} that the model of an archion automatically avoids the serious problems related to the necessity of a stable relic supermassive quark existence, predicted by the hadronic axion model. As a result, the so-called "wild isotopes" should exist as a consequence of a corresponding nucleosynthesis process, and they are not observed today, as is known. On the other hand, including the hadronic axion model into the Grand Unified model should lead to an inevitable existence of a supermassive lepton associated with the axion. This is due to the fact that within such unification, a composite of a supermassive quark with an ordinary light quark would cause a supermassive quark instability as well as the axion interaction with leptons, which is incompatible with the hadronic axion properties~\cite{ref210}. In other words, the axion just would not be hadronic in this case.

Everything is completely different within the archion model. In contrast to the hadronic axion model, the archion model (i.e. the Berezhiani-Saharov-Khlopov (BSK) model of horizontal unification~\cite{ref209}) naturally gives rise to a supermassive quark instability and a strong suppression of the axion-like archion interaction with leptons~\cite{ref203,ref204,ref205,ref206,ref207,ref208,ref209,ref210}. At the same time, because of the suppression of the archion interaction with the lightest generation of leptons ($u$,$d$,$e$,$\nu_e$), the existing limitations on the corresponding scale (the energy scale $f_{PQ}$ associated with the break-down of the U(1) PQ-symmetry) of the invisible axion are reduced to

\begin{equation}
f_a \equiv f_{PQ} \sim 10^6 ~~ GeV,
\label{eq140}
\end{equation}

\noindent which makes the BSK-model of an archion rather close to the hadronic axion model~\cite{ref032,ref032a,ref033,ref033a,ref102,ref103}.

It is interesting that for such values of the energy $f_a$ in the BSK-model of horizontal unification the stable massive neutrinos dominance becomes possible, or more strictly speaking, the stable sterile neutrinos with the mass about several keV and small mixing angles with the active neutrinos. If we also take into account that one or more of these gauge-singlet fermions can have Majorana masses below the electroweak scale, in which case they appear as sterile neutrinos in the low-energy theory~\cite{ref218}, it becomes clear that these particles have fundamental importance extending the Standard Model by gauge singlet fermions,  which can accommodate the neutrino masses. Furthermore, if one of the sterile neutrinos has mass in the 1--20 keV range and has small mixing angles with the active neutrinos, such particle is a plausible candidate for dark matter~\cite{ref219}. The same particle could be produced in the supernova explosion, and its emission from a cooling neutron star could explain the pulsar kicks, facilitate core collapse supernova explosions, and affect the formation of the first stars and black holes (e.g.,~\cite{ref218} and Refs. therein). Therefore, there is a strong motivation to search for signatures of sterile neutrinos in this mass range, especially as the first results were obtained recently indicating the detection of the sterile neutrino mass of 5.0$\pm$0.2~keV and a mixing angle in a narrow range for which neutrino oscillations can produce all of the dark matter~\cite{ref220}.

We tried to find a trace of the radiative decay of the sterile neutrino in the form $\nu _s \to \nu + \gamma$ as well.  As we noted before, one should not be embarrassed by the fact that the corresponding life-time of the sterile neutrino is many orders of magnitude larger than the age of the universe (see~(\ref{eq103}) and (\ref{eq105})), because sterile neutrinos are produced in the early universe by neutrino oscillations~\cite{ref219} and, possibly, by other mechanisms as well, and therefore every dark matter halo should contain some fraction of these particles. In other words, given a large number of particles in these astrophysical systems, even a small decay width can make them observable via the photons produced in the radiative decay.

\begin{figure}[tb]
    \begin{center}
        \includegraphics[width=18cm]{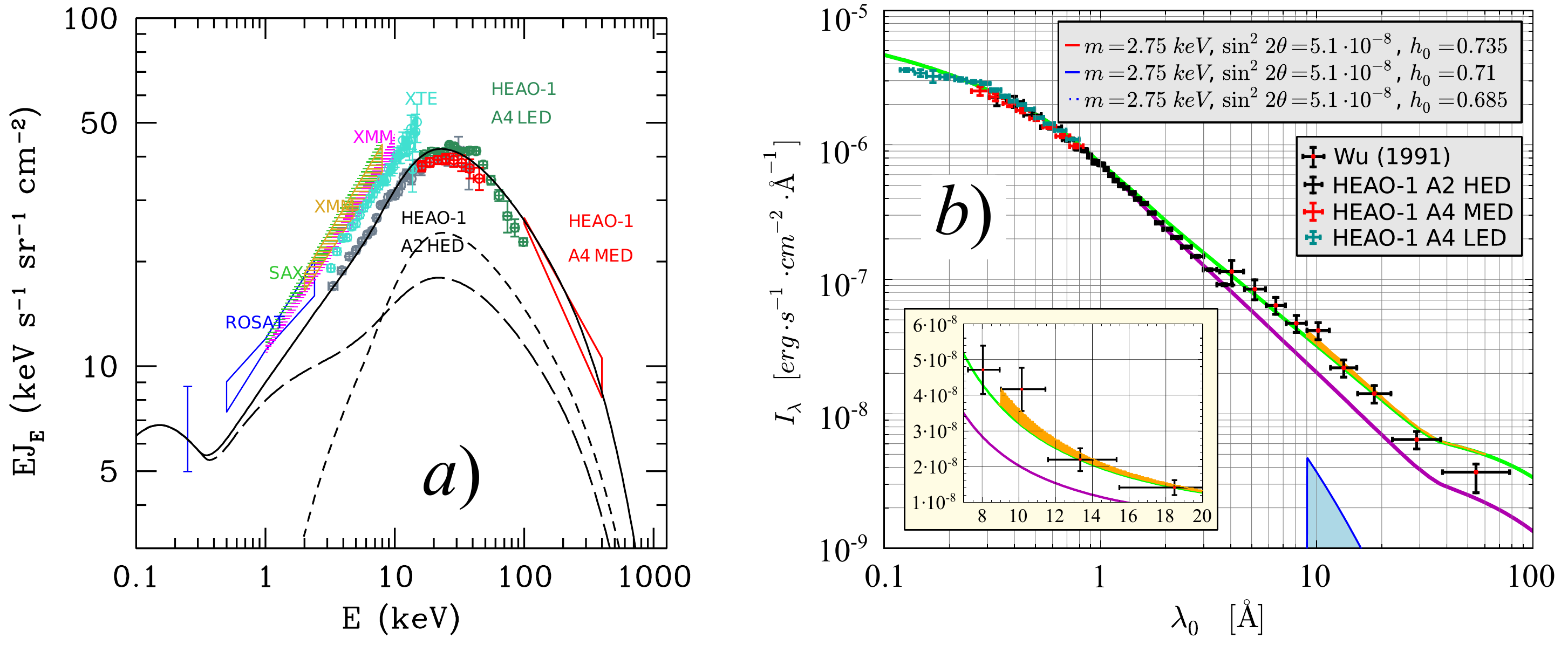}
    \end{center}
\caption[caption]{\textbf{(a)} The cosmic XRB spectrum and the predicted contribution from the active galactic nuclei (AGNs), that gives origin to the X-ray background. Gray points: HEAO-1 A2 HED data (Gruber~et~al.~\cite{ref221}). Dark green points: HEAO-1 A4 LED (Gruber~et~al.~\cite{ref221}). Cyan points: Rossi-XTE (Revnivtsev~et~al.~\cite{ref222}). Red bowtie: HEAO-1 A4 MED (Kinzer~et~al.~\cite{ref223}). Blue bowtie: ROSAT PSPC data (Georgantopoulos~et~al.~\cite{ref224}). Light green bowtie: BeppoSAX (Vecchi~et~al.~\cite{ref225}). Purple and yellow bowties: Newton-XMM (Lumb~et~al.~\cite{ref226}; De Luca \& Molendi~\cite{ref227}). Solid line: synthesis model spectrum by Haardt-Madau~\cite{ref228}, produced by a mixture of absorbed (the short-dashed black line) and unabsorbed (the long-dashed black line) AGNs. Adopted from~\cite{ref228}.\\
\textbf{(b)} The cosmic XRB spectrum as a function of the wavelength $\lambda_0$. Solid purple line: synthesis model spectrum by Haardt-Madau~\cite{ref228}, produced by a mixture of absorbed and unabsorbed AGNs. Black points with bars (crosses): HEAO A-2 results (Wu~et~al.~\cite{ref229}). Solid green line: our fit of cosmic XRB spectrum. The orange region is a sum of our fit of spectrum (the green line) and the spectrum of $\gamma$-quanta (the blue shaded region) born in the supposed radiative decay of the sterile neutrinos.}
\label{fig25}
\end{figure}

We limited our search for possible $\gamma$-spectra of the sterile neutrino decay to the analysis of the experimental results obtained by different authors during their study of the diffuse X-ray background (Fig.~\ref{fig25}a). The cosmic XRB spectrum in the 0.1 -- 1~keV range was particularly interesting in this sense, because the X-ray spectrum here is not "distorted" by the absorption processes, since unabsorbed AGNs dominate in this energy region (see predictions by the Haardt-Madau model on Fig.~\ref{fig25}a). Allowing for the fact that in this energy region the problem of Lyman-$\alpha$ forest may be neglected, it becomes possible to use Eq.~(\ref{eq086}) for theoretical calculation of the intensity of sterile neutrino contributions to the diffuse X-ray background as a function of the wavelength $\lambda_0$.

It is necessary to mention some specifics of cosmic XRB spectrum calculation for this region\footnote{The observations performed using the X-ray spectroscopy of locations in the Universe are not used here, because we calculate the spectrum of $\gamma$-quanta born in radiative decays $\nu_S \to \nu + \gamma$ for different $z$ (integral method~(86)~\cite{ref166}) in contrast to the X-ray spectroscopy of emission lines from the cooling stars (so-called differential method (e.g.~\cite{ref157})}. First, the XRB spectrum is highly "damaged" (see Fig.~\ref{fig25}a) by multiplicity of observations with different measuring bases, and possibly, different methodologies. For this reason we decided to use the observational data obtained in the single HEAO experiment. It covers the whole spectrum in 1 -- 400~keV range, and also provides the measurements of the cosmic XRB spectrum in 0.1 -- 1~keV region~\cite{ref229}. Second, when calculating the $\gamma$-spectrum formed by the radiative decay of sterile neutrinos, we used Eqs.~(\ref{eq086})-(\ref{eq088}) adjusted for expressions~(\ref{eq103})-(\ref{eq105}), which reflect the radiative features of sterile neutrinos, and assumed their velocity distribution (see~(\ref{eq088})) to be Gaussian with a variance of $\upsilon_c \sim 30 ~km / s$~\cite{ref230,ref230a} and their density to be equal~\cite{ref231,ref232}

\begin{equation}
\Omega_S h_0^2 \approx 0.3 \left( \frac{\sin^2 2 \theta}{10^{-10}} \right) \left( \frac{m_S}{100 ~keV} \right)^2.
\label{eq141}
\end{equation}

We also introduced a factor of 3 for the cosmic XRB spectrum calculation~(\ref{eq086}), which manually takes into account that the bulk of the EBL contributions in the decaying-neutrino scenario comes from neutrinos which are distributed on larger scales. We will refer to these collectively as free-streaming neutrinos, though some of them may actually be associated with more massive systems such as clusters of galaxies, and a possible impact of the astrophysical data non-gaussianity. In support of this assumption let us quote a remark made in paper~\cite{ref236}, where a factor of 10 was used: "...the fact that a significant part of the dark matter at redshifts $z \leqslant 10$ is concentrated in galaxies and clusters of galaxies just means that the strongest signal from the dark matter decay should come from the sum of the signals from the compact sources at $z \leqslant 10$. Taking into account that the DM decay signal from $z \leqslant 10$ is some two orders of magnitude stronger than that from $z \geqslant 10$, while the subtraction of resolved sources reduces the residual X-ray background maximum by a factor of 10, we argue that it would be wrong to subtract the contribution from the resolved sources from the XRB observations when looking for the DM decay signal".

Our theoretical fit (the yellow curve in Fig.~\ref{fig25}b) of spectral intensity of the background radiation~(\ref{eq086}) produced by sterile neutrino decays describes the experimental data from HEAO A-2 in near X-ray band~\cite{ref221} adequately. However, this result requires a serious and thorough verification, since the data by Wu~et~al.~\cite{ref229} are, in fact, the only cosmic XRB spectrum measurements of the Large Magellanic Clouds in the 0.16 -- 3.5~keV region. From this point of view, this result is more likely to serve as a demonstration of the possibilities and peculiarities of the sterile neutrino search in the diffuse extragalactic\footnote{Wu's analysis~\cite{ref229} of the background  data reveals a limit on the mean absolute intensity for the extragalactic emission $I_{\lambda}(0.16-3.5~keV) \sim 5.6 \cdot 10^{-8} ~ergs \cdot cm^{-2}s^{-1}sr^{-1}$.} X-ray background which sometimes may be more suitable than the flux (in the form of a narrow band) from any particular region of the sky.

Turning back to the archion properties, let us point out that such sterile neutrino ($m_S = 2.75~keV$, $\sin^2 2 \theta = 5.1 \cdot 10^{-8}$) representing the dark matter with density $\Omega _S \approx 0.20$ (see~(\ref{eq141}) and Fig.~\ref{fig26}) fits into the archion model very well. This is due to the fact that it is the sterile neutrino appearance that makes it possible to solve the problem of missing dark matter (see~(\ref{eq100})) which is associated with the archions model (in a case of $m_S > m_a$) and emerges in LTR cosmology:

\begin{figure}[tb!]
    \begin{center}
        \includegraphics[width=15cm]{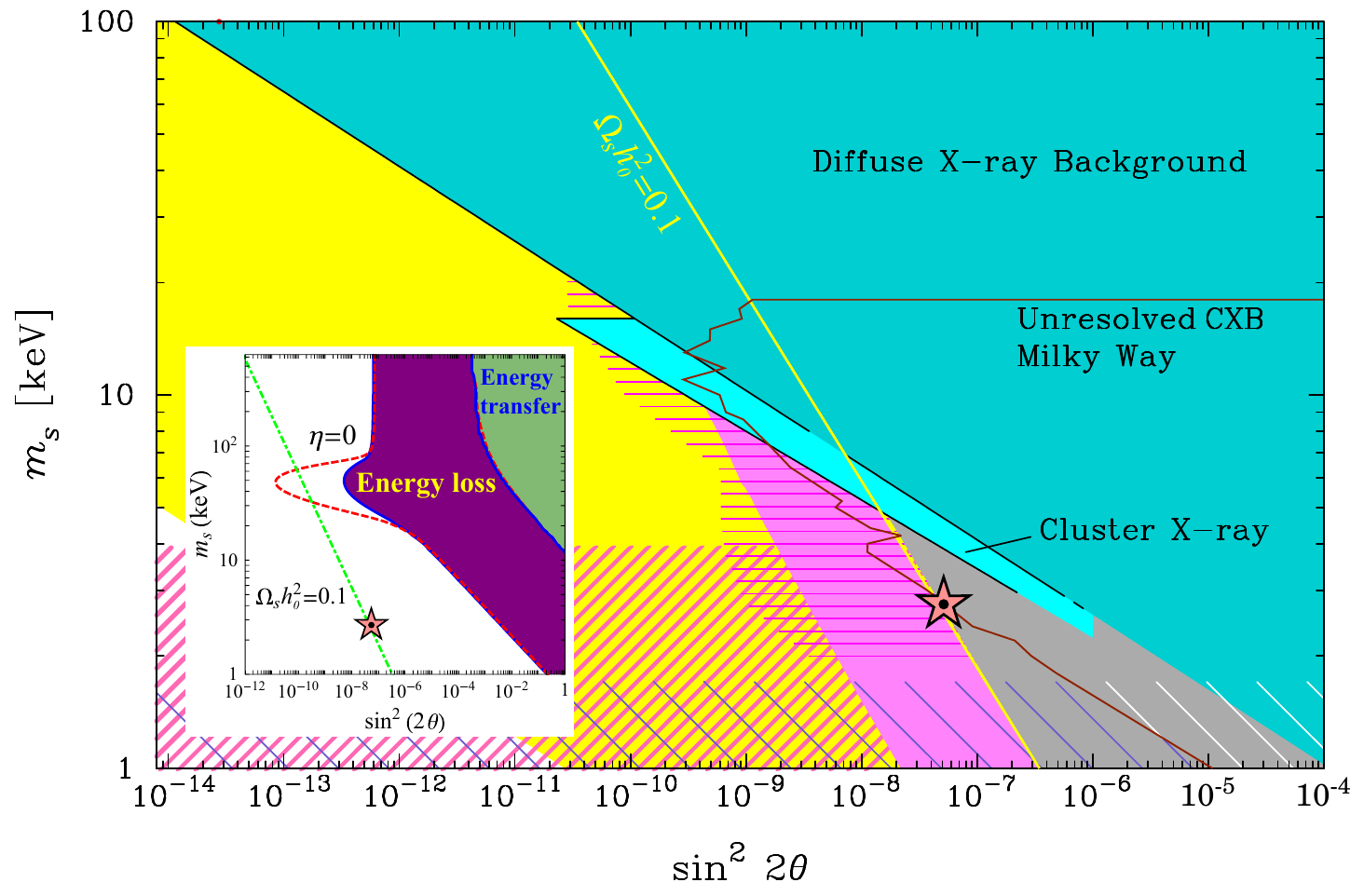}
    \end{center}
\caption[caption]{Full parameter space constraints for the sterile neutrino production model based on the diffuse extragalactic X-ray background, assuming sterile neutrinos constitute the dark matter. To facilitate comparisons, we adopt many of the conventions used by Abazajian~\cite{ref233}. Favored regions are in red/magenta colors, disfavored and excluded regions are in blue/turquoise colors. The favored parameters consistent with pulsar kick generation are in horizontal hatching (Kusenko~et~al.~\cite{ref234,ref235}). Constraints from X-ray observations include the diffuse X-ray background (turquoise) (Boyarsky~et~al.~\cite{ref236}). Also shown is the best current constraint from Chandra, from observations of contributions of dark matter X-ray decay in the cosmic X-ray background through the CDFN and CDFS (brown contour, "Unresolved CXB Milky Way"). Also shown is an estimate of the sensitivity of a 1~Ms observation of M31 with IXO (yellow). The region at $m_S < 1.7~keV$ is disfavored by conservative application of constraints from the Lyman-$\alpha$ forest.\\
\textbf{Inset:} Supernova bound on sterile neutrino masses $m_S$ and mixing angles $\sin^2 (2 \theta)$, where the purple region is excluded by the energy-loss argument while the green one by the energy-transfer argument~\cite{ref231}. The excluded region will be extended to the dashed (red) line if the build-up of degeneracy parameter is ignored, i.e., $\eta (t) = 0$. The dot-dashed (green) line represents the sterile neutrinos as dark matter with the correct relic abundance $\Omega _S h_0 ^2 = 0.1$. The red star ($m_S = 2.75~keV$, $\sin^2 2 \theta = 5.1 \cdot 10^{-8}$) marks our result of sterile neutrino parameters identification.}
\label{fig26}
\end{figure}

\begin{equation}
\Omega_{DM} = \Omega_S + \Omega_a ^{nth} \sim 0.25,
\label{eq142}
\end{equation}

\noindent where $\Omega _a ^{nth} \approx 0.04$ is a non-standard density of the axion dark matter~(\ref{eq100}).

Briefly summarizing all said above about the problem of hadronic axion and axion-like archion, in our opinion, the archion model is preferable as compared to the hadronic axion model, but as the saying is, time will tell.

And finally let us emphasize two the most painful points of the present paper, at least from its authors' point of view.

Regardless of the model type, the simultaneous existence of two axion-like particles (the hadronic axion with the mass 17~eV and the axion-like arhion with the mass 2~eV) rises a natural question whether there may be two different energy scales $f_a$ associated with the breakdown of the U(1) PQ-symmetry, one of them coming from the hadronic axion ($f_a = 3.5 \cdot 10^5 ~GeV^{-1}$), and another relic one -- from the axion-like archion ($f_a = 3.0 \cdot 10^6 ~GeV^{-1}$). Or otherwise stated, can the energy have two scales, or in general, be hierarchical? If it can, then what kind of consequences of such fundamental symmetry violations hierarchy may be observed today? However, if one just winks at the possible existence of the axion-like particle with the 2~eV mass, the problem just vanishes without any consequences for the results of the present paper.

The second painful point is related to the key problem of the axion mechanism of Sun luminosity and is stated rather simply: "Is the process of axion conversion into $\gamma$-quanta by the Primakoff effect really possible in the Solar tachocline magnetic field?" This question is directly connected to the problem of the hollow magnetic flux tubes existence in the convective zone of the Sun, which are supposed to connect the tachocline with the photosphere. So, both the theory and experiment have to answer the question of whether there are the waveguides in the form of the hollow magnetic flux tubes in the convective zone of the Sun, which are perfectly transparent for $\gamma$-quanta, or our model of the axion mechanism of Sun luminosity and solar dynamo -- geodynamo connection is built around simply guessed rules of calculation which do not reflect any real nature of things.

\putbib[Axion_luminosity-references]
\end{bibunit}

\appendix
\numberwithin{equation}{section}
\numberwithin{figure}{section}

\begin{bibunit}[unsrt]

\section{Appendix I. Effect of  $\gamma$-quanta channeling in periodic structures}
\label{A1}

It is known \cite{ref-a1.01} that the real part of dielectric susceptibility $\chi (\omega )$ for the photons with energies exceeding the $K$-electrons binding energy has the form

\begin{equation}
Re \chi (\omega) = - \omega_e ^2 / \omega^2,
\label{eq-a1.01}
\end{equation}

\noindent where $\omega _e = (4 \pi N e^2 / m_e )^{1/2}$ is the electron plasma frequency, $N$ is the electrons density, $m_e$ is the electron mass. Since $\omega_e \sim 10 ~eV$ for the majority of materials, the susceptibility is small and negative in the X-ray band. It means that X-ray photons may experience the total reflection on the border of two materials from the one with the larger value of $\vert Re~ \chi (\omega ) \vert$, and the angle of photons arrival to the border should not exceed $\theta _c = \vert Re~ \chi \vert ^{1/2}$, where $\Delta ( Re~ \chi )$ is a dielectric susceptibility step. It is appropriate to mention that the phenomenon of a small-angle X-ray reflection has been used in X-ray optic elements for a long time~\cite{ref-a1.02}, and also for transporting (e.g.~\cite{ref-a1.03}) and turning (e.g.~\cite{ref-a1.04}) the X-ray bundles by cylindrical tubes.

On the other hand, it is also known that according to the optical-mechanical analogy (Hamilton and Fermat principles identity), in certain cases the radiation propagation may be described in terms of ray trajectories obeying the principle of least action. The refractive index, defining a ray trajectory, plays a role of an external potential in this case.  Based on this analogy, one may conjecture that there exists an effect of high-energy chargeless particles channeling that leads to the substantial anisotropy after passing the periodic structures. This assumption is validated using the example Vinecky and Finegold model problem \cite{ref-a1.51,ref-a1.52} related to calculation of the ray trajectories and absorption coefficients for the hard photons in geometrical optics approximation.

\subsection{Statement of a problem}

Let us follow \cite{ref-a1.51,ref-a1.52} and consider $\gamma$-radiation with frequency $\omega = 2 \pi c / \lambda$ transmission through a medium under the following condition:

\begin{equation}
\lambda_e < \lambda \ll a_0,
\label{eq-a1.02}
\end{equation}

\noindent where $\lambda _e = \hbar / mc$  is the electron Compton wavelength, $a_0$ is a typical interatomic distance in the medium.

The left-hand side of the inequality (\ref{eq-a1.02}) lets one consider the quanta as propagating in the continuous medium with a refraction index $n$. As long as the right-hand side of (\ref{eq-a1.02}) holds, the incident wave frequency is substantially higher than the lattice vibration eigenfrequencies (at least for the higher electron shells of the atoms that form the lattice). Therefore the scattering electrons may be considered free, and a crystal in the large may be treated as a frozen spatially-inhomogeneous electron plasma (the impact of nuclei on the scattering is negligible).

The electron plasma dielectric constant\footnote{For a justification of the term "dielectric constant" applicability for the X-ray and $\gamma$-bands in the absence of the Lorentz field averaging over the physically small volume elements see~\cite{ref-a1.05}.} may be written down in the form

\begin{equation}
\varepsilon = \varepsilon_1 + i \varepsilon_2 = 1 - \frac{\omega_e ^2 ( \vec{r})}{\omega^2 + i \omega \omega_{\gamma}},
\label{eq-a1.03}
\end{equation}

\noindent where $\omega _{\gamma}$ is the damping parameter which defines the wave absorption in the system.

For the refraction index

\begin{equation}
\nonumber n = \sqrt{\varepsilon} = n_1 + i n_2
\end{equation}

\noindent from (\ref{eq-a1.03}) taking into account that $\omega \gg \omega _e \gg \omega _{\gamma}$ in the $\gamma$-band, we derive

\begin{equation}
n_1 \approx 1 - \frac{1}{2} \left( \frac{\omega_e}{\omega} \right)^2, ~~ n_2 \approx \frac{1}{2} \frac{\omega_e ^2 \omega_{\gamma}}{\omega^3}.
\label{eq-a1.04}
\end{equation}

\begin{figure}
    \begin{center}
        \includegraphics[width=10cm]{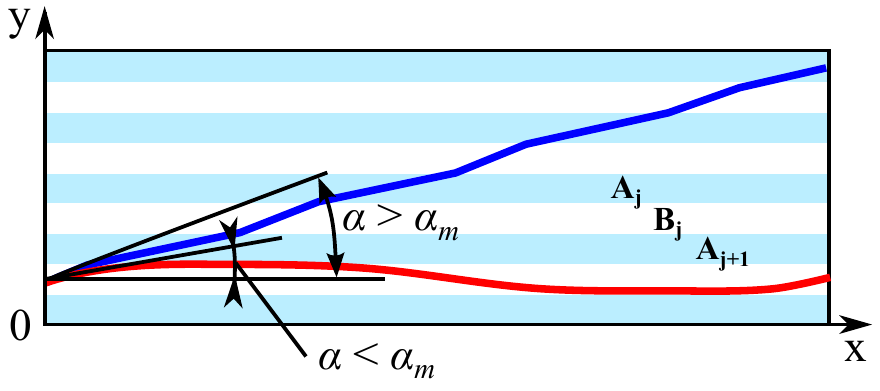}
    \end{center}
\caption{Ray trajectory dependence on the angle $\alpha$ in a long-period structure: the channeling happens when $\alpha < \alpha _m$; when $\alpha > \alpha _m$, the ray crosses the surfaces.}
\label{fig-a1.1}
\end{figure}

In order to perform the quantitative calculations, let us consider a simplified model of the one-dimensional lattice (a stack of alternating layers with different, but constant, densities and uniform $N(\vec{r})$ distribution along the layer (Fig.~\ref{fig-a1.1})).

Let us choose the distribution in the $y$ direction (perpendicular to the layers) in the form:

\begin{equation}
N (y) = N_0 \left( 1 + \beta^2 \sin^2 \pi \frac{y}{a} \right).
\label{eq-a1.05}
\end{equation}

Corresponding expressions for $n_1$, $n_2$ are

\begin{equation}
n_1 = 1 - \frac{1}{2} \xi ^2 \left( 1 + \beta^2 \sin^2 \frac{1}{2} k_a y, ~~ n_2 = \frac{1}{2} \zeta \xi^2 \left( 1 + \beta^2 \sin^2 \frac{1}{2} k_a y \right) \right),
\label{eq-a1.06}
\end{equation}

\noindent where

\begin{equation}
\xi ^2 \equiv \frac{\omega_0 ^2}{\omega^2} = \frac{4 \pi N_0 e^2}{m \omega^2} \ll 1, ~~ \zeta \equiv \frac{\omega_{\gamma}}{\omega} \ll \xi, ~~ k_a \equiv \frac{2\pi}{a}.
\label{eq-a1.07}
\end{equation}

So the problem reduces to determining the intensity of the $\gamma$-quanta beam hitting a sample of the thickness $x$ with the angle $\alpha$ to the $O_x$ axis (Fig.~\ref{fig-a1.1}).

We shall use the equation for the intensity of a ray that passed a path $l$ in the absorbing medium:

\begin{equation}
J_1 = J_0 \exp \left \lbrace - 2 k \int n_2 (l) dl \right \rbrace , ~~ k = \frac{2\pi n_1}{\lambda} \approx \frac{2 \pi}{\lambda},
\label{eq-a1.08}
\end{equation}

The integral in (\ref{eq-a1.08}) is taken along the ray trajectory, while the trajectory is determined by means of the Fermat principle \cite{ref-a1.06}

\begin{equation}
\nonumber \delta \int n_1 dl = 0,
\end{equation}

\noindent variation of which yields a system of three nonlinear second-order differential equations describing the ray equation in a parametric form $x = x(l)$, $y = y(l)$, $z = z(l)$:

\begin{equation}
n_1 \frac{d^2 x_i}{dl^2} + \left( \frac{\partial n_1}{\partial x_j} \frac{dx_j}{dl} \right) \frac{dx_i}{dl} = \frac{\partial n_1}{\partial x_i}, ~~ i,j = 1,2,3
\label{eq-a1.09}
\end{equation}

\noindent ($x_1$, $x_2$, $x_3$ correspond to $x$, $y$, $z$). It follows from the identical relation $dx_j dx_j = dl^2$ that only two equations of (\ref{eq-a1.09}) are independent.

\subsection{Solution of the model problem}

For the model (\ref{eq-a1.05}) the ray trajectory lies in one plane $xy$, and the system (\ref{eq-a1.09}) reduces to the equation

\begin{equation}
n_1 \frac{d^2 y}{dl^2} + \frac{\partial n_1}{\partial y} \left( \frac{dy}{dl} \right) ^2 = \frac{\partial n_1}{\partial y}.
\label{eq-a1.10}
\end{equation}

Let us point out some features of the channeling process that directly follow from Eq.~(\ref{eq-a1.10}). Assuming $dy / dl = 0$ everywhere along the ray, we derive the particular solution~(\ref{eq-a1.10}) which describes the propagation in the planes parallel to the planes of constant density. According to~(\ref{eq-a1.10}), such motion is only possible when $dn_1 / dy = 0$, i.e. within the planes $A_j$ and $B_j$, corresponding to the minimum and maximum values of $n_1 (y)$. Otherwise, the straight motion of a ray along the planes $n_1 (y) = const$ is impossible. Particularly, if the ray is initially parallel to the planes, but lies neither in $A_j$, nor in $B_j$, it is bent towards the nearest $B_j$, since $n_1$ grows in this direction. Therefore, the motion within the plane $B_j$ is stable, while within $A_j$ it is not. In other words, the rays entering the layers along the $O_x$ axis are pushed out into the "channels" of the low electron density (regions with minimum $N(y)$).

The first integral of Eq.~(\ref{eq-a1.10}) is as follows:

\begin{equation}
\frac{d y}{d l} = \left[ 1 - \left( \frac{n_1 (y_0)}{n_1 (y)} \cos \alpha \right)^2 \right]^{1/2}, ~~ \frac{d y}{d x} = \left[ \left( \frac{n_1 (y)}{n_1 (y_0) \cos \alpha} \right)^2 - 1 \right]^{1/2}.
\label{eq-a1.11}
\end{equation}

Let us use (\ref{eq-a1.11}) for determining the area of the ray motion depending on initial values of $\alpha$ and $y_0$. By taking $dy / dx = 0$ in the ray's turning point $y = y_m$ in (\ref{eq-a1.11}), we write down

\begin{equation}
n_1 ^2 (y_m) = n_1 ^2 (y_0) \cos^2 \alpha.
\label{eq-a1.12}
\end{equation}

By substituting (\ref{eq-a1.06}) we derive

\begin{equation}
\sin^2 \left( \frac{1}{2} k_a y_m \right) = \frac{1 - \xi ^2}{( \beta \xi )^2} \sin^2 \alpha + \cos^2 \alpha \cdot \sin^2 \left( \frac{1}{2} k_a y_0 \right).
\label{eq-a1.13}
\end{equation}

Eq.~(\ref{eq-a1.13}), obviously, has a real root $y_m$ only in the case

\begin{equation}
\frac{1 - \xi^2}{( \beta \xi )^2} \sin^2 \alpha + \cos ^2 \alpha \cdot \sin^2 \left( \frac{1}{2} k_a y_0 \right) \leqslant 1,
\label{eq-a1.14}
\end{equation}

\noindent whence taking into account (\ref{eq-a1.07}) we obtain

\begin{equation}
\alpha \leqslant \alpha_m (y_0), ~~ \alpha_m (y_0) = \arctan \left[ \beta \xi \cos \left( \frac{1}{2} k_a y_0 \right) \right] \ll 1.
\label{eq-a1.15}
\end{equation}

The $dy / dx$ function zeros have the form

\begin{equation}
y_m = \pm k_a ^{-1} \arccos \left( \cos k_a y_0 - 2 (\beta \xi )^{-2} \sin^2 \alpha \right).
\label{eq-a1.16}
\end{equation}

For each $y_0$ a ray with the angle of arrival $\alpha < \alpha _m (y_0)$ "oscillates" between the end points $[ y_m, - y_m]$ when passing through a sample, i.e. it "channels" between two neighbouring planes $A_j$  and  $A_{j+1}$. Hence, $\alpha _m (y_0)$ is the maximum angle of channeling for the given $y_0$. According to (\ref{eq-a1.15}), the angle $\alpha _m = 0$ corresponds to $y_0 - y_m = (1/2) a$, which means that the values of $y_0$ that make the channeling possible lie in the region

\begin{equation}
\nonumber \left( - \frac{1}{2} a, \frac{1}{2} a \right),
\end{equation}

\noindent thus embracing the whole sample facet exposed to the beam.

If $\alpha > \alpha _m (y_0)$, the $dy / dx$ function has no zeros -- a ray that entered under the angle $\alpha > \alpha _m (y_0)$ does not channel -- instead it crosses the layers one by one and is heavily absorbed by them. Fig.~\ref{fig-a1.1} shows the trajectories of both the "captured" ray that propagates between two planes throughout the sample and the non-channeling one.

From this analysis the following picture of the channeling emerges. When a beam hits the sample in the $\alpha = 0$ direction, virtually all the rays pass through the sample without any substantial absorption -- including the ones that propagate within the strongly absorbing layers as they are pushed out back into the channels. Under small $\alpha \neq 0$ the thin bands of $\delta y_0$ values appear on both sides of the low density planes, where the incident rays quit channeling by switching to the crossing trajectories. Since each crossing is accompanied by a loss of some amount of energy, we obtain two sets of rays -- experiencing the weak and strong absorption respectively. With $\alpha$ growth (and, consequently, the $\delta y_0$ bands growth) first set of rays shrinks, while the second set grows.  When $\alpha = \alpha _m (y_0 = 0)$, the channeling region collapses and all the rays are strongly absorbed. Therefore the value

\begin{equation}
\alpha _M \equiv \alpha_m (y_0 = 0) = \arctan \beta \xi \approx \beta \xi - \beta \frac{\omega_e}{\omega}
\label{eq-a1.17}
\end{equation}

\noindent characterizes the maximum possible channeling angle, thus determining the divergence of a beam for a given sample. As seen from (\ref{eq-a1.17}), the value of $\alpha_M$ decreases with the beam rigidity growth!

Let us estimate the value of $\alpha _M$ for the plasma medium in the Solar tachocline zone. Let $\beta \sim 1$ for simplicity. Taking into account that the plasma frequency in the tachocline is equal to $\omega _e \sim 4.6 \cdot 10^{16} ~s^{-2}$ [see~(\ref{eq008}) in the main part], and the frequency of $\gamma$-quanta with the energy $\langle E \rangle = 4.2 ~keV$ (or $\lambda \sim 4.6 \cdot 10^{-9} ~cm$) is $\omega = w \pi c / \lambda = 2 \pi \langle E \rangle / \hbar \sim 4 \cdot 10^{19} ~s^{-1}$, from (\ref{eq-a1.17}) we obtain $\alpha _m \sim 10^{-3}$. For the sake of channeling effect illustration, Fig.~\ref{fig-a1.1} shows the model results of the photon ($\lambda \sim 10^{-9} ~cm$) beam propagation through the layered media with $\alpha _M \sim 10^{-4}$ (see Section~\ref{sec-A1.4}).

Let us now integrate the Eqs. (\ref{eq-a1.11}). For the case of (\ref{eq-a1.04}) these equations are reducible to the following form (up to the first vanishing terms of the order $\xi ^2$ under the radical sign):

\begin{equation}
\frac{d y}{d l} = p \sin \alpha \sqrt{1 - q^2 \sin^2 \left( \frac{1}{2} k_a y \right)}, ~~ \frac{d y}{d x} = p \tan \alpha \sqrt{1 - q^2 \sin^2 \left( \frac{1}{2} k_a y \right)},
\label{eq-a1.18}
\end{equation}

\noindent where

\begin{equation}
p \equiv \sqrt{1 + (\beta \xi)^2 \cot ^2 \alpha \sin^2 \left( \frac{1}{2} k_a y_0 \right)}, ~~ q \equiv \frac{\beta \xi \cot \alpha}{p} = \left( \sin^2 \frac{\pi y_0}{a} + \frac{\tan ^2 \alpha}{( \beta \xi )^2} \right)^{-1/2}.
\label{eq-a1.19}
\end{equation}

By integrating (\ref{eq-a1.18}) we obtain the following trajectory equation:

\begin{equation}
x = x_j + \frac{2}{k_a \beta \xi}
\begin{cases}
F \left( \arcsin q \sin \left( \frac{1}{2} k_a y \right), q^{-1} \right) & at ~~ q \geqslant 1, \\
qF \left( \frac{1}{2} k_a y,q \right) & at ~~ q < 1
\end{cases}
\label{eq-a1.20}
\end{equation}

Here $F( \varphi , q )$ is an elliptic integral of the first kind, and the inverse function amF (the Jacobian elliptic function) period is equal to $4 K (q)$, where $K (q) \equiv F (\pi /2 , q)$ \cite{ref-a1.07}. Correspondingly, the function $y(x)$ inverse to (\ref{eq-a1.20}a) is periodical with the period

\begin{equation}
x _{\tau} = \frac{8}{k_a \beta \xi} K ( q^{-1} )
\label{eq-a1.21}
\end{equation}

\noindent and is associated with the channeling trajectories. For the crossing trajectories~(\ref{eq-a1.29}b) which are translation-invariant under the simultaneous transformations

\begin{equation}
x \rightarrow x + j x_{\tau}, ~~ y \rightarrow y + ja, ~~ j = 0,1,2,...,
\label{eq-a1.22}
\end{equation}

\noindent we have

\begin{equation}
x _{\tau} = \frac{8 q}{k_a \beta \xi} K (q).
\label{eq-a1.23}
\end{equation}

The regions of channeling and crossing trajectories are delimited by the value of the angle $\alpha = \alpha _m (y_0)$ corresponding to the value $q = 1$. When $\alpha = \alpha _m (y_0)$, the rays asymptotically approach the "repelling" planes $A_j$ and $x_{\tau} \to \infty$.

A number of trajectory oscillations along the distance $x$ for the channeling and crossing rays is, respectively

\begin{equation}
N_{\tau} = \frac{x}{x_{\tau}} = \frac{1}{8} k_a \beta \xi 
\begin{cases}
K^{-1} (q^{-1}) & at ~~ q > 1, \\
q^{-1} K^{-1}(q) & at ~~ q<1.
\end{cases}
\label{eq-a1.24}
\end{equation}

\subsection{Determination of the absorption coefficient angular dependence}

Let us move on to the absorption coefficient calculation in the exponent (\ref{eq-a1.08}). By means of (\ref{eq-a1.18}), integration over trajectory reduces to integration over a variable $ u = k_a y / 2$ within the region

\begin{equation}
\nonumber 
\begin{cases}
0 \leqslant u \leqslant u_m & at ~~ q > 1, \\
0 \leqslant u \leqslant \pi / 2 & at ~~ q<1.
\end{cases}
\end{equation}

Let us remind that $u = k_a y_m / 2 = \arcsin (q^{-1})$ by virtue of (\ref{eq-a1.15}) and (\ref{eq-a1.19}) corresponds to the turning points $y_m$ of the channeling trajectories. Further, substituting (\ref{eq-a1.08}) into the expression for $n_2$ (\ref{eq-a1.06}), performing integration and multiplying the obtained integrals, which correspond to the above mentioned regions, by $N_{\tau}$ (\ref{eq-a1.24}), we obtain

\begin{equation}
\sigma = k \zeta \xi ^2 \frac{x}{\cos \alpha}
\begin{cases}
1 + \beta^2 \left( 1 - \frac{E(q^{-1})}{K (q^{-1})} \right)   & at ~~ q > 1, \\
1 + \beta^2 \left( 1 - \frac{E(q)}{K (q)} \right)   & at ~~ q < 1, \\
\end{cases}
\label{eq-a1.25}
\end{equation}

\noindent where $E(q)$ is the complete elliptic integral of the second kind \cite{ref-a1.08}.

According to (\ref{eq-a1.25}), the intensity $J(x)$ (see (\ref{eq-a1.08})) of a ray that passed through a sample of the thickness $x$ may be written down as

\begin{equation}
J (x) = J_0 \exp (- \sigma x ) = J_0 \exp \left( - \frac{\chi_0}{\cos \alpha} x \right) \cdot Q (\alpha, y_0, x),
\label{eq-a1.26}
\end{equation}

\noindent where

\begin{equation}
Q (\alpha, y_0, x) = 
\begin{cases}
\exp \left[ -\frac{\chi_0}{\cos \alpha} \beta^2 x \left( 1 - \frac{E(q^{-1})}{K (q^{-1})} \right) \right]   & at ~~ q > 1, \\
\exp \left[ -\frac{\chi_0}{q^2 \cos \alpha} \beta^2 x \left( 1 - \frac{E(q)}{K (q)} \right) \right] & at ~~ q < 1.
\end{cases}
\label{eq-a1.27}
\end{equation}

\begin{figure}
    \begin{center}
        \includegraphics[width=9cm]{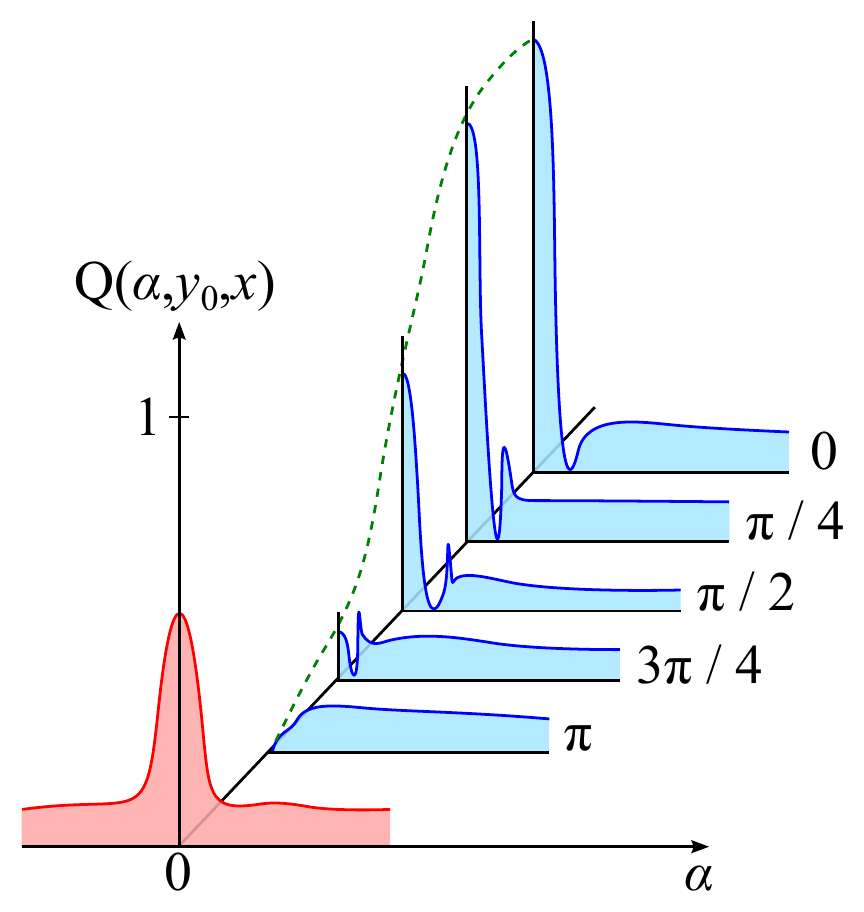}
    \end{center}
\caption{Transmitted beam relative intensity dependence on the arrival angle $\alpha$ for different values of $y_0$ (blue curves; the symmetric parts of the curves for $ \alpha < 0$ are not shown). Red curve represents $Q(\alpha, y_0, x)$ averaged over all values of $y_0$.}
\label{fig-a1.2}
\end{figure}

Here the multiplier $J_0 \exp ( - \chi _0 x / \cos \alpha )$ corresponds to $\gamma$-rays propagation through a homogeneous medium with the electron density $N_e$ and the absorption coefficient $\chi _0 = k \zeta \xi ^2$. An additional multiplier $Q(\alpha ,y_0,x)$ characterizes the influence of the medium layering. Fig.~\ref{fig-a1.2} shows the $Q(\alpha,y_0,x)$ curves for

\begin{equation}
\frac{y_0}{a} = 0, ~ 1/8, ~ 1/4, ~ 3/8, ~ 1/2
\label{eq-a1.28}
\end{equation}

\noindent with $\lambda = 10^{-9} ~cm$, $\beta = 1$ and $\chi _0 x = 2$, obtained by a numerical calculation using the elliptic integral value tables~\cite{ref-a1.08}.

As seen in the figure, the "total transmission" $(Q=1)$ is observed when $\alpha=0$, $y_0=0$ (the propagation in the $B_j$ plane), and the maximum absorption depending on $\beta$ and $(Q \to \exp (- \chi _0 \beta x  \cos \alpha ))$ is observed when $\alpha=0$, $y_0=(1/2)a$ (the propagation in the $A_j$ plane).

The dips on the decaying parts of the transmission curves when $y_0 \neq (1/2) a$ correspond to the critical values of the angle $\alpha = \alpha _m (y_0)$ (\ref{eq-a1.15}), for which the $\gamma$-quanta beam, that entered a sample, asymptotically approaches the nearest $A_j$ plane and is absorbed in it. With somewhat bigger values of $\alpha$ the rays pass on to the neighboring inter-plane space and moving within it leave a sample before they enter the next $A_{j+1}$ plane. The additional narrow maxima on blue curves (Fig.~\ref{fig-a1.2}) correspond to this type of rays.

Finally, when $\alpha > \alpha _M$ and, particularly, when $\alpha \to \pi / 2$ (propagation across the layers) from (\ref{eq-a1.18}) we obtain $q \to 0$, which corresponds to $Q \to \exp (-\chi _0 \beta ^2 x / 2 \cos \alpha)$, i.e. there is an additional absorption due to the averaged additional density of the layers $N_1 = N_0 \beta ^2$.

Fig.~\ref{fig-a1.2} shows a curve $Q(\alpha ,y_0,x)$ averaged over all values of $y_0$. The side dips and maxima are smoothed out, and there is only a central maximum. It corresponds to a clearly defined primary $\gamma$-quanta propagation with their angle of arrival within $\alpha _M \sim 2 \cdot 10^{-4}$. Therefore, the dense layers "modulating" a material play a role of an anisotropic filter which passes the $\gamma$-radiation for a narrow range of angles $\alpha < \alpha _M$ only. The physical mechanism of the arising transparency anisotropy consists in $\gamma$-quanta channeling between the layers because of the radiation "refraction" in a heterogeneous electron plasma.

\subsection{Channeling effect onset conditions}
\label{sec-A1.4}

By definition, channeling occurs for the particles the motion of which in the transversal phase plane is limited to a region

\begin{equation}
\Delta y \Delta p \sim a \alpha \hbar k,
\label{eq-a1.29}
\end{equation}

\noindent where $\alpha$ is a channeling angle. According to the uncertainty principle, the size of this region cannot be less than $\hbar$, i.e.

\begin{equation}
a k_0 \geqslant \alpha ^{-1}.
\label{eq-a1.30}
\end{equation}

By letting $\alpha \sim \omega _e / \omega$ for the estimation, from (\ref{eq-a1.30}) we derive

\begin{equation}
a k_0 \geqslant 1, ~~ k_0 \equiv \omega_0 / c .
\label{eq-a1.31}
\end{equation}

For example, in a monocrystal with $a = a_0 \sim 3 \cdot 10^{-8} ~cm$, $\omega_0 \sim 5 \cdot 10^{16} ~s^{-1}$ we have $a_0 k_0 \sim 5 \cdot 10^{-2} \ll 1$, i.e. the condition (\ref{eq-a1.31}) does not hold and, consequently, the channeling is impossible. This prohibition is true for any quanta, since in the region $\omega \gg \omega_0$, where the approximation (\ref{eq-a1.04}) and (\ref{eq-a1.05}) is applicable, the frequency $\omega$ falls out from the condition (\ref{eq-a1.31}).

However, for a layered system with $a \gg a_0$ the condition~(\ref{eq-a1.31}) may prove to be true, and the channeling may be possible. A more strict condition $a k_0 \gg 1$ makes the classical description of this effect possible. This is the case for a long-period structure with $a / a_0 \gg 10^2$ considered above.  It is appropriate to emphasize that we consider here the structures based on the amorphous matrices which are not monocrystals.  Otherwise the long-period structure would have been overlapped by the short-period oscillations with the amplitude equal or greater than $\beta N_0$, which would have lead to a noticeable tunnel effect. The question about channeling possibilities in this case requires a special research.

Thus, the papers \cite{ref-a1.51,ref-a1.52} show that the phenomenon of X-ray and $\gamma$-radiation channeling exists in layered structures under conditions when it is possible to use geometric optics.  The essence of this phenomenon lies in the fact that the rays are reflected form the layers with the higher electron density if they propagate under the small enough angle ($\alpha < \alpha _M$) to the layers plane.  The initially uniform intensity distribution in the transversal plane becomes non-uniform, since the rays concentrate in the "channels" -- the layers with the lower electron density.  It leads to the substantial absorption decrease and deeper radiation penetration into the sample \textit{along} the layers than in the case of an arbitrary arrival angle.

\subsection{On the account of an absorption impact on X-ray intensity when channeling through the solar layered structures}

As it was shown above, the process of $\gamma$-rays channeling through a homogeneous medium with the electron density $N_e$ and the absorption coefficient $\chi _0 = k \zeta \xi ^2$ may be described by the multiplier $J_0 \exp ( - \chi _0 x / \cos \alpha )$ in~(\ref{eq-a1.26}).

Calculation of this multiplier for an arbitrary point in the solar convective zone, obviously, requires the estimation of the average Rosseland free path or so-called Rosseland opacity (the photon absorption coefficient averaged according to Rosseland) in these points. On the other hand, it is necessary to know the radial profiles of the temperature and density in the solar convective zone in order to calculate the Rosseland free path or Rosseland opacity.

\begin{figure}
    \begin{center}
        \includegraphics[width=10cm]{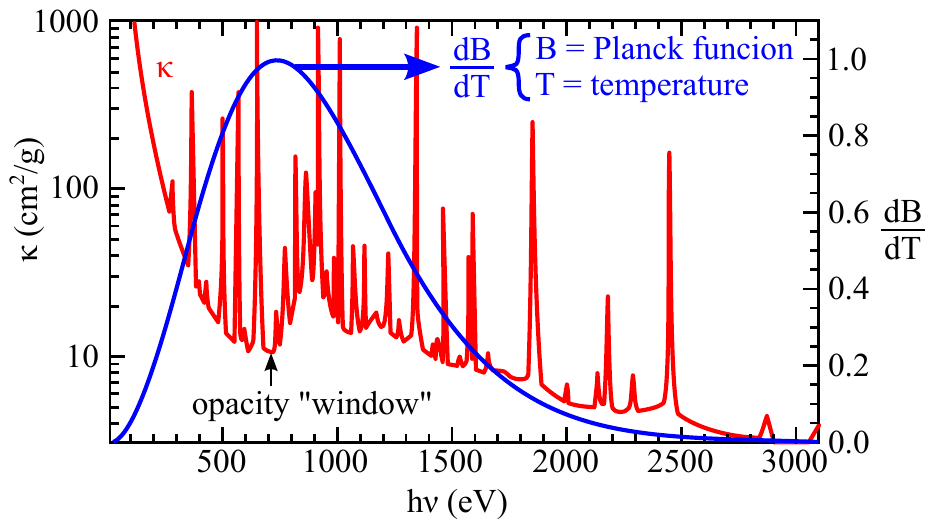}
    \end{center}
\caption{Frequency dependent opacity \cite{ref-a1.12,ref-a1.12a,ref-a1.13} for the 17 element solar composition \cite{ref-a1.14} near the base of the solar convection zone compared to $dB / dT$. The electron temperature and density were 193~eV and $1 \cdot 10^{23} ~cm^{-3}$, respectively. Adopted from \cite{ref-a1.15}.}
\label{fig-a1.3}
\end{figure}

The transmission of photons with the intensity $J_0$ normally incident on a uniform plasma~(\ref{eq-a1.26}) is given by

\begin{equation}
T (\nu ) = J (\nu ) / J_0 (\nu ) = \exp (- \chi_0 x ) = \exp \left[ - k( \nu ) \rho x \right],
\label{eq-a1.32}
\end{equation}

\noindent where $h \nu$ is the photon energy and $J(\nu)$ is the attenuated photon intensity emerging from the plasma, $k(\nu)$ is the opacity per unit mass typically measured in units of $cm^2 / g$, $\rho$ is the density, and $x$ is the optical path length. For plasmas such as the Sun that are much larger than the photon mean free path, radiation transport is usually described by the diffusion approximation \cite{ref-a1.09,ref-a1.10,ref-a1.11} using the Rosseland mean opacity $k_R$,

\begin{equation}
\frac{1}{k_R} = \int d \nu \frac{1}{k ( \nu )} \frac{dB}{dT} \Bigg/ \int d \nu \frac{dB}{dT},
\label{eq-a1.33}
\end{equation}

\noindent where $B$ is the Planck function, $T$ is the plasma temperature, and the weighting function $dB / dT$ peaks at roughly 3.8~kT. Near the convective zone (CZ) boundary $T \sim 190 ~ eV$ and $dB / dT$ peaks at $ h\nu  \sim 750  ~eV$ (Fig.~\ref{fig-a1.3}). The frequency dependent opacity near the CZ boundary calculated using the opacity project model \cite{ref-a1.12,ref-a1.12a,ref-a1.13} is displayed in Fig.~\ref{fig-a1.3}. Comparison with the weighting function for the Rosseland mean shows that the most important photon energies are approximately $300 < h \nu < 1300  ~eV$.

However, it is necessary to point out some essential peculiarities of the absorption impact on the X-ray intensity when it channels inside the solar layered structures based on, e.g. magnetic flux tubes superlattices (see Appendix~\ref{A2}).

\underline{First}, as it was noted in the body text, the total energy balance of the Sun is not violated in the framework of the axion mechanism of Sun luminosity, but the radiation transport changes substantially relative to the standard model of the Sun -- the part of radiation transport not related to axions is very small ($\sim 0.015 \Lambda_{Sun}$). It means that the thermodynamic parameters (temperature, pressure, plasma density, electron density etc.) are \textbf{\textit{considerably smaller}} in the axion model of the Sun as compared to the standard model.  On the other hand, the Rosseland free path or Rosseland opacity calculation inside the thin magnetic flux tubes, which play the role of $\gamma$-quanta waveguides formed in the tachocline, \textbf{\textit{must take into account the new values for the mentioned parameters}} in the framework of the axion mechanism of Sun luminosity.

\underline{Second}, we suppose that the decrease of the pressure in the convective zone, for example, by an order of magnitude may lead to such decrease of the pressure inside the magnetic flux tubes that it virtually \textit{does not influence the radiation transport in these tubes}. In other words, the Rosseland free paths of $\gamma$-quanta in the thin magnetic flux tubes is so big that the corresponding Rosseland opacity tends to zero, and so do the absorption coefficients in~(\ref{eq-a1.32}). The low refractivity, or equally the high transparency of the thin magnetic flux tubes is achieved due to the high magnetic pressure (see~(\ref{eq013}) and Fig.~\ref{fig09}) able to compensate the outer pressure of the convective zone completely (see Appendix~\ref{A2} for details).


\putbib[Axion_luminosity_appendix1-references]
\end{bibunit}

\begin{bibunit}[unsrt]

\section{Appendix II. On a possibility of the layered structures formation in the solar convective zone on the basis of the magnetic flux tubes superlattices}
\label{A2}

It is natural to ask a question, whether there is a physical possibility for the above mentioned long-period structures formation in the magnetohydrodynamical plasma media typical for the solar dynamo evolution. Curiously enough, such mechanisms do exist. Let us examine some of them briefly.

\subsection{Zonal jet streams}

The long-period structures in magnetohydrodynamical plasma media may show as the so-called zonal jet streams, spontaneously generated in turbulent systems. In fact, zonal jets are very common in nature. Well-known examples are those in the atmospheres of giant planets and the alternating jet streams found in the Earth's world ocean~\cite{ref-a2.01}. Zonal flow formation in nuclear fusion devices are also well studied~\cite{ref-a2.02}. 

As we have already pointed out above, a common feature of these zonal flows is that they are spontaneously generated in turbulent systems. Because the Earth's outer core is believed to be in a turbulent state, it is possible that there is a zonal flow in the liquid iron of the outer core. It is interesting that a previously unknown convective regime of the outer core that has a dual structure comprising inner, sheet-like radial plumes and an outer, westward cylindrical zonal flow\footnote{Computer simulations have been playing an important role in the development of our understanding of the geodynamo, but the direct numerical simulation of the geodynamo with a realistic parameter regime is still beyond the power of today's supercomputers. Difficulties in simulating the geodynamo arise from the extreme conditions of the core, which are characterized by very large and very small values of the non-dimensional parameters of the system. Along them, the Ekman number, $E$, has been adopted as a barometer of the distance of simulations from real core conditions, in which $E$ is of the order of 10$^{-15}$. Following the initial computer simulations of the geodynamo, the Ekman number achieved has been steadily decreasing, with recent geodynamo simulations performed with $E$ of the order of 10$^{-6}$~\cite{ref-a2.04}. In work by Miyagoshi~et~al.~\cite{ref-a2.03,ref-a2.04} they present a geodynamo simulation with the Ekman number of the order of 10$^{-7}$ -- the highest resolution yet achieved, making use of 4096 processors of the Earth Simulator. And what is ahead when the magnitude of $E$ becomes closer to its real value?!}, was recently found~\cite{ref-a2.03} (Fig.~\ref{fig06} in the body text).

Fig.~\ref{fig06} in the main body of the present paper shows snapshots of the same data of zonal flow formation in the Earth's core. The sheet convection structure is visualized as isosurfaces of the axial vorticity, which are almost straight in the $z$ direction. The blue curves surrounding the $\omega _z$ isosurfaces are streamlines of the velocity in the outer part of the dual-convection structure. The ring-like shape of each streamline indicates that the azimuthal component is dominant.

However, according to~(\ref{eq-a1.26}) in \ref{A1}, for the ideal (without absorption) photon channeling the long-period structures, in which one of the interlaced media has almost zero density, are necessary. Surprisingly, such long-period (in terms of density) structures may appear in the plasma media in general, and in the solar convective zone in particular. We mean here the so-called magnetic flux tubes, the properties of which are discussed below.

\subsection{Some properties of the magnetic flux tubes in Sun convective zone}

\begin{figure}
    \begin{center}
        \includegraphics[width=10cm]{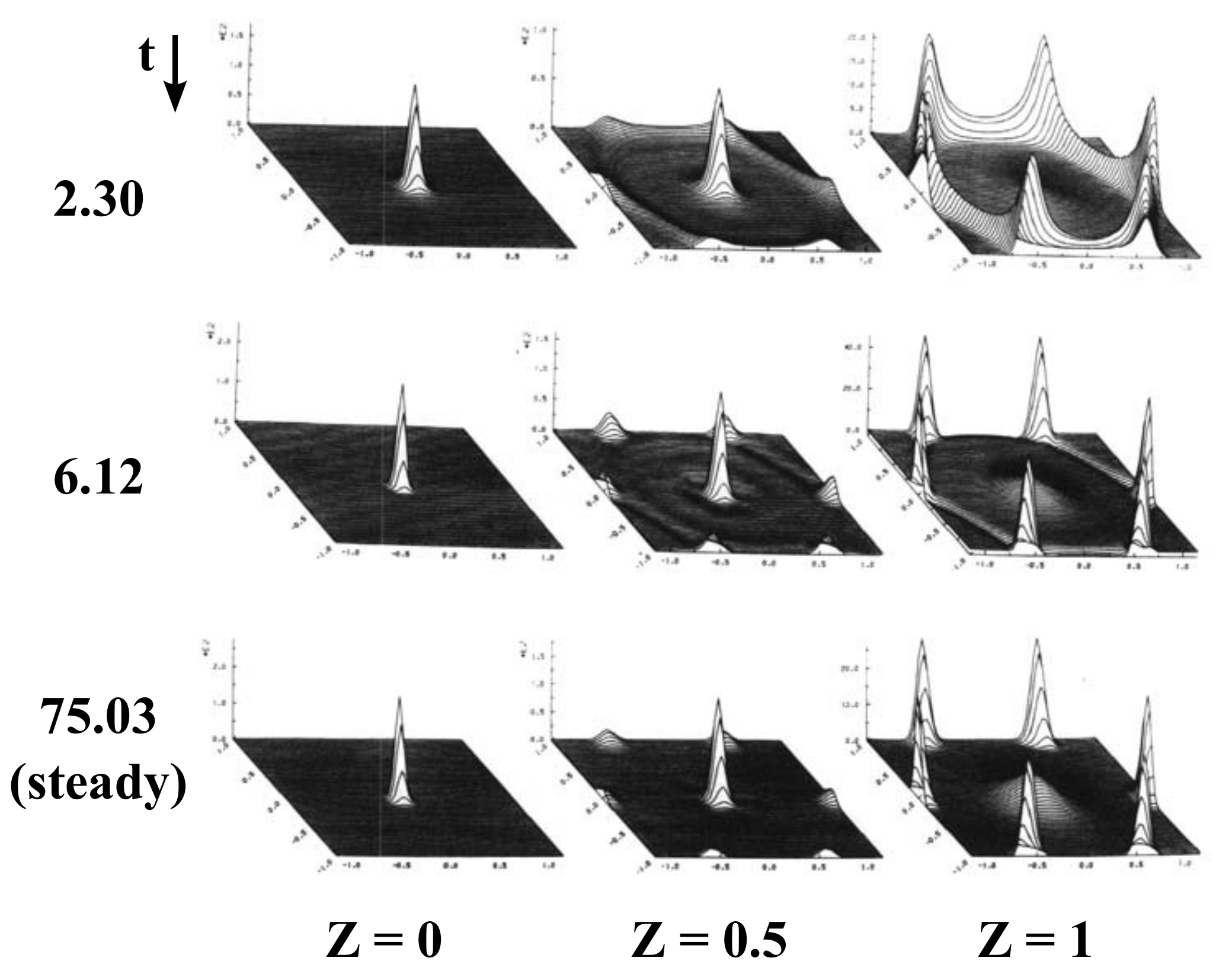}
    \end{center}
\caption{Vertical field component for $R_m = 400$ when the imposed field is vertical, displayed by perspective plots. Four times are shown, for the three levels $z$ = 0, 1/2, and 1. In this and subsequent plots the time unit is (cell height)/(max. vertical velocity), i.e. $\leqslant$ 1.4 turnover time. Adopted from~\cite{ref-a2.05}.}
\label{fig-a2.1}
\end{figure}

Fig.~\ref{fig-a2.1} shows the results of the three-dimensional solar hexagonal magnetoconvection simulation~\cite{ref-a2.05}.  In addition to a mere fact of the long-period layers formation based on the magnetic flux tubes superlattices, it is necessary to make sure that the matter density inside the tubes is much smaller than the density of the outer plasma.

At the same time, as the analysis in Appendix~\ref{A1} shows, for the purposes of photon channeling it is not necessary to guarantee the layers periodicity. In other words, the channeling requires a large number of necessarily interlaced, although not necessarily periodic, layers of different density. The total crosscut area of such interlaced layers must also be comparable to the photon beam cross-section.

In this connection below we shall examine some properties of the magnetic flux tubes in the Sun convective zone, which may serve as a basis for estimating the temperature, pressure and matter density inside the tubes depending on the similar plasma parameters outside the tube.  We shall also derive the zero refractivity (or absolute transparency) condition for the effective photon channeling along the magnetic tubes.

\subsection{The self-confinement of force-free magnetic fields and energy conservation law.}

Magnetic field $\vec{B}$ alternating along the vertical axis $z$ induces the vortex electric field in the magnetic flux tube containing a dense plasma. The charged particles rotation in plasma with the angular velocity $\omega$ leads to a centrifugal force per unit volume

\begin{equation}
\vert \vec{F} \vert \sim \rho \vert \vec{\omega} \vert ^2 r.
\label{eq-a2.01}
\end{equation}

If we consider this problem in the rotating noninertial reference frame, such noninertiality, according to the equivalence principle, is equivalent to "introducing" a radial non-uniform gravitational field with the "free fall acceleration"

\begin{equation}
g (r) = \vert \vec{\omega} \vert ^2 r.
\label{eq-a2.02}
\end{equation}

In such case the pressure difference inside ($p_{int}$) and outside ($p_{ext}$) the rotating "liquid" of the tube may be treated by analogy with the regular hydrostatic pressure (Fig.~\ref{fig-a2.1}).

\begin{figure}
    \begin{center}
        \includegraphics[width=10cm]{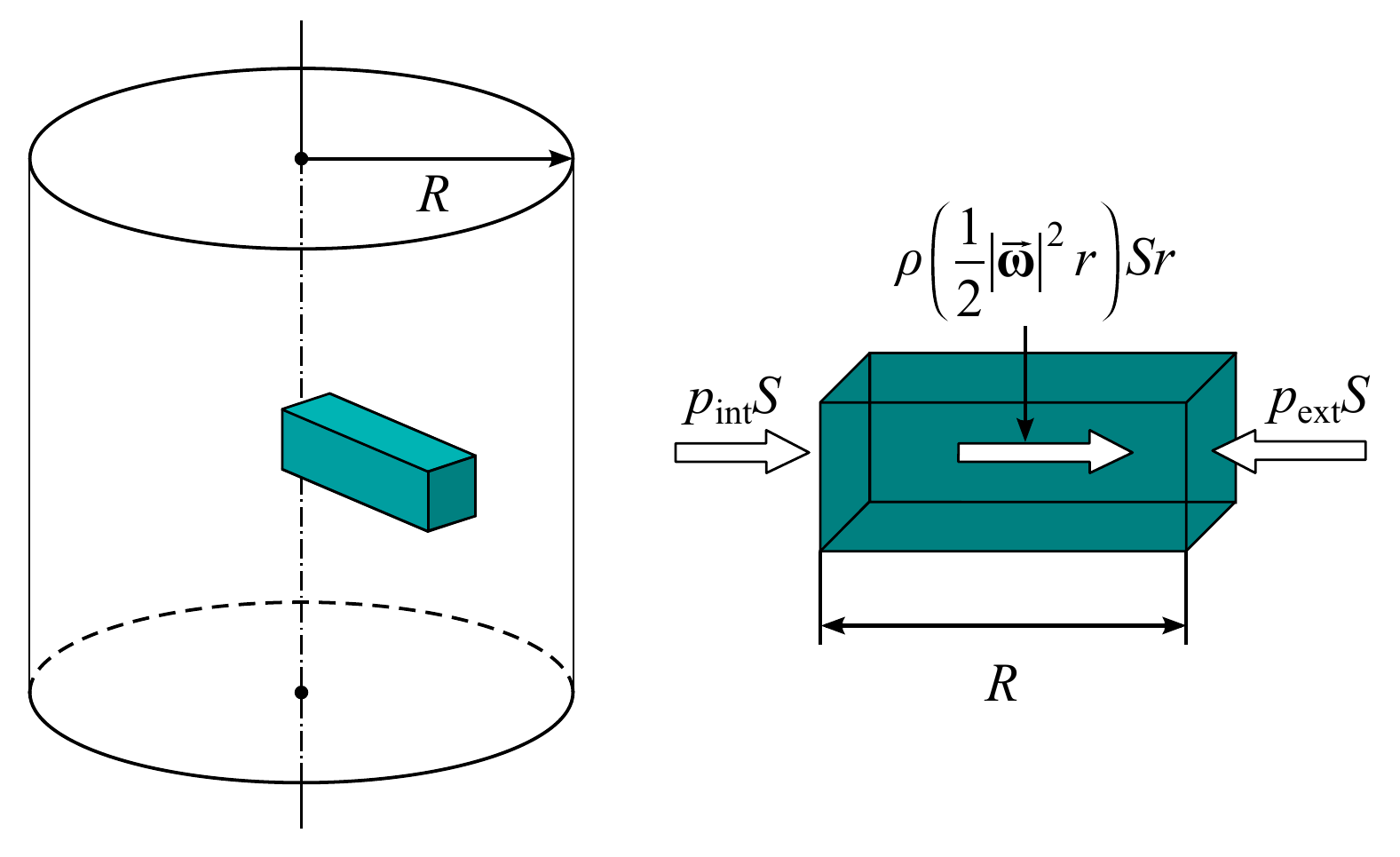}
    \end{center}
\caption{Representation of the "hydrostatic equilibrium" in the rotating "liquid" of a magnetic flux tube.}
\label{fig-a2.2}
\end{figure}

Let us pick a radial "liquid column" inside the tube as it is shown in Fig.\ref{fig-a2.1}. Since the "gravity" is non-uniform in this column, it is equivalent to a uniform field with the "free fall acceleration"

\begin{equation}
\left\langle g(r) \right\rangle = \frac{1}{2} \vert \vec{\omega} \vert ^2 R.
\label{eq-a2.03}
\end{equation}

\noindent where $R$ is the tube radius which plays a role of the "liquid column height" in our analogy.

By equating the forces acting on the chosen column similar to the hydrostatic pressure (Fig.~\ref{fig-a2.1}), we derive:

\begin{equation}
p_{ext} = p_{int} + \frac{1}{2} \rho \vert \vec{\omega} \vert ^2 R^2.
\label{eq-a2.04}
\end{equation}

This rises the natural question as to what physics is hidden behind the "centrifugal" pressure. In this relation, let us consider the magnetic field energy density

\begin{equation}
w _B = \frac{\vert \vec{B} \vert ^2}{2 \mu_0},
\label{eq-a2.05}
\end{equation}

\noindent where $\mu _0$ is the magnetic permeability of vacuum.

Suppose that the total magnetic field energy of the "growing" tube grows linearly between the tachocline and the photosphere. In this case if the average total energy of the magnetic field in the tube transforms into the kinetic energy of tube matter rotation completely, it is easy to show that

\begin{equation}
\left\langle E_B \right\rangle = \frac{1}{2} w_B V = \frac{1}{2} \frac{\vert \vec{B} \vert ^2}{2 \mu_0} V = \frac{I \vert \vec{\omega} \vert ^2}{2},
\label{eq-a2.06}
\end{equation}

\noindent where $I = mR^2 / 2$ is the tube's moment of inertia about the rotation axis, $m$ and $V$ are the mass and the volume of the tube medium respectively.

Then from~(\ref{eq-a2.06}) it follows that

\begin{equation}
\frac{\rho \vert \vec{\omega} \vert R^2}{2} = \frac{\vert \vec{B} \vert ^2}{2 \mu_0}.
\label{eq-a2.07}
\end{equation}

Finally, substituting~(\ref{eq-a2.07}) into~(\ref{eq-a2.04}) we obtain the desired relation

\begin{equation}
p_{ext} = p_{int} + \frac{\vert \vec{B} \vert ^2}{2 \mu_0},
\label{eq-a2.08}
\end{equation}

\noindent which is exactly equal to a well-known expression by Parker~\cite{ref-a2.06}, describing the so-called self-confinement of force-free magnetic fields.

\subsection{Hydrostatic equilibrium and a sharp tube medium cooling effect}

Assume that the tube has the length $l(t)$ by the time $t$. Then its volume is equal to $S \cdot l(t)$ and the heat capacity is

\begin{equation}
c \cdot \rho (t) \cdot S l (t),
\label{eq-a2.09}
\end{equation}

\noindent where $S$ is the tube cross-section, $\rho (t)$ is the density inside the tube, $c$ is the specific heat capacity.

If the tube becomes longer by $\upsilon (t) dt$ for the time $dt$, then the magnetic field energy increases by

\begin{equation}
\frac{1}{2} \frac{\vert \vec{B} \vert}{2 \mu_0} S \upsilon (t) dt .
\label{eq-a2.10}
\end{equation}

\noindent where $\upsilon (t)$ is the tube propagation speed.

The matter inside the tube, obviously, has to cool by the temperature $dT$ so that the internal energy release maintained the magnetic energy growth. Therefore, the following equality must hold:

\begin{equation}
c \rho (t) l (t) \frac{dT}{dt} S = -\frac{1}{2} \frac{\vert \vec{B} \vert ^2}{2 \mu_0} \upsilon (t) S.
\label{eq-a2.11}
\end{equation}

Taking into account the fact that the tube grows practically linearly~\cite{ref-a2.07}, i.e. $\upsilon t = l$, the Parker relation~(\ref{eq-a2.08}) and the tube's equation of state

\begin{equation}
\nonumber p_{int} (t) = \frac{\rho}{\mu_{*}} R_{*} T(t) ~~ \Longleftrightarrow ~~ \rho = \frac{\mu_{*}}{R_{*}} \frac{p_{int}}{T(t)},
\end{equation}

\noindent the equality~(\ref{eq-a2.11}) may be rewritten (by separation of variables) as follows:

\begin{equation}
\frac{dT}{T} = - \frac{R_{*}}{2 c \mu_{*}} \left[ \frac{p_{ext}}{p_{int} (t)} - 1 \right] \frac{dt}{t},
\label{eq-a2.12}
\end{equation}

\noindent where $\mu_{*}$ is the tube matter molar mass, $R_{*}$ is the universal gas constant.

After integration of~(\ref{eq-a2.12}) we obtain

\begin{equation}
\ln \left[ \frac{T (t)}{T(0)} \right] = - \frac{R_{*}}{2 c \mu_{*}} \int \limits _0 ^t \left[ \frac{p_{ext}}{p_{int} (\tau)} - 1 \right] \frac{d\tau}{\tau}.
\label{eq-a2.13}
\end{equation}

It is easy to see that the multiplier ($1 / \tau$) in~(\ref{eq-a2.13}) assigns the region near $\tau = 0$ in the integral. The integral converges since

\begin{equation}
\nonumber \lim \limits _{\tau \rightarrow 0} \left[ \frac{p_{ext}}{p_{int} (\tau)} - 1 \right] = 0.
\end{equation}

Expanding $p_{int}$ into Taylor series and taking into account that $p_{int}(\tau = 0) = p_{ext}$, we obtain

\begin{equation}
p_{int} (\tau ) = p_{ext} + \frac{d p_{int} (\tau = 0)}{d \tau} = p_{ext} (1 - \gamma \tau ),
\label{eq-a2.14}
\end{equation}

\noindent where

\begin{equation}
\gamma = - \frac{1}{p_{ext}} \frac{d p_{int} (\tau = 0)}{d \tau} = \frac{1}{p_{int} (\tau = 0)} \frac{d p_{int} (\tau = 0)}{d \tau}.
\label{eq-a2.15}
\end{equation}

From~(\ref{eq-a2.14}) it follows that

\begin{equation}
\frac{p_{ext}}{p_{int} ( \tau )} - 1 = \frac{1}{1 - \gamma \tau} - 1 \approx \gamma \tau ,
\label{eq-a2.16}
\end{equation}

\noindent therefore, substituting~(\ref{eq-a2.16}) into~(\ref{eq-a2.13}), we find its solution in the form

\begin{equation}
T (t) = T(0) \exp \left( - \frac{R_{*}}{2 c \mu _{*}} \gamma t \right).
\label{eq-a2.17}
\end{equation}

It is extremely important to note here that the solution~(\ref{eq-a2.17}) points at the remarkable fact that at least at the initial stages of the tube formation its temperature \textbf{\textit{decreases exponentially}}, i.e. \textbf{\textit{very sharply}}. The same conclusion can be made for the pressure and the matter density in the tube. Let us show it.

Assuming that the relation~(\ref{eq-a2.15}) holds not only for $\tau = 0$, but also for small $\tau$ close to zero, we obtain

\begin{equation}
p_{int} (t) = p_{int} (0) \exp ( - \gamma t ),
\label{eq-a2.18}
\end{equation}

\noindent i.e. the inner pressure also decreases exponentially, but with the different exponential factor.

Further, assuming that the heat capacity of a molecule in the tube is

\begin{equation}
c = \frac{i}{2} \frac{R_{*}}{\mu _{*}}
\label{eq-a2.19}
\end{equation}

\noindent where $i$ is a number of molecule's degrees of freedom, and substituting the expressions~(\ref{eq-a2.17}) and~(\ref{eq-a2.18}) into the tube's equation of state, we derive the expression for the matter density in the tube

\begin{equation}
\rho (t) = \frac{\mu _{*}}{R_{*}} \frac{p_{int} (0)}{T(0)} \exp \left[ - \left( 1 - \frac{1}{i} \right) \gamma t \right],
\label{eq-a2.20}
\end{equation}

Taking into account that $i \geqslant 3$, we can see that the density decreases exponentially just like the temperature (\ref{eq-a2.17}) and pressure (\ref{eq-a2.18}).

\subsection{Ideal photon channeling (without absorption) conditions inside the magnetic flux tubes}

Let us calculate the inner pressure $p_{int}$ of the magnetic flux tube in the solar tachocline. In the framework of the standard model of the Sun the pressure in the tachocline zone is about $\sim 6 \cdot 10^{12} ~Pa$, while the magnetic field strength reaches $400 ~T$, according to our estimates (Fig.~\ref{fig09} in the body text). Then from the hydrostatic condition by Parker~(\ref{eq-a2.08}) it follows that the inner pressure of the magnetic flux tube in the solar tachocline is equal to

\begin{equation}
p_{int} \sim 6 \cdot 10^{12} - 4 \cdot 10^{10} \simeq 6 \cdot 10^{12} ~~ [Pa],
\label{eq-a2.21}
\end{equation}

\noindent which is comparable to the external pressure.

However, the situation changes drastically within the framework of the "axion" model of the Sun. We pointed out earlier (\ref{A1}) that the values of thermodynamic parameters (temperature, pressure and plasma density) in the "axion" model are \textbf{\textit{substantially smaller}} than the corresponding parameters in the standard model of the Sun. Let us suppose that the pressure in the tachocline zone in the "axion" model falls by an order of magnitude and is about $ \sim 10^{11} ~Pa$. In such case it is easy to see that the magnetic field as strong as $\sim 500 ~T$ compensates the outer pressure almost completely, and the inner pressure

\begin{equation}
p_{int} \sim 10^{11} - O (10^{11}) \rightarrow 0 ~~ [Pa]
\label{eq-a2.22}
\end{equation}

\noindent becomes ultralow.

The result~(\ref{eq-a2.22}) means that  the temperature and the matter density decrease together with the inner pressure, and the decrease is sharply exponential, because the exponential factor in~(\ref{eq-a2.16}) becomes very large

\begin{equation}
\gamma \tau  =  \frac{p_{ext}}{p_{int} (\tau )} - 1 \rightarrow \infty .
\label{eq-a2.23}
\end{equation}

Because of the fact that the density, pressure and temperature in the tube are ultralow, they virtually do not influence the radiation transport in these tubes at all. In other words, the Rosseland free paths for $\gamma$-quanta inside the tubes are so big that the Rosseland opacity and the absorption coefficients in~(\ref{eq-a1.32}) tend to zero. The low refractivity (or high transparency) is achieved for the following limiting condition:

\begin{equation}
p_{ext} \simeq \frac{\vert \vec{B} \vert ^2}{2 \mu_0}.
\label{eq-a2.24}
\end{equation}

The obtained results should not be considered as a proof, but rather as some trial estimates that validate the substantiation of an almost ideal (without absorption) photon channeling mechanism along the magnetic flux tubes.

A major disadvantage of our reasoning is that, first of all, our estimates apply to the initial magnetic tube formation stages, and second, that it lacks the model calculations of the temperature, pressure and density in the framework of the "axion" model of the Sun, and consequently, there is no comparison with the corresponding parameters of the standard model. On the other hand, we believe that the idea of $\gamma$-quanta channeling along the magnetic flux tubes is so physically natural and promising, that these disadvantages will be overcome in the near future in spite of the obvious serious difficulties.

\putbib[Axion_luminosity_appendix2-references]
\end{bibunit}

\end{document}